\documentclass[aps,rmp,twocolumn,eqsecnum,amssymb]{revtex4}

\usepackage{graphicx}
\usepackage{dcolumn}
\usepackage{bm}
\usepackage{amsmath,amssymb,amsthm,amscd}

\begin{document}
\preprint{CERN-PH-TH/}
\title{Chern-Simons Theory and Topological Strings }
\author{Marcos Mari\~no}
 \altaffiliation[Also at ]{Departamento de Matem\'atica, IST, Lisboa, Portugal}
\email{marcos@mail.cern.ch}
\affiliation{
Department of Physics, CERN, Geneva 23, Switzerland
}


\begin{abstract}
We review the relation between Chern-Simons gauge theory and
topological string theory on noncompact Calabi-Yau spaces. This relation has made possible
to give an exact solution of topological string theory on these spaces to all orders in the string
coupling constant. We focus on the construction of this solution, which is encoded in the
topological vertex, and we emphasize the implications of the physics of string/gauge theory
duality for knot theory and for the geometry of Calabi-Yau manifolds.
\end{abstract}

\pacs{Valid PACS appear here}
\maketitle
\tableofcontents

\newcommand{\eps}{\epsilon}
\newcommand{\ra}{\rangle}
\newcommand{\la}{\langle}
\newcommand{\T}{\chi_{T}(k)}
\newcommand{\Tm}{\chi_{T}(k')}
\newcommand{\Cn}{{\cal C}_n}
\newcommand{\vp}{\varphi}
\newcommand{\ve}{\varepsilon}
\newcommand{\tl}{\wt\lambda}

\newcommand{\vv} {\bar v}
\newcommand{\uu} {\bar u}
\newcommand{\IB}{\relax{\rm I\kern-.18em B}}
\newcommand{\DE}{\relax{\rm I\kern-.18em E}}
\newcommand{\IP}{\relax{\rm I\kern-.18em P}}
\newcommand{\IR}{\relax{\rm I\kern-.18em R}}
\newcommand{\B}{b'}
\newcommand{\C}{c'}
\newcommand{\bB}{\bar b'}
\newcommand{\bC}{\bar c'}
\newcommand{\VV}{{\cal V}}
\newcommand{\BB}{{\cal B}}
\newcommand{\II}{{\cal I}}
\newcommand{\GG}{{\cal G}}
\newcommand{\HH}{{\cal H}}
\newcommand{\MM}{{\cal M}}
\newcommand{\CC}{{\cal C}}
\newcommand{\CL}{{\cal L}}
\newcommand{\CH}{{\cal H}}
\newcommand{\CO}{{\cal O}}
\newcommand{\CQ}{{\cal Q}}
\newcommand{\PP}{{\cal P}}
\newcommand{\EE}{{\cal E}}
\newcommand{\LL}{{\cal L}}
\newcommand{\Tr}{{\rm Tr}}
\newcommand{\CK}{{\cal K}}

\newcommand{\be}{\begin{equation}}
\newcommand{\ee}{\end{equation}}
\newcommand{\ben}{\begin{eqnarray}\displaystyle}
\newcommand{\een}{\end{eqnarray}}

\newdimen\tableauside\tableauside=1.0ex
\newdimen\tableaurule\tableaurule=0.4pt
\newdimen\tableaustep
\def\phantomhrule#1{\hbox{\vbox to0pt{\hrule height\tableaurule width#1\vss}}}
\def\phantomvrule#1{\vbox{\hbox to0pt{\vrule width\tableaurule height#1\hss}}}
\def\sqr{\vbox{%
  \phantomhrule\tableaustep
  \hbox{\phantomvrule\tableaustep\kern\tableaustep\phantomvrule\tableaustep}%
  \hbox{\vbox{\phantomhrule\tableauside}\kern-\tableaurule}}}
\def\squares#1{\hbox{\count0=#1\noindent\loop\sqr
  \advance\count0 by-1 \ifnum\count0>0\repeat}}
\def\tableau#1{\vcenter{\offinterlineskip
  \tableaustep=\tableauside\advance\tableaustep by-\tableaurule
  \kern\normallineskip\hbox
    {\kern\normallineskip\vbox
      {\gettableau#1 0 }%
     \kern\normallineskip\kern\tableaurule}%
  \kern\normallineskip\kern\tableaurule}}
\def\gettableau#1{\ifnum#1=0\let\next=\null\else
\squares{#1}\let\next=\gettableau\fi\next}

\tableauside=1.0ex
\tableaurule=0.4pt
\def\IE{\mathbb{E}}

%
\newcommand{\twoVgraph}{\raisebox{0pt}{
                 \begin{picture}(18,18)(-9,-5)
                 \put(0,0){\circle{16}} \put(-8,0){\line(1,0){16}}
                 \end{picture}}}

\newcommand{\fourVgraph}{\raisebox{0pt}{
                 \begin{picture}(18,26)(-9,-9)
                 \put(0,0){\oval(16,24)} \put(-8,4){\line(1,0){16}}
                 \put(-8,-4){\line(1,0){16}}
                 \end{picture}}}

\newcommand{\sixVgraph}{\raisebox{0pt}{
                 \begin{picture}(26,24)(-13,-8)
                 \put(-9,8){\circle{6}} \put(9,8){\circle{6}}
                 \put(-6,8){\line(1,0){12}} \put(0,-8){\circle{6}}
                 \put(-9,5){\line(2,-3){7}} \put(9,5){\line(-2,-3){7}}
                 \end{picture}}}

\newcommand{\eightVgraphI}{\raisebox{0pt}{
                 \begin{picture}(26,26)(-13,-9)
                 \put(-9,9){\circle{6}} \put(9,9){\circle{6}}
                 \put(-9,-9){\circle{6}} \put(9,-9){\circle{6}}
             \put(-6,9){\line(1,0){12}}
                 \put(-9,6){\line(0,-1){12}}
                 \put(-6,-9){\line(1,0){12}}
                 \put(9,6){\line(0,-1){12}}
                 \end{picture}}}

\newcommand{\eightVgraphII}{\raisebox{0pt}{
                 \begin{picture}(28,28)(-14,-10)
                 \put(-13,13){\line(1,0){26}}
                 \put(-13,-13){\line(1,0){26}}
                 \put(-13,-13){\line(0,1){26}}
                 \put(13,-13){\line(0,1){26}}
                 \put(-3,3){\line(1,0){6}}
                 \put(-3,-3){\line(1,0){6}}
                 \put(-3,-3){\line(0,1){6}}
                 \put(3,-3){\line(0,1){6}}
                 \put(-13,13){\line(1,-1){10}}
                 \put(-13,-13){\line(1,1){10}}
                 \put(13,13){\line(-1,-1){10}}
                 \put(13,-13){\line(-1,1){10}}
                 \end{picture}}}

\newcommand{\tenVgraphI}{\raisebox{0pt}{
                 \begin{picture}(26,29)(-13,-9)
                 \put(-9,9){\circle{6}} \put(9,9){\circle{6}}
                 \put(-9,-9){\circle{6}} \put(9,-9){\circle{6}}
                 \put(0,9){\circle{6}}
                 \put(-6,9){\line(1,0){3}}
                 \put(6,9){\line(-1,0){3}}
                 \put(-9,6){\line(0,-1){12}}
                 \put(-6,-9){\line(1,0){12}}
                 \put(9,6){\line(0,-1){12}}
                 \end{picture}}}

\newcommand{\tenVgraphII}{\raisebox{0pt}{
                 \begin{picture}(28,31)(-14,-10)
                 \put(-13,13){\line(1,0){10}}
                 \put(0,13){\circle{6}}
                 \put(13,13){\line(-1,0){10}}
                 \put(-13,-13){\line(1,0){26}}
                 \put(-13,-13){\line(0,1){26}}
                 \put(13,-13){\line(0,1){26}}
                 \put(-3,3){\line(1,0){6}}
                 \put(-3,-3){\line(1,0){6}}
                 \put(-3,-3){\line(0,1){6}}
                 \put(3,-3){\line(0,1){6}}
                 \put(-13,13){\line(1,-1){10}}
                 \put(-13,-13){\line(1,1){10}}
                 \put(13,13){\line(-1,-1){10}}
                 \put(13,-13){\line(-1,1){10}}
                 \end{picture}}}

\section{\label{sec:level1}Introduction}

Even though string theory has not found yet a clear place in our
understanding of Nature, it has already established itself as a source
of fascinating results and research directions in mathematics. In recent
years, string theory and some of its
close cousins (like conformal field theory and topological field theory) have had
an enormous impact in representation theory, differential geometry, low-dimensional
topology, and algebraic geometry.

One mathematical area which has been deeply influenced by conformal field theory
and topological field theory is knot theory. Witten (1989) found that many topological
invariants of knots and links
discovered in the 1980s (like the Jones and the HOMFLY polynomials) could be
reinterpreted as correlation functions of Wilson loop operators in Chern-Simons
theory, a gauge theory in three dimensions with topological invariance.
Witten also showed that the partition function of this theory
provided a new topological invariant of three-manifolds, and by
working out the exact solution of Chern-Simons gauge theory he made a connection
between these knot and three-manifold invariants and
conformal field theory in two dimensions (in particular, the
Wess-Zumino-Witten model).

In a seemingly unrelated development, it was found that the study of
string theory on Calabi-Yau manifolds (which was triggered by the
phenomenological interest of the resulting four-dimensional models) provided new insights in the
geometry of these spaces. Some correlation functions
of string theory on Calabi-Yau manifolds turn out to compute numbers of
holomorphic maps from the string worldsheet to the target, therefore they
contain information about the
enumerative geometry of the Calabi-Yau spaces.
This led to the introduction of Gromov-Witten
invariants in mathematics as a way to capture this information. Moreover, the
existence of a powerful duality symmetry of string theory in Calabi-Yau spaces
-mirror symmetry- allowed the computation of generating functions
for these invariants, and made possible to solve with physical techniques difficult
enumerative problems (see Hori {\it et al.}, 2003, for a review of these
developments). The existence of a topological sector in string theory
which captured the enumerative geometry of the target space led also to
the construction of simplified models of string theory
which kept only the topological information of the more complicated,
physical theory. These models are called topological string theories and
turn out to provide in many cases exactly solvable models of string dynamics.

The key idea that allowed to build a bridge between topological string
theory and Chern-Simons theory was the gauge theory/string theory
correspondence. It is an old idea, going back to `t Hooft (1974) that gauge theories
can be described in the $1/N$ expansion by string theories. This idea has been difficult to
implement, but in recent years some spectacular progress was made
thanks to the work of Maldacena (1998), who found a duality between type IIB string
theory on ${\rm AdS}_5 \times {\bf S}^5$ and ${\cal N}=4$ super Yang-Mills with gauge
group $U(N)$. It is then natural to ask if gauge theories which are simpler than ${\cal N}=4$
Yang-Mills -like for example Chern-Simons theory-
also admit a string theory description. It was shown by Gopakumar and Vafa (1999) that
Chern-Simons gauge theory on the three-sphere
has in fact a closed string description in terms of topological
string theory propagating on a particular Calabi-Yau target, the so-called resolved conifold.

The result of Gopakumar and Vafa has three important consequences. First of all, it
provides a toy model of the gauge theory/string theory correspondence which
makes possible to test in detail general ideas about this duality. Second, it gives
a stringy interpretation of invariants of knots in the three-sphere. More
precisely, it establishes a relation between invariants of knots based on quantum
groups, and Gromov-Witten invariants of open strings propagating
on the resolved conifold. These are {\it a priori} two very different mathematical objects, and in this
way the physical idea of a correspondence between gauge theories and strings gives
new and fascinating results in mathematics that we are only starting to unveil.
Finally, one can use the results of Gopakumar and Vafa to completely solve
topological string theory on certain Calabi-Yau threefolds in a closed form. As we will see,
this gives the all-genus answer for certain string amplitudes, and it is in fact one of the
few examples in string theory where such an answer is available.
The all-genus solution to the amplitudes
also encodes the information about all the Gromov-Witten invariants for those threefolds. Since the solution
involves building blocks from Chern-Simons theory, it suggests yet another bridge between
knot invariants and Gromov-Witten theory.

In this review we will focus on this last aspect. The organization of the review is the following: in section II we
give an introduction to the relevant aspects of Chern-Simons theory that will be needed for the
applications to Calabi-Yau geometry. In particular, we give detailed results for the computation
of the relevant knots and link invariants. In section III we give a short review on the $1/N$ expansion
of Chern-Simons theory, which is the approach that makes possible the connection to string theory.
Section IV contains a review of closed and open topological string theory on Calabi-Yau threefolds, and we
construct in full detail the geometry of non-compact, toric Calabi-Yau spaces, since these are the
manifolds that we will be able to study by using the gauge theory/string theory correspondence.
In section V we establish the correspondence between Chern-Simons theory on the three-sphere
and closed topological string theory on a resolved conifold. In section VI we show how the arguments
of section V can be extended to construct gauge theory duals of topological string theory on more
complicated non-compact, toric Calabi-Yau manifolds. In section VII we complete this program by
defining the topological vertex, an object that allows to solve topological string
theory on all non-compact, toric Calabi-Yau threefolds by purely combinatorial methods. We also give
a detailed derivation of the topological vertex from Chern-Simons theory, and we give various
applications of the formalism. The last section contains some conclusions and open directions for further research.
A short Appendix contains some elementary facts about the theory of symmetric polynomials that are used in the review.

There are many issues that we have not analyzed in detail in this review. For example, we have not discussed
the mirror-symmetric side of the story, and we do not address in detail the relation between topological
string amplitudes and type II superstring amplitudes. We refer the reader to the excellent book by Hori {\it et al.} (2003)
for an introduction to these topics. Other reviews of the topics discussed here can be found in 
Grassi and Rossi (2002) and Mari\~no (2002b).

\section{Chern-Simons theory and knot invariants}

\subsection{\label{sec:level2}Chern-Simons theory: basic ingredients}

In a groundbreaking paper, Witten (1989) showed that Chern-Simons gauge theory, which
is a quantum field theory in three dimensions, provides a physical
description of a wide class of invariants of three-manifolds
and of knots and links in three-manifolds\footnote{This was also conjectured in Schwarz (1987).}. The Chern-Simons action with gauge group $G$
on a generic three-manifold $M$ is defined by
\begin{equation}
S={k \over 4\pi} \int_M {\rm Tr} \Bigl( A\wedge d A + {2 \over 3} A
\wedge A \wedge A \Bigr)
\label{csact}
\end{equation}
Here, $k$ is the coupling constant, and $A$ is a $G$-gauge connection on the
trivial bundle over $M$. In this review we will mostly consider Chern-Simons
theory with gauge group $G=U(N)$. As noticed by Witten (1989),
since this action does not involve the metric, the resulting
quantum theory is topological, at least formally. In particular,
the partition function
\begin{equation}
Z (M)= \int [{\cal D} A]  {\rm e}^{iS}
\label{partcs}
\end{equation}
should define a topological invariant of the manifold $M$.
A detailed analysis shows
that this is in fact the case, with an extra subtlety: the invariant
depends not only on the three-manifold but also on a choice of framing ({\it i.e.} a
choice of trivialization of the bundle $TM \oplus TM$). As explained by Atiyah (1990),
for every three-manifold there is a canonical choice of framing, and the
different choices are labelled by an integer $s \in {\bf Z}$ in such a way
that $s=0$ corresponds to the canonical framing. In the following all the
results for the partition functions
will be presented in the canonical framing.

\begin{figure}
\scalebox{.6}{\includegraphics{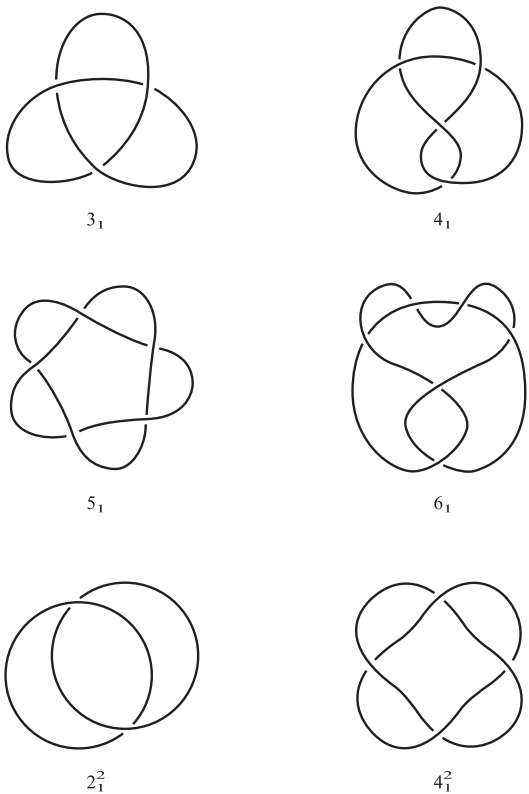}}
\caption{\label{nudosf}Some knots and links. In the notation $x_n^L$, $x$
indicates the number of crossings, $L$ the number of components (in case it
is a link with $L>1$) and $n$ is a number used to enumerate knots and links
in a given set characterized by $x$ and $L$. The knot $3_1$ is also known
as the trefoil knot, while $4_1$ is known as the figure-eight knot. The
link $2_1^2$ is called the Hopf link. }
\end{figure}

\begin{figure}
\scalebox{.4}{\includegraphics{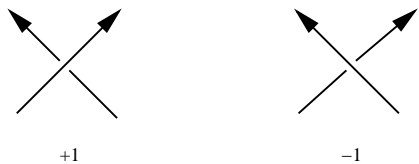}}
\caption{\label{linkingf}When computing the linking number of two knots, the
crossings are assigned a sign $\pm 1$ as indicated in the figure.}
\end{figure}

Besides providing invariants of three-manifolds, Chern-Simons theory also
provides invariants of knots and links inside three-manifolds
(for a survey of modern knot theory, see Lickorish (1998) and Prasolov and
Sossinsky (1997)). Some examples
of knots and links are depicted in Fig. \ref{nudosf}.
Given an oriented knot ${\cal K}$ in ${\bf S}^3$, we can
consider the trace of the holonomy of the gauge connection around
$\cal K$ in a given irreducible representation $R$ of $U(N)$, which gives
the Wilson loop operator:
\begin{equation}
W^{\cal K}_R(A)={\rm Tr}_R U_{\cal K}
\label{wilson}
\end{equation}
where
\begin{equation}
U_{\cal K}= {\rm P}\,\exp\, \oint_{\cal K} A
\end{equation}
is the holonomy around the knot.
(\ref{wilson}) is a gauge invariant operator whose definition does not
involve the metric on the three-manifold. The irreducible representations
of $U(N)$ will be labelled by highest weights or equivalently by
the lengths of rows in a Young tableau, $l_i$,
where $l_1 \ge l_2 \ge \cdots$.
If we now consider a link ${\cal L}$ with components ${\cal K}_{\alpha}$, $\alpha=1,
\cdots, L$, we can in principle compute the correlation function,
\ben
W_{R_1 \cdots R_L}({\cal L})&=&\langle W^{{\cal K}_1}_{R_1}\cdots
W^{{\cal K}_L}_{R_L}\rangle \nonumber\\&=&
{1\over Z(M)}\int [{\cal D} A] \Bigl( \prod_{\alpha=1}^L W_{R_{\alpha}}^{{\cal K}_{\alpha}}
\Bigr) {\rm e}^{iS}.
\label{vevknot}
\een
The topological character of the action, and the fact that the Wilson loop
operators can be defined without using any metric on the three-manifold,
indicate that (\ref{vevknot}) is a topological
invariant of the link ${\cal L}$. Notice that we are taking the knots and links to be
oriented, and this makes a difference. If ${\cal K}^{-1}$ denotes
the knot obtained from ${\cal K}$ by inverting its orientation, we have that
\be
\label{reorient}
{\rm Tr}_R U_{{\cal K}^{-1}}={\rm Tr}_R U^{-1}_{\cal K} ={\rm Tr}_{\overline R}
U_{\cal K},
\end{equation}
where ${\overline R}$ denotes the conjugate representation. For further use
we notice that, given two linked oriented knots ${\cal K}_1$, ${\cal K}_2$, one
can define a elementary topological
invariant, the {\it linking number}, by
\be
\label{lknumber}
{\rm lk}({\cal K}_1, {\cal K}_2)={1\over 2} \sum_{\rm p} \epsilon (p),
\end{equation}
where the sum is over all crossing points, and $\epsilon(p)=\pm1$ is a sign associated
to the crossings as indicated in Fig. ~\ref{linkingf}. The linking number of a link
${\cal L}$ with components ${\cal K}_{\alpha}$, $\alpha=1, \cdots, L$, is defined by
\be
{\rm lk}({\cal L})=\sum_{\alpha<\beta} {\rm lk}({\cal K}_\alpha, {\cal K}_\beta).
\ee

Some of the correlation functions of Wilson loops in Chern-Simons theory turn out
to be closely related to important polynomial invariants of knots and links.
For example, one of the most important polynomial invariants of a link ${\cal L}$ is
the HOMFLY polynomial $P_{\cal L}(q, \lambda)$, which depends on two variables
$q$ and $\lambda$ and was introduced by Freyd {\it et al.} (1985). This polynomial turns out to
be related to the correlation function (\ref{vevknot}) when the gauge group
is $U(N)$ and all the components are in the fundamental representation $R_{\alpha}=\tableau{1}$.
More precisely, we have
\begin{equation}
W_{\tableau{1} \cdots \tableau{1}}({\cal L})=
\lambda^{{\rm lk}({\cal L})}
\biggl( {\lambda^{1\over 2} -\lambda^{-{1\over 2}}
\over q^{1\over 2} -q ^{-{1\over 2}}} \biggr) P_{\cal L}(q, \lambda)
\label{homflyrel}
\end{equation}
where ${\rm lk}({\cal L})$ is the linking number of ${\cal L}$, and the variables
$q$ and $\lambda$ are related to the Chern-Simons variables as
\be
q= \exp \biggl( {2 \pi i \over k+N}\biggr),
\quad \lambda=q^N.
\label{polvars}
\ee
When $N=2$ the HOMFLY polynomial reduces to a one-variable
polynomial, the Jones polynomial. When the gauge group of Chern-Simons
theory is $SO(N)$, $W_{\tableau{1} \cdots \tableau{1}}(\CL)$ is closely related
to the Kaufmann polynomial.
For the mathematical definition and properties of these polynomials, see
for example Lickorish (1998).

\subsection{Perturbative approach}

The partition function and correlation functions of Wilson loops in
Chern-Simons theory can be computed in a variety
of ways. One can for example use standard perturbation theory.
In the computation of the partition function in perturbation theory, we
have to find first the classical solutions of the Chern-Simons equations of
motion. If we write $A=\sum_a A^a T_a$, where $T_a$ is a basis of the Lie algebra, we find
$$
{\delta S \over \delta A^a_{\mu}}={k\over 4 \pi}
\epsilon^{\mu \nu \rho}F^a_{\nu \rho},$$
therefore
the classical solutions are just flat connections on $M$. Flat connections are in
one-to-one correspondence with group homomorphisms
\be
\pi_1(M) \rightarrow G.
\end{equation}
For example, if $M={\bf S}^3/{\bf Z}_p$ is the lens space $L(p,1)$, one has
$\pi_1(L(p,1))={\bf Z}_p$, and flat connections are labeled by homomorphisms
${\bf Z}_p \rightarrow G$. Let us assume that
these are a discrete set of points (this happens, for example, if $M$ is a
rational homology sphere, since in that case $\pi_1(M)$ is a finite group).
In that situation, one expresses $Z(M)$ as a
sum
of terms associated to stationary points:
\begin{equation}
Z (M)= \sum_c Z^{(c)}(M),
\end{equation}
where $c$ labels the different flat connections $A^{(c)}$ on $M$.
Each of the $Z^{(c)}(M)$ will be an asympotic series in $1/k$ of the form
\begin{equation}
\label{perts}
Z^{(c)}(M)=Z^{(c)}_{\rm 1-loop}(M). \exp \Biggl\{ \sum_{\ell=1}^\infty
S^{(c)}_\ell  x^\ell
\Biggr\}.
\end{equation}
In this equation, $x$ is the effective expansion parameter:
\begin{equation}
\label{coupling}
x = { 2 \pi i \over k+y},
\end{equation}
and $y$ is the dual Coxeter number of the group (for $G=U(N)$, $y=N$).
The one-loop correction $Z^{(c)}_{\rm 1-loop}(M)$ was first analyzed by Witten (1989), and has been
studied in great detail since then (Freed and Gompf, 1991; Jeffrey, 1992; Rozansky, 1995).
It has the form
\be
Z^{(c)}_{\rm 1-loop}(M) \propto {{\sqrt {| \tau^{(c)}_R|}
\over {\rm vol} (H_c)}},
\label{asym}
\ee
where $ \tau^{(c)}_R$ is the Reidemeister-Ray-Singer torsion of $A^{(c)}$ and $H_c$ is the
isotropy group of $A^{(c)}$. Notice that, for the trivial flat connection
$A^{(c)}=0$, $H_c=G$.

The terms $S^{(c)}_\ell$ in (\ref{perts}) correspond to connected diagrams with
$2 \ell$ vertices. In Chern-Simons theory the vertex is trivalent, so $S^{(c)}_\ell$
is the contribution to the free energy at $\ell+1 $ loops. The contributions of the
trivial connection will be denoted simply by $S_\ell$.
Notice that the computation of $S^{(c)}_\ell$ involves the evaluation of group factors of
Feynman diagrams, and therefore they depend explicitly on the gauge group $G$. When $G=U(N)$, they are
polynomials in $N$. For example, $S_1$ contains the group factor $2 N (N^2-1)$.

The perturbative evaluation of Wilson loop correlators can also be done
using standard procedures. First of all one has to expand the holonomy
operator as
\ben
\label{holop}
& & W_R^{\cal K}(A) =
{\rm Tr}_R \biggl[ {\bf 1} + \oint_{\cal K} dx^{\mu} A_{\mu}(x) \nonumber\\
& +&\oint_{\cal K} dx^{\mu} \int^x dy^{\nu} A_{\nu}(y) A_{\mu}(x) + \cdots\biggr],
\een
%
where $A_{\mu}=\sum_a A_{\mu}^a T_a$. Then, after gauge fixing, one can proceed and
evaluate the correlation functions in standard perturbation theory. The perturbative
study of Wilson loops was started by Guadagnini, Martellini, and Mintchev (1990). A nice review of its development can be
found in Labastida (1999). Here we will rather focus on the nonperturbative approach to
Chern-Simons theory, which we now explain.

\subsection{Canonical quantization and surgery}

As it was shown by Witten (1989),
Chern-Simons theory is exactly solvable by using nonperturbative methods and the relation to the
Wess-Zumino-Witten (WZW) model. In order to present this solution, it
is convenient to recall some basic facts about the canonical quantization
of the model.

Let $M$ be a three-manifold with boundary given by a Riemann surface $\Sigma$.
We can insert a general operator ${\cal O}$ in $M$, which will be in general a
product of Wilson loops along different knots and in arbitrary representations
of the gauge group. We will consider the case in
which the Wilson loops do not intersect the surface $\Sigma$.
The path integral over the three-manifold with boundary $M$ gives a
wavefunction $\Psi_{M,\cal O} ({\cal A})$ which is a functional of the values of the field
at $\Sigma$. Schematically we have:
\begin{equation}
\Psi_{M, \cal O}({\cal A})=\langle {\cal A} | \Psi_{M, \cal O} \rangle =
\int_{A|_{\Sigma}={\cal A}}{\cal D}A \,{\rm e}^{iS}\,  {\cal O}.
\end{equation}
In fact, associated to the Riemann surface $\Sigma$ we have a
Hilbert space ${\cal H}(\Sigma)$, which can be obtained by doing
canonical quantization of Chern-Simons theory on $\Sigma \times {\bf R}$. Before
spelling out in detail the structure of these Hilbert spaces, let us
make some general considerations about the computation of physical quantities.

\begin{figure*}
\scalebox{.6}{\includegraphics{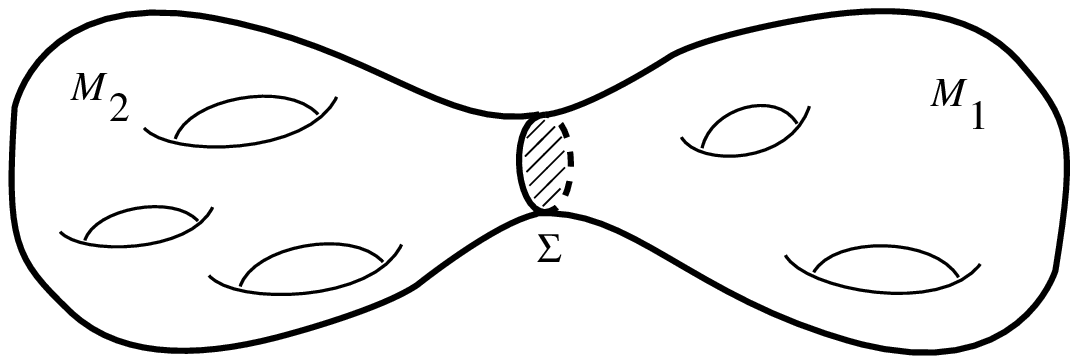}}
\caption{\label{heegard}Heegaard splitting of a three-manifold $M$ into two three manifolds
$M_1$ and $M_2$ with a common boundary $\Sigma$.}
\end{figure*}

In the context of canonical quantization, the partition function can be computed as follows.
We first perform a
Heegaard splitting of the three-manifold, {\it i.e.} we represent it as the
connected sum of two three-manifolds $M_1$ and $M_2$ sharing a common
boundary $\Sigma$, where $\Sigma$ is a Riemann surface. If $f: \Sigma
\rightarrow \Sigma$ is an homeomorphism, we will write $M=M_i \cup_f
M_2$, so that $M$ is obtained by gluing $M_1$ to $M_2$ through their common
boundary and using the homeomorphism $f$. This is represented in
Fig.~\ref{heegard}. We can then compute the full path integral (\ref{partcs})
over $M$ by
computing first the path integral over $M_1$ to obtain a state $|\Psi_{M_1}\rangle$
in ${\cal H}(\Sigma)$. The boundary of $M_2$ is also $\Sigma$, but with opposite orientation, so
its Hilbert space is the dual space ${\cal H}^*(\Sigma)$. The path integral
over $M_2$ produces then a state $\langle \Psi_{M_2}| \in {\cal H}^*(\Sigma)$. The homeomorphism
$f: \Sigma \rightarrow \Sigma$ will be
represented by an operator acting on ${\cal H}(\Sigma)$,
\be
U_f: {\cal H}(\Sigma) \rightarrow {\cal H}(\Sigma).
\end{equation}
and the partition function can be finally evaluated as
\be
\label{surg}
Z(M)=\langle \Psi_{M_2}| U_f | \Psi_{M_1}\rangle.
\end{equation}
Therefore, if we know explicitly what the wavefunctions and the operators associated
to homeomorphisms are, we can compute the partition function.

One of the most fundamental
results of Witten (1989) is in fact a precise description of ${\cal H}(\Sigma)$:
it is the space of conformal
blocks of a WZW model on $\Sigma$ with gauge group $G$ and level $k$ (for
an extensive review of the WZW model, see for example Di Francesco {\it et al.}, 1997). In particular,
${\cal H}(\Sigma)$ has finite dimension. We will
not review here the derivation of this fundamental result. Instead we will use
the relevant information from the WZW model in order to solve Chern-Simons theory.

The description of the space of conformal blocks on Riemann surfaces can be made
very explicit when $\Sigma$ is a sphere or a torus. For $\Sigma={\bf S}^2$, the space of
conformal blocks is one-dimensional, so ${\cal H}({\bf S}^2)$
is spanned by a single element. For $\Sigma={\bf T}^2$,
the space of conformal blocks is in one
to one correspondence with
the integrable representations of the affine Lie algebra associated to $G$ at
level $k$. We will use the following notations:
the fundamental weights of $G$ will be denoted by
$\lambda_i$, and the simple roots by $\alpha_i$, with $i=1, \cdots, r$,
and $r$ denotes the rank of $G$.
The weight and root lattices of $G$ are
denoted by $\Lambda^w$ and $\Lambda^r$, respectively, and
$|\Delta_+|$ denotes the number of positive roots. The fundamental chamber
${\cal F}_l$ is given by $\Lambda^w/l
\Lambda^r$, modded out by the action of the Weyl group. For example,
in $SU(N)$ a
weight $p=\sum_{i=1}^r p_i \lambda_i$ is in ${\cal F}_l$ if
\begin{equation}
\sum_{i=1}^r p_i < l,\,\,\,\,\,\, {\rm and} \,\,\ p_i >0, \, i=1, \cdots, r.
\end{equation}
We recall that a representation given by a highest weight $\Lambda$ is integrable if
$\rho + \Lambda$ is in the
fundamental chamber ${\cal F}_l$, where $l=k+y$ ($\rho$ denotes as usual the Weyl vector,
given by the sum of the fundamental weights). In the following, 
the states in the Hilbert state of the torus
${\cal H}({\bf T}^2)$ will be denoted by $
|p\rangle = | \rho + \Lambda \rangle$
where $\rho + \Lambda$, as we have stated, is an integrable representation
of the WZW model at level $k$. We will also denote these states by $|R\rangle$, where
$R$ is the representation associated to $\Lambda$. The state $|\rho \rangle$ will be denoted
by $| 0\rangle$. The states $|R\rangle$ can be chosen to be orthonormal (Witten, 1989; Elitzur {\it et. al.},
1989; Labastida and Ramallo, 1989), so we have
\be
\langle R | R'\rangle =\delta_{R R'}.
\end{equation}
There is a special class of homeomorphisms of ${\bf T}^2$ that have a
simple expression as operators in ${\cal H}({\bf T}^2)$; these are the ${\rm
Sl}(2, {\bf Z})$ transformations. Recall that the group ${\rm
Sl}(2, {\bf Z})$ consists of $2\times 2$ matrices with integer entries and
unit determinant. If $(1,0)$ and $(0,1)$ denote the
two one-cycles of ${\bf T}^2$, we can specify the action of an ${\rm
Sl}(2, {\bf Z})$ transformation on the torus by giving its action on this homology basis.
The ${\rm Sl}(2, {\bf Z})$ group is generated by the transformations $T$ and $S$, which are given by
\begin{equation}
T=\left(\begin{array}{cc} 1 & 1\\
0 &1\end{array} \right), \,\,\,\,\,\,\, S=\left(\begin{array}{cc} 0 & -1\\
1 &0\end{array} \right).
\end{equation}
Notice that the $S$ transformation exchanges the one-cycles of the torus. These
transformations can be lifted to ${\cal H}({\bf T}^2)$, and they have the
following matrix elements in the basis of integrable representations:
\ben
\label{st}
T_{pp'}&=& \delta_{p,p'} {\rm e}^{2\pi i (h_p -c/24)},\nonumber\\
S_{pp'}&=& {i^{|\Delta_+|} \over (k+y)^{r/2}} \Biggl( {{\rm Vol} \,
\Lambda^w \over{\rm Vol} \,
\Lambda^r} \Biggr)^{1\over 2}\nonumber\\
& & \,\,\, \times \sum_{w \in {\cal W}} \epsilon (w) \exp \Bigl(
-{2 \pi i \over k+y} p \cdot w(p')\Bigr).
\een
In the first equation, $c$ is the central charge of the WZW model, and
$h_p$ is the conformal weight of the primary field
associated to $p$:
\be
h_p = {p^2 -\rho^2 \over 2(k+y)},
\label{confweight}
\end{equation}
where we remind that $p$ is of the form $\rho + \Lambda$.
In the second equation, the sum over $w$ is a sum over the elements of the
Weyl group ${\cal W}$, $\epsilon(w)$ is the signature of the element $w$, and ${\rm Vol} \,
\Lambda^{w(r)}$ denote respectively the volume of the weight (root) lattice. 
We will often write $S_{RR'}$ for $S_{p p'}$, where $p=\rho + \Lambda$, 
$p'=\rho + \Lambda'$ and $\Lambda$, $\Lambda'$ are the highest weights corresponding 
to the representations $R$, $R'$.   

\begin{figure}
\scalebox{.4}{\includegraphics{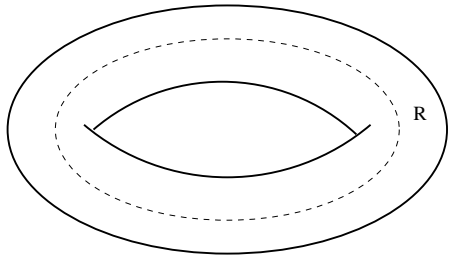}}
\caption{\label{torbasis}Performing the path integral on a solid torus with a Wilson line
in representation $R$ gives the state $|R\rangle$ in
${\cal H}({\bf T}^2)$.}
\end{figure}

What is the description of the states $|R\rangle$ in ${\cal H}({\bf T}^2)$ from the point of view of canonical quantization?
Consider the solid torus ${\cal T}=D \times {\bf S}^1$, where $D$ is a disc in ${\bf R}^2$. This is
a three-manifold whose boundary is a ${\bf T}^2$, and it has a noncontractible cycle
given by the ${\bf S}^1$. Let us now consider the Chern-Simons path integral on the solid torus,
with the insertion of the operator
${\cal O}_{R}={\rm Tr}_R U$ given by a Wilson loop in the representation $R$ around the
noncontractible cycle, as shown in Fig.~\ref{torbasis}. In this way one obtains a state
in ${\cal H}({\bf T}^2)$, and one has
\begin{equation}
|\Psi_{{\cal T}, {\cal O}_{R}}\rangle =| R \rangle.
\end{equation}
In particular, the path integral over the solid torus with no operator insertion gives
$|0 \rangle$, the ``vacuum'' state.

These results allow us to compute the partition function of
any three-manifold that
admits a Heegaard splitting along a torus. Imagine for example that
we take two solid tori and we glue them along their boundary with the
identity map. Since a solid torus
is a disc times a circle $D \times {\bf S}^1$, by performing this operation we
get a manifold which is ${\bf S}^1$ times the two discs glued together along their
boundaries. Therefore, with this surgery we obtain ${\bf S}^2 \times {\bf S}^1$, and
(\ref{surg}) gives then
\be
Z({\bf S}^2 \times {\bf S}^1)=\langle 0| 0 \rangle =1.
\end{equation}
If we do the gluing, however, after performing an $S$-transformation on the ${\bf T}^2$
the resulting manifold is instead ${\bf S}^3$. To see this, notice that the complement to
a solid torus inside ${\bf S}^3$ is indeed another solid torus whose noncontractible
cycle is homologous to the contractible cycle in the first torus. We then find
\begin{equation}
Z({\bf S}^3)=\langle 0 |S | 0 \rangle = S_{00}.
\label{pfsphere}
\end{equation}
By using Weyl's denominator formula,
\be
\sum_{w \in {\cal W}} \epsilon (w) {\rm e}^{w(\rho)} =\prod_{\alpha>0}
2 \sinh {\alpha \over 2},
\label{wdf}
\end{equation}
where $\alpha >0$ are positive roots, one finds
\be
Z({\bf S}^3)= {1 \over (k+y)^{r/2}} \Biggl( {{\rm Vol} \,
\Lambda^w \over{\rm Vol} \,
\Lambda^r} \Biggr)^{1\over 2} \prod_{\alpha>0}
2 \sin \Bigl( {\pi (\alpha \cdot \rho) \over k+y} \Bigr).
\label{css}
\end{equation}
The above result can be generalized in order to compute path
integrals in ${\bf S}^3$ with some knots and links. Consider a solid torus
where a Wilson line in representation $R$ has been inserted. The corresponding
state is $|R\rangle$, as we explained before. If we
now glue this to an empty solid torus after an
$S$-transformation, we obtain a trivial knot, or {\it unknot}, 
in ${\bf S}^3$. The path integral with the
insertion is then,
\be
Z({\bf S}_3, {\cal O}_{R})=\langle 0| S| R \rangle.
\end{equation}
It follows that the normalized vacuum expectation value for the unknot in
${\bf S}^3$, in representation $R$, is given by
\be
W_{R}({\rm unknot})  ={S_{0R} \over S_{00}}=
{\sum_{w \in {\cal W}} \epsilon(w) {\rm e}^{-{2 \pi i \over k+y} \rho
\cdot w(\Lambda + \rho)} \over \sum_{w \in {\cal W}}
\epsilon(w) {\rm e}^{-{2 \pi i \over k+y} \rho
\cdot w( \rho)}}.
\end{equation}
Remember that the character of the representation $R$, evaluated on
an element $a \in \Lambda_{\rm w}\otimes {\bf R}$ is defined by
\be
{\rm ch}_R (a) = \sum_{\mu \in M_{R}} e^{a \cdot \mu}
\label{chardef}
\ee
where $M_R$ is the set of weights associated to the irreducible
representation $R$. By using Weyl's character formula we can write
\be
\label{qdimchar}
W_{R}({\rm unknot})=
{\rm ch}_R\Bigl[ -{2 \pi i \over k+y} \rho \Bigr].
\end{equation}
Moreover, using (\ref{wdf}), we finally obtain
\be
W_{R}({\rm unknot})=\prod_{\alpha >0} {\sin \Bigl( { \pi \over k+y}
\alpha\cdot (\Lambda + \rho) \Bigr) \over
\sin \Bigl( { \pi \over k+y}
\alpha\cdot \rho \Bigr)}
\end{equation}
This quantity is often called the {\it quantum dimension} of $R$, and it is denoted
by ${\rm dim}_q R$.

We can also consider a solid torus with Wilson loop in representation $R$, glued to
another solid torus with the representation $R'$ through an $S$-transformation. What
we obtain is clearly a link in ${\bf S}^3$
with two components, which is the Hopf link shown in Fig.~\ref{nudosf}. Taking into
account the orientation carefully, we find that this is the Hopf link with linking
number $+1$. The
path integral with this insertion is:
\be
Z({\bf S}^3, {\cal O}_{R} {\cal O}_{R'})=\langle R' | S| R \rangle,
\end{equation}
so the normalized vacuum expectation value is
\be
\label{hopfinv}
{\cal W}_{R R'}\equiv W_{R R'}({\rm Hopf}^{+1})= { S_{{\overline R}' R} \over S_{0 0}}=
{ S^{-1}_{R' R} \over S_{0 0}},
\end{equation}
where the superscript $+1$ refers to the linking number. Here we have used 
that the bras $\langle R|$ are canonically associated to conjugate 
representations $\overline R$, and that $S_{{\overline R}' R}=
S^{-1}_{R' R}$ (see for example Di Francesco {\it et al.}, 1997).
Therefore, the Chern-Simons invariant of the Hopf link is essentially an $S$-matrix
element. In order to obtain the invariant of the Hopf link with linking number $-1$, we
notice that the two Hopf links can be related by changing the orientation of one of
the components. We then have
\be
\label{hopfplus}
W_{R R'}({\rm Hopf}^{-1}) ={ S_{R' R} \over S_{0 0}},
\end{equation}
where we have used the property (\ref{reorient}). 

When we take $G=U(N)$, the above vacuum expectation values for unknots and Hopf links
can be evaluated
very explicitly in terms
of Schur polynomials. It is well known that the character of the unitary group in the
representation $R$ is given by the
Schur polynomial $s_R$ (see for example Fulton and Harris, 1991). There is a precise relation between
the element $a$ where one evaluates the character in (\ref{chardef})
and the variables entering the Schur polynomial.
Let $\mu_i$, $i=1, \cdots, N$,
be the weights associated to the fundamental representation
of $U(N)$. Notice that, if $R$ is given by a Young tableau whose rows have
lengths $l_1 \ge \cdots
\ge l_N$, then $\Lambda_R = \sum_i l_i \mu_i$. We also have
\be
\rho = \sum_{i=1}^N {1\over 2} (N-2i+1) \mu_i.
\end{equation}
Let $a \in \Lambda^w \otimes {\bf R}$
be given by
\be
a = \sum_{i=1}^N a_i \mu_i.
\end{equation}
Then,
\be
{\rm ch}_R [a]=s_R (x_i=e^{a_i}).
\end{equation}
For example, in the case of the quantum dimension, one has
${\rm dim}_q R={\rm dim}_q {\overline R}$, and we find
\be
{\rm dim}_q R=s_R (x_i=q^{{1 \over 2}(N-2i+1)}),
\end{equation}
where $q$ is given in (\ref{polvars}). By using that $s_R$ is homogeneous of degree $\ell(R)$ in the coordinates
$x_i$ we finally obtain
$$
{\rm dim}_q R=\lambda^{\ell(R)/2} s_R (x_i=q^{-i+{1\over 2}})
$$
where $\lambda=q^N$ as in (\ref{polvars}), and there are $N$ variables $x_i$.
The quantum dimension can be written very explicitly in terms of the $q$-numbers:
\be
\label{qnumbers}
[x]=q^{x\over 2} -q^{-{x\over 2}},\,\,\,\ [x]_{\lambda} = \lambda^{1\over 2}
q^{x\over 2} -\lambda^{-{1\over 2}} q^{-{x\over 2}}.
\end{equation}
If $R$ corresponds to a Young tableau with $c_R$ rows of lengths $l_i$,
$i=1, \cdots, c_R$,
the quantum dimension is given by:
\begin{equation}
\label{expf}
{\rm dim}_q R = \prod_{1\le i < j \le c_R} {[l_i -l_j +j-i]
\over [j-i]} \prod_{i=1}^{c_R} { \prod_{v=-i+1}^{l_i -i} [v]_{\lambda}
\over \prod_{v=1}^{l_i} [v-i + c_R]}.
\end{equation}
It is easy to check that in the limit $k+N \rightarrow \infty$ ({\it i.e.} in the
semiclassical limit) the quantum dimension becomes the dimension of the representation
$R$. Notice that the quantum dimension is a rational function of $q^{\pm {1\over 2}}$,
$\lambda^{\pm {1\over 2}}$.
This is a general property of all normalized vacuum expectation values of knots and links in ${\bf S}^3$.

The $S$-matrix elements that appear in (\ref{hopfinv}) and (\ref{hopfplus}) can be evaluated
through the explicit expression (\ref{st}), by using the
relation between $U(N)$ characters and
Schur functions that we explained above. Notice first that
\be
\label{sweyl}
{S_{R_1 R_2}^{-1} \over S_{00}}={\rm ch}_{R_1}\Bigl[ {2 \pi i \over k+y} (\Lambda_{R_2}+\rho) \Bigr]
{\rm ch}_{R_2}\Bigl[ {2 \pi i \over k+y} \rho \Bigr].
\end{equation}
If we denote by $l_i^{R_2}$, $i=1, \cdots, c_{R_2}$ the lengths of rows for the Young tableau
corresponding to $R_2$, it is easy to see that
\be
\label{mlfor}
{\cal W}_{R_1 R_2}(q, \lambda)=(\lambda q)^{\ell(R_1)\over 2} s_{R_1} (x_i= q^{l^{R_2}_i -i}) {\rm dim}_q R_2.
\ee
where we set $l_i^{R_2}=0$ for $i > c_{R_2}$. A convenient way to evaluate $s_{R_1} (x_i= q^{l^{R}_i -i})$
for a partition $\{l^R_i\}_{\{i=1, \cdots, c_R\}}$ associated to $R$
is to use the Jacobi-Trudy formula (\ref{jt}). It is easy to show that the generating functional of
elementary symmetric functions (\ref{esgen}) for this specialization is given by
\be\label{eser}
E_{R}(t)= E_{\emptyset}(t)
\prod_{j=1}^{c_R} { {1 + q^{l^{R}_j -j} t}
\over {1 + q^{- j} t}},
\ee
where
\be
\label{trive}
E_{\emptyset}(t)=1+ \sum_{n=1}^{\infty} a_n t^n,
\ee
and the coefficients $a_n$ are defined by
 \be
\label{cnml}
a_n= \prod_{r=1}^n {1 - \lambda^{-1} q^{r-1} \over q^r-1}.
\ee
The formula (\ref{mlfor}), together with the expressions above for $E_R(t)$,
provides an explicit expression for (\ref{hopfinv}) as a rational function of
$q^{\pm{1\over 2}}$, $\lambda^{\pm{1\over 2}}$, and it
was first written down by Morton and Lukac (2003).

\subsection{Framing dependence}

In the above discussion on the correlation functions of Wilson loops
we have missed an important ingredient. We already mentioned that, in order to
define the partition function of Chern-Simons theory at the quantum level,
one has to specify a framing of the three-manifold. It turns out that the
evaluation of correlation functions like (\ref{vevknot}) also involves a
choice of framing of the knots, as discovered by Witten (1989). Since
this is important in the context of topological strings, we will explain
it in some detail.

A good starting point to understand the framing is to take
Chern-Simons theory with gauge group $U(1)$.
The Abelian Chern-Simons theory turns out to be extremely simple,
since the cubic term in (\ref{csact}) drops out, and we are left
with a Gaussian theory (Polyakov, 1988). $U(1)$
representations are labelled by integers, and the correlation function (\ref{vevknot})
can be computed exactly. In order to do that, however, one has to
choose a framing for each of the knots ${\cal K}_{\alpha}$. This arises as follows:
in evaluating the correlation function, contractions of the holonomies
corresponding to different ${\cal K}_{\alpha}$ produce the following integral:
\begin{equation}
\label{linking}
{\rm lk} ({\cal K}_{\alpha}, {\cal K}_{\beta})=
{1 \over 4 \pi} \oint_{{\cal K}_{\alpha}}dx^{\mu} \oint_{{\cal K}_{\beta}} dy^{\nu}
\epsilon_{\mu \nu \rho} { (x-y)^{\rho} \over |x-y|^3}.
\end{equation}
This is a topological invariant, {\it i.e.} it is invariant
under deformations of the knots ${\cal K}_{\alpha}$, ${\cal K}_{\beta}$, and it is in fact
the Gauss integral representation of their linking number ${\rm lk}({\cal K}_{\alpha}, {\cal K}_{\beta})$
defined in (\ref{lknumber}).
On the other hand, contractions of the holonomies corresponding to the
same knot $\cal K$ involve the integral
\begin{equation}\label{cotor}
\phi ({\cal K})={1 \over 4 \pi} \oint_{\cal K}dx^{\mu} \oint_{\cal K} dy^{\nu}
\epsilon_{\mu \nu \rho} { (x-y)^{\rho} \over |x-y|^3}.
\end{equation}
This integral is well-defined and finite (see,
for example, Guadagnini, Martellini, and Mintchev, 1990), and it is
called the {\it cotorsion} or {\it writhe} of $\cal K$. It gives
the self-linking number of ${\cal K}$: if we project ${\cal K}$ on a plane,
and we denote by $n_{\pm}({\cal K})$ the number of positive (negative) crossings
as indicated in Fig.~\ref{linkingf}, then we have that
\be
\phi({\cal K})=n_+ ({\cal K}) -n_{-}({\cal K}).
\end{equation}
The problem is that the cotorsion
is not invariant under deformations of the knot. In order to
preserve topological invariance of the correlation function,
one has to choose another
definition of the composite operator $(\oint_{\cal K}A)^2$ by means of a
framing. A framing of the knot consists of
choosing another knot ${\cal K}^f$ around $\cal K$,
specified by a normal vector
field $n$. The cotorsion $\phi({\cal K})$ becomes then
\begin{equation}
\label{regul}
\phi_f ({\cal K})={1 \over 4 \pi}
\oint_{\cal K}dx^{\mu} \oint_{{\cal K}^f} dy^{\nu}
\epsilon_{\mu \nu \rho} { (x-y)^{\rho} \over |x-y|^3} =
{\rm lk} ({\cal K}, {\cal K}^f).
\end{equation}
The correlation function that we obtain in this way is
a topological invariant (since it only involves linking numbers) but the
price that we have to pay is that our regularization depends on a set
of integers $p_{\alpha} ={\rm lk} ({\cal K}_{\alpha}, {\cal K}^f_{\alpha})$ (one for each knot).
The correlation function (\ref{vevknot}) can now be computed, after choosing the framings, as
follows:
\begin{widetext}
\be
\label{vevans}
\bigg\langle \prod_{\alpha} \exp \bigl( n_{\alpha} \oint_{\CK_{\alpha}} A \bigr) \bigg\rangle =
\exp \biggl\{ { \pi i \over k} \Bigl(\sum_{\alpha} n_{\alpha}^2 p_{\alpha} +
\sum_{\alpha \not= \beta} n_{\alpha} n_{\beta} \, {\rm lk} ({\cal K}_{\alpha}, {\cal K}_{\beta}) \Bigr) \biggr\}.
\ee
\end{widetext}
This regularization is nothing but the `point-splitting' method
familiar in the context of quantum field theory.

Let us now consider Chern-Simons theory with gauge group $U(N)$, and
suppose that we are interested in the computation of (\ref{vevknot}), in the context of perturbation theory.
It is easy to see that self-contractions of the holonomies lead to
the same kind of ambiguities that we found in the abelian case, {\it i.e.}
a choice of framing has to be made for each knot ${\cal K}_{\alpha}$. The
only difference with the Abelian case is that the self contraction of ${\cal K}_{\alpha}$
gives a group factor ${\rm
Tr}_{R_{\alpha}}(T_a T_a)$,
where $T_a$ is a basis of the Lie algebra (see for example Guadagnini,
Martellini, and Mintchev, 1990).
The precise result
can be better stated as the effect on the correlation function
(\ref{vevknot}) under a change of framing, and it says that,
under a change
of framing of ${\cal K}_{\alpha}$ by $p_{\alpha}$ units, the vacuum expectation value of the product of
Wilson loops changes as follows (Witten, 1989):
\begin{equation}
\label{naframing}
W_{R_1 \cdots R_L}  \rightarrow \exp \biggl[ 2\pi i \sum_{\alpha=1}^L
p_{\alpha}  h_{R_{\alpha}} \biggr]W_{R_1 \cdots R_L}.
\end{equation}
In this equation, $h_R$ is
the conformal weight of the Wess-Zumino-Witten primary field corresponding
to the representation $R$. One can write (\ref{confweight}) as
\begin{equation}
\label{cweight}
h_R = {C_R\over 2(k+N)},
\end{equation}
where $C_R={\rm Tr}_{R}(T_a T_a)$
is the quadratic Casimir in the
representation $R$. For $U(N)$
one has
\begin{equation}
\label{explcas}
C_R = N \ell(R) + \kappa_R.\end{equation}
where $\ell (R)$ is the total number of boxes in the tableau, and
\begin{equation}\label{kapar}
\kappa_R =\ell(R) +  \sum_i \bigl( l_i^2 -2il_i  \bigr).
\end{equation}
In terms of the variables (\ref{polvars})
the change under framing (\ref{naframing}) can be
written as
\begin{equation}
 \label{unframing}
W_{R_1 \cdots R_L} \rightarrow  q^{{1 \over 2}\sum_{\alpha=1}^L
 \kappa_{R_{\alpha}} p_{\alpha}}  \lambda^{{1\over 2}
\sum_{\alpha=1}^L \ell (R_{\alpha}) p_{\alpha}} W_{R_1 \cdots R_L}.
\end{equation}
Therefore, the evaluation of vacuum
expectation values of Wilson loop operators in Chern-Simons theory depends
on a choice of framing for knots. It turns out that for
knots and links in ${\bf S}^3$, there is a {\it standard} or canonical
framing, defined by requiring that the self-linking number is zero. The
expressions we have given before for the Chern-Simons invariant of the unknot and
the Hopf link are all in the
standard framing. Once the value of the invariant
is known in the standard framing, the value in any other framing specified
by nonzero integers $p_{\alpha}$ can be easily obtained from (\ref{naframing}).

\subsection{More results on Wilson loops}

In this subsection we discuss some useful results for the
computation of vacuum expectation values of Wilson loops. Most of
these results can be found for example in Guadagnini (1992).

The first property we want to state is the {\it factorization property}
for the vacuum expectation values of disjoint links, which says the following. Let
$\CL$ be a link with $L$ components $\CK_1, \cdots, \CK_L$ which are {\it
disjoint} knots, and let us attach the representation $R_\alpha$ to the
$\alpha$-th component. Then one has
\be
\label{factoriz}
W_{R_1 \cdots R_L}(\CL) =\prod_{\alpha=1}^L W_{R_{\alpha}}(\CK_{\alpha}).
\ee
This property is easy to prove in Chern-Simons theory. It only involves some
elementary surgery and the fact that ${\CH}({\bf S}^2)$ is one-dimensional. A
proof can be found in Witten (1989).

The second property we will consider is
{\it parity symmetry}.
Chern-Simons theory is a theory of oriented links, and under a parity transformation
a link ${\cal L}$ will transform into its mirror ${\cal L}^*$. The mirror of ${\cal L}$ is
obtained from its planar projection simply by changing undercrossings by overcrossings, and
viceversa. On the other hand, parity changes the sign of the Chern-Simons action, in other words
$k+N \rightarrow -(k+N)$. We then find that vacuum expectation values transform as
\be
W_{R_1 \cdots R_L}({\cal L}^*)(q, \lambda)= W_{R_1 \cdots R_L}({\cal L})(q^{-1}, \lambda^{-1}).
\end{equation}
This is interesting from a knot-theoretic point of view, since it implies that
Chern-Simons invariants of links can distinguish in principle a link from its mirror image.
As an example of this property, notice
for example that the unknot is identical to its mirror image,
therefore quantum dimensions satisfy:
\be
({\rm dim}_q R)(q^{-1}, \lambda^{-1})=({\rm dim}_q R)(q, \lambda).
\end{equation}

Let us now discuss the simplest example of a {\it fusion rule} in
Chern-Simons theory. Consider a vacuum expectation value of the form
\be
\langle {\Tr}_{R_1} U {\Tr}_{R_2} U \rangle,
\end{equation}
where $U$ is the holonomy of the gauge field around a knot $\CK$. The
classical operator ${\Tr}_{R_1} U {\Tr}_{R_2}U$ can always be written as
\be
{\Tr}_{R_1} U {\Tr}_{R_2}U ={\Tr}_{R_1 \otimes R_2} U =\sum_{R} N_{R_1 R_2}^R {\Tr}_{R} U,
\end{equation}
where $R_1 \otimes R_2$ denotes the tensor product, and 
$N_{R_1 R_2}^R$ are tensor product coefficients. In Chern-Simons theory, the quantum
Wilson loop operators satisfy a very similar relation, with the only difference that
the coefficients become the fusion coefficients for integrable representations
of the WZW model. This can be
easily understood if we take into account that the admissible representations that appear
in the theory are the integrable ones, so one has to truncate the list of
``classical'' representations, and this implies in particular that the
product rules of classical traces have to be modified. However, in the computation of knot invariants in
$U(N)$ Chern-Simons theory it is natural to work in a setting in
which both $k$ and $N$ are much larger than any of the representations involved.
In that case, the vacuum expectation
values of the theory satisfy
\be
\label{fusiontwo}
\langle {\Tr}_{R_1} U {\Tr}_{R_2} U \rangle =\sum_{R} N_{R_1 R_2}^R
\langle {\Tr}_{R} U \rangle.
\end{equation}
where $N_{R_1 R_2}^R$ are the Littlewood-Richardson coefficients of $U(N)$.
As a simple application of the fusion rule, imagine that we want to compute
$\langle {\Tr}_{R_1} U_1 {\Tr}_{R_2} U_2 \rangle$, where
$U_{1,2}$ are holonomies around disjoint unknots with zero framing. We can take the unknots
to be very close, in such a way that the paths along which we compute the holonomy are
the same. In that case, this vacuum expectation value becomes
exactly the l.h.s of (\ref{fusiontwo}). Using
also the factorization property (\ref{factoriz}), we deduce the following fusion rule:
\be
{\rm dim}_q R_1 \,{\rm dim}_q R_2 =\sum_R N_{R_1 R_2}^R  {\rm dim}_q R.
\end{equation}

\begin{figure}
\scalebox{.7}{\includegraphics{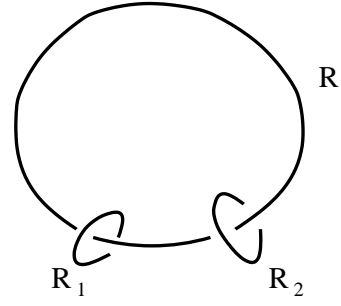}}
\caption{\label{fusion}A link with two disjoint unknots.}
\end{figure}

The last property we will state is the behaviour of correlation
functions under {\it direct sum}. This operation is defined as
follows. Let us consider two links ${\cal L}_1$, ${\cal L}_2$
with components ${\cal K}_{1}, {\cal K}$
and ${\cal K}_{2}, {\cal K}$, respectively, {\it i.e.} the component knot ${\cal K}$ is the 
same in ${\cal L}_1$ and ${\cal L}_2$. The
{\it direct sum} ${\cal L}={\cal L}_1\sharp {\cal L}_2$ is a link of three components
which is obtained by joining ${\cal L}_1$ and ${\cal L}_2$ through ${\cal K}$.
It is not difficult to prove that the Chern-Simons
invariant of ${\cal L}$ is given by (Witten, 1989)
\be
W_{R_1 R_2 R}({\cal L})= {W_{R_1 R}({\cal L}_1)W_{R_2 R}({\cal L}_2)
\over W_R ({\cal K})}.
\label{dsum}
\ee
As an application of this rule, let us consider the three-component link in Fig.~\ref{fusion}.
This link is a direct sum of two Hopf links whose common component is an
unknot in representation $R$, and the knots ${\cal K}_1$, ${\cal K}_2$ are unknots
in representations $R_1$, $R_2$. The equation (\ref{dsum}) expresses the Chern-Simons invariant
of ${\cal L}$ in terms of invariants of Hopf links and quantum dimensions. Notice that the invariant of
the link in Fig.~\ref{fusion} can be also computed by using the fusion rules. If we
fuse the two parallel unknots with representations $R_1$, $R_2$, we find 
\be
\label{threefus}
W_{R_1 R_2 R}({\cal L}) =\sum_{R'} N_{R_1 R_2}^{R'}
\langle{\Tr}_{R'} U' {\Tr}_R U \rangle,
 \end{equation}
where $U$ is the holonomy around the unknot in representation $R$, and 
$U'$ is the holonomy around the unknot which is obtained by fusing the 
two parallel unknots in Fig.~\ref{fusion}. (\ref{threefus}) 
expresses the invariant (\ref{dsum}) in terms of the invariants of a Hopf link
with representations $R'$, $R$.

\subsection{Generating functionals for knot and link invariants}

In the applications of Chern-Simons theory, we will need the invariants of
knots and links in arbitrary representations of the gauge group. It is then
natural to consider generating functionals for Wilson loop operators in arbitrary representations.

There are two natural basis for the
set of Wilson loop operators: the basis labelled by representations $R$, which is
the one that we have considered so far, and the basis
labelled by conjugacy classes of the symmetric group. Wilson
loop operators in the basis of conjugacy classes are constructed as
follows. Let $U$ be the holonomy of the gauge connection around the knot
${\cal K}$. Let $\vec k=(k_1, k_2, \cdots)$ be a vector of infinite entries, almost
all of which are zero. This vector defines naturally a conjugacy class $C(\vec k)$ of the
symmetric group $S_{\ell}$ with
\be
\ell=\sum_j j \, k_j.
\ee
We will also denote
\be
\label{ksum}
|\vec k| =\sum_j k_j.
\ee
The conjugacy class $C(\vec k)$ is simply the class that has $k_j$ cycles of length $j$.
We now define the operator
\be
\Upsilon_{\vec k}(U) =\prod_{j=1}^{\infty} ({\rm Tr} U^j)^{k_j},
\label{ups}
\end{equation}
which gives the Wilson loop operator in the conjugacy class basis. 
It is a linear combination of the operators ${\rm Tr}_R\, U$ labelled by representations:
\begin{equation}
\label{trek}
\Upsilon_{\vec k} (U)=
\sum_{R} \chi_R (C(\vec k)){\rm Tr}_R \,U,
\end{equation}
where $\chi_R (C(\vec k))$ are the characters of the symmetric group $S_{\ell}$
in the representation $R$ evaluated at the conjugacy class $C(\vec k)$. The above
formula can be inverted as
\begin{equation}
{\rm Tr}_R(U)= \sum_{\vec k} {\chi_R (C(\vec k)) \over z_{\vec k}}
\Upsilon_{\vec k}(U), \label{frob}
\end{equation}
with
\be
\label{zk}
z_{\vec k}=\prod_j k_j! j^{k_j}.
\ee
If $U$ is a diagonal matrix $U={\rm diag}(x_1, \cdots, x_N)$, it is an elementary
result in the representation theory of the unitary group that ${\rm Tr}_R \, U$
is the Schur polynomial in $x_i$:
\be
{\rm Tr}_R \, U =s_R (x)
\ee
It is immediate to see that
\be
\Upsilon_{\vec k}(U)=P_{\vec k}(x),
\ee
where the Newton polynomials are defined in (\ref{newt}). The relation (\ref{trek}) is
nothing but Frobenius formula, which relates the two basis of symmetric polynomials given by
the Schur and the Newton polynomials. The vacuum expectation values of the operators
(\ref{ups}) will be denoted by
\be
W_{\vec k}=\langle \Upsilon_{\vec k }(U) \rangle.
\ee

If $V$ is a
$U(M)$ matrix (a ``source'' term), one can define the following
operator, which was introduced by Ooguri and Vafa (2000) and is known in this context
as the Ooguri-Vafa operator:
\begin{equation}
Z(U,V)=\exp\Bigl[ \sum_{n=1}^\infty {1 \over n} {\rm Tr}\, U^n\,
{\rm Tr }\, V^n\Bigr].
\label{ovop}
\end{equation}
When expanded, this operator can be written in the $\vec k$-basis as follows,
\begin{equation}
Z(U,V)=1 + \sum_{\vec k} {1 \over z_{\vec k}} \Upsilon_{\vec k}(U)
\Upsilon_{\vec k}(V).
\end{equation}
We see that $Z(U,V)$
includes all possible Wilson loop operators $\Upsilon_{\vec k}(U)$
associated to a knot ${\cal K}$. One can also use Frobenius formula
to show that
\begin{equation}
\label{ovrep}
Z(U,V)=\sum_{R} {\rm Tr}_R(U) {\rm Tr}_R (V),
\end{equation}
where the sum over representations starts with the trivial one. 
Notice that in the above equation $R$ is regarded as a Young tableau, and since 
we are taking both $N$ and $M$ to be large, it can be regarded as a representation of 
both $U(N)$ and $U(M)$. In $Z(U,V)$ we assume that
$U$ is the holonomy of a dynamical gauge field and
that $V$ is a source. The vacuum expectation value $Z_{\rm CS}(V)=\langle Z(U,V)
\rangle$ has then information about the vacuum expectation values of the Wilson loop operators,
and by taking its logarithm one can define the connected vacuum
expectation values $W^{(c)}_{\vec k}$:
\begin{equation}
F_{\rm CS}(V)=\log Z_{\rm CS}(V)= \sum_{\vec k}{1 \over z_{\vec k}!}
W^{(c)}_{\vec k} \Upsilon_{\vec k}(V) \label{convev}
\end{equation}
One has, for example:
$$
W^{(c)}_{(2,0,\cdots)}=\langle ({\rm Tr} U)^2\rangle
-\langle {\rm Tr} U\rangle^2=W_{\tableau{2}} + W_{\tableau{1 1}}
-W_{\tableau{1}}^2.
$$
The free energy $F_{\rm CS}(V)$, which is a generating functional for
connected vacuum expectation values $W^{(c)}_{\vec k}$, is an important quantity when one considers the
string/gauge theory correspondence, as we will see.

\section{The $1/N$ expansion and Chern-Simons theory}

\subsection{The $1/N$ expansion}

In quantum field theory, the usual perturbative expansion gives a series in
powers of the coupling constants of the model. However, in theories with
a $U(N)$ or $SU(N)$ gauge symmetry there is an extra parameter that enters into the game,
namely $N$, and it turns out that there is a way to express the free energy and
the correlation functions of the theory as power series in $1/N$. This is the $1/N$ expansion
introduced by `t Hooft (1974) in the context of QCD.

\begin{figure}[!ht]
\scalebox{.6}{\includegraphics{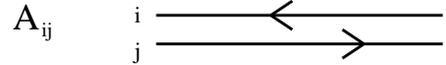}}
\caption{\label{fat} The index structure of the fields in the adjoint representation of
$U(N)$ is represented by a double line.}
\end{figure}

A good starting point to construct the $1/N$ expansion is the usual perturbative expansion.
The $N$ dependence of the perturbative expansion comes from the group factors of Feynman diagrams,
but it is clear that a single Feynman diagram gives rise to a polynomial in $N$ involving
different powers of $N$. Therefore, the standard Feynman diagrams, which are good in order
to keep track of powers of the coupling constants, are not good in order to keep track of powers
of $N$. What we have to do is to ``split'' each diagram into different pieces which
correspond to a definite power of $N$. To do that, one
writes the Feynman diagrams of the theory as ``fatgraphs'' or ribbon
graphs (`t Hooft, 1974).

\begin{figure}[!ht]
\scalebox{.6}{\includegraphics{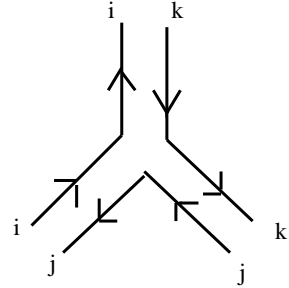}}
\caption{\label{cubicvertex} The cubic vertex in the double line notation.}
\end{figure}

\begin{figure}[!ht]
\scalebox{.2}{\includegraphics{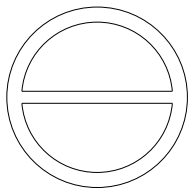}}
\caption{\label{ncube} A planar diagram with $h=3$.}
\end{figure}

\begin{figure}[!ht]
\scalebox{.2}{\includegraphics{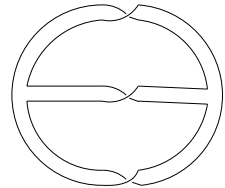}}
\caption{\label{justn} A nonplanar diagram with $g=h=1$.}
\end{figure}

In the fatgraph approach to perturbation theory, the propagator of the
gluon field $A_{ij}$ is represented by a double line, as shown in Fig.~\ref{fat}. The indices $i,j=1, \cdots, N$ are
gauge indices for the adjoint representation. Similarly,
the trivalent vertex of Chern-Simons theory is represented in this notation
as in Fig. \ref{cubicvertex}.
Once we have rewritten the Feynman rules in the double-line notation, we can construct the
corresponding graphs, which look like ribbons and are called ribbon graphs or fatgraphs.
A usual Feynman diagram can give rise to many different fatgraphs.
For example, the two-loop vacuum diagram $\twoVgraph$ in Chern-Simons theory, which
comes from contracting two cubic vertices, gives rise to two fatgraphs. The first
one, which is shown in Fig. \ref{ncube}, gives a group factor $2 N^3$, while the second one, which is shown in 
Fig. \ref{justn}, gives $-2 N$. The advantage
of introducing fatgraphs is precisely that each of them gives a definite power of $N$: fatgraphs are
characterized topologically by the number of propagators $E$, the number of
vertices $V$, and finally the
number of closed loops, $h$. If we denote by $x$ the coupling constant, each
propagator gives a power of $x$, each interaction vertex
gives a power of $x^{-1}$, and
each closed loop gives a power of $N$, so that every
fatgraph will give a contribution in $x$ and $N$ given by
\begin{equation}
\label{fatfactor}
x^{E-V}N^h.
\end{equation}
The key point now is to regard the fatgraph as a Riemann surface with holes, in which each
closed loop represents the boundary of a hole. The genus of such a surface is determined by the
elementary topological relation
\be
2g-2 =E-V-h
\end{equation}
therefore we can write (\ref{fatfactor}) as
\be
\label{topfactor}
x^{2g-2+h}N^h=x^{2g-2}t^h
\end{equation}
where we have introduced the 't Hooft parameter $t=Nx$. Fatgraphs with
$g=0$ are called {\it planar}, while the
ones with $g>0$ are called {\it nonplanar}. The diagram in Fig. \ref{ncube}, for example, 
is planar: it has $E=3$, $V=2$ and $h=3$, therefore $g=0$. The diagram in Fig. \ref{justn} is 
nonplanar: it has $E=3$, $V=2$ and $h=1$, therefore $g=1$.

We can now organize the computation of the different quantities in the field
theory in terms of fatgraphs. For example, the computation of the free energy is
given in the usual perturbative expansion by connected vacuum bubbles. When the
vacuum bubbles are written in the double line notation, we find that the
perturbative expansion of the free energy can be written as
\begin{equation}
F^{\rm p}=
\sum_{g=0}^{\infty} \sum_{h=1}^{\infty} F^{\rm p}_{g,h} x^{2g-2} t^h,
\label{openf}
\end{equation}
where $F^{\rm p}_{g,h}$ is simply a
number that can be computed by the usual rules
of perturbation theory. The superscript p refers to the perturbative free energy. As we will see, the
total free energy may have a nonperturbative contribution which is not
captured by Feynman diagrams. In (\ref{openf}) we have written the diagrammatic series as an
expansion in $x$ around $x=0$, keeping the 't Hooft parameter $t=xN$ fixed. Equivalently, we can
regard it as an
expansion in $1/N$, keeping $t$ fixed, and then the $N$ dependence appears
as $N^{2-2g}$. Therefore, for $t$ fixed and $N$ large, the leading contribution comes
from planar diagrams, which go like ${\cal O}(N^2)$. The nonplanar diagrams give subleading
corrections. Notice that the contribution of a given order in $N$ (or in $x$) is given by an
infinite series where we sum over all possible numbers of holes $h$, weighted by $t^h$.

In Chern-Simons theory we are also interested in computing
the vacuum expectation values of Wilson loop operators. The $1/N$ expansion
of Wilson loops can be easily analyzed (see for example Coleman, 1988),
and it turns out that the correlation functions that have a well-defined
behaviour are the connected vacuum expectation values $W_{\vec k}^{(c)}$
introduced in (\ref{convev}). They admit an expansion
of the form,
\begin{equation}
  W_{\vec k}^{(c)}=\sum_{g=0}^{\infty} W_{\vec k, g} (t)  x^{2g-2 + |\vec k|},
\label{opencon}
\end{equation}
where $W_{\vec k, g} (t)$ is a function of the 't Hooft parameter and $|\vec k|$ is defined
in (\ref{ksum}).

\subsection{The $1/N$ expansion in Chern-Simons theory}

The above considerations on the $1/N$ expansion are valid
for any $U(N)$ gauge
theory, and in particular one should be able to expand the free energy of Chern-Simons theory
on the three-sphere as in (\ref{openf}) \footnote{For earlier work on the $1/N$ expansion of 
Chern-Simons theory, see Camperi {\it et al.} (1990), Periwal (1993) and Correale and 
Guadagnini (1994).}. Of course,
the computation of $F_{g,h}$ in Chern-Simons theory directly from perturbation theory
is difficult, since it involves the evaluation of integrals
of products of propagators over the three-sphere. However, since we know the exact answer for the
partition function, we just have to expand it to obtain (\ref{openf}) and the explicit expression
for $F_{g,h}$.

The partition function of CS with gauge group $U(N)$ on the
three-sphere can be obtained from the formula (\ref{css}) for $SU(N)$ after
multiplying it by an overall $N^{1/2}/(k+N)^{1/2}$, which is the
partition function of the $U(1)$ factor. The final result is
\be
Z={1 \over (k+N)^{N/2}} \prod_{\alpha>0}
2 \sin \Bigl( {\pi (\alpha \cdot \rho) \over k+N} \Bigr).
\label{css3}
\end{equation}
Using the explicit description of the positive roots of
$SU(N)$, one gets
\be
F=\log Z= -{N \over 2} \log (k+N) +
\sum_{j=1}^{N-1} (N-j) \log \biggl[ 2 \sin { \pi j \over k+N} \biggr].
\label{csfs3}
\end{equation}
We can now write the $\sin$ as
\be
\sin \pi z = \pi z \prod_{n=1}^{\infty}\biggl( 1 -{ z^2 \over n^2} \biggr),
\end{equation}
and we find that the free
energy is the sum of two pieces. We will call the first one the
{\it nonperturbative} piece:
\ben
& & F^{\rm np}=-{N^2 \over 2}\log (k+N) \nonumber \\
&+& {1 \over 2}N(N-1) \log 2\pi +
 \sum_{j=1}^{N-1} (N-j) \log j,
\label{np}
\een
and the other piece will be called the {\it perturbative} piece:
\be
F^{\rm p}
=\sum_{j=1}^{N} (N-j) \sum_{n=1}^{\infty} \log \biggl[ 1 - {j^2  g_s^2
\over 4 \pi^2 n^2}\biggr],
\label{compli}
\end{equation}
where we have denoted
\be
g_s ={2\pi \over k+ N}
\label{oscc}
\ee
which, as we will see later, coincides with the open string coupling constant
under the gauge/string theory duality.

To see that (\ref{np}) has a nonperturbative origin, we notice
that (see for example Ooguri and Vafa, 2002):
\be
\label{volun}
{\rm vol}(U(N))={ (2\pi)^{ {1 \over 2}N(N+1)} \over G_2(N+1)},
\ee
where $G_2(z)$ is the Barnes function, defined by
\be
G_2 (z+1)=\Gamma (z) G_2(z), \,\,\,\,\, G_2(1)=1.
\ee
One then finds
\be
F^{\rm np}=\log \, { (2 \pi g_s)^{{1 \over 2}N^2}
\over {\rm vol}(U(N))}.
\end{equation}
This indeed corresponds to the volume of the gauge group in the
one-loop contribution (\ref{asym}), where $A^{(c)}$
is in this case the trivial flat connection. Therefore, $F^{\rm np}$ is the log of
the prefactor of the path integral, which is not captured by Feynman diagrams.

Let us work out now the perturbative piece (\ref{compli}), following
Gopakumar and Vafa (1998a and 1999).
By expanding the $\log$, using that $\sum_{n=1}^{\infty} n^{-2k}=\zeta
(2k)$, and the formula
\be
\sum_{j=1}^{N} j^{k} = {1\over k+1}\sum_{l=1}^{k+1}(-1)^{\delta_{lk}}{k +1
\choose l} B_{k+1-l}N^{l},
\end{equation}
where $B_n$ are Bernoulli numbers, we find that (\ref{compli}) can be written as
\be
F^{\rm p} = \sum_{g=0}^{\infty} \sum_{h=2}^{\infty} F^{\rm p}_{g,h}
g_s^{2g-2+h} N^h,
\end{equation}
where $F^{\rm p}_{g,h}$ is given by:
\ben
F^{\rm p}_{0,h}&=& -{|B_{h-2}| \over (h-2) h!}, \quad h\ge 4, \nonumber\\
F^{\rm p}_{1,h}&=&{1 \over 12}{ |B_h| \over h\, h!},
\label{fgzerone}
\een
($F^{\rm p}_{0,h}$ vanishes for $h\le 3$) and for $g\ge 2$ one finds:
\be
F^{\rm p}_{g,h}= {\zeta (2g-2+h) \over (2 \pi)^{2g-3+h}}
{2g-3+h \choose h}{ B_{2g} \over 2g (2g-2)}.
\label{fghs3}
\ee
This gives the contribution of connected diagrams with
two loops and beyond to the free energy of Chern-Simons on the sphere.
The nonperturbative piece also admits an asymptotic expansion that
can be easily worked out by expanding the Barnes function
(Periwal, 1993; Ooguri and Vafa 2002). One finds:
\ben
F^{\rm np}&= &{N^2 \over 2} \Bigl( \log (Ng_s) -{3 \over 2} \Bigr)
-{1 \over 12}\log N + \zeta' (-1) \nonumber \\
&+& \sum_{g=2}^{\infty} {B_{2g} \over
2g (2g-2)} N^{2-2g}.
\een

\subsection{The string interpretation of the $1/N$ expansion}

The expansion (\ref{openf}) of the free energy in a $U(N)$ gauge theory looks very much
like the perturbative expansion of an {\it open}
string theory with $U(N)$ Chan-Paton factors, where $x$ is the
open string coupling constant, and
$F_{g,h}$ corresponds to
some amplitude on a Riemann surface of genus $g$ with $h$ holes. There is in fact a way
to produce a {\it closed} string theory interpretation of the free energy of gauge theories. Let
us introduce the function
\begin{equation}
F^{\rm p}_g (t) =\sum_{h=1}^{\infty} F^{\rm p}_{g,h} t^h.
\label{opencl}
\end{equation}
The perturbative free energy can be written now as
\begin{equation}
F^{\rm p}=\sum_{g=0}^{\infty} x^{2g-2} F^{\rm p}_g(t),
\label{closedf}
\end{equation}
which looks like a {\it closed} string expansion where $t$ is some
modulus of the theory. Notice that (\ref{opencl}) contains the
contribution of all open Riemann surfaces that appear in the perturbative series
with the same bulk topology (specified by the genus $g$), but with
different number of holes. Therefore, by ``summing over all holes'' we are
``filling up the holes'' to produce a closed Riemann surface of genus $g$. This leads to the
't Hooft idea (1974) that, given a gauge theory, one should be able to find a string
theory interpretation in the way we have described, namely, the fatgraph
expansion of the free energy is resummed to give a function of the 't
Hooft parameter $F^{\rm p}_g(t)$ at every genus, which is then interpreted as a
closed string amplitude. For example, the planar sector of the gauge
theory corresponds to a closed string theory at tree level ({\it i.e.} at genus $g=0$).
Although we are only considering
here the perturbative piece of the free energy, we will see that in the Chern-Simons case
the nonperturbative piece is crucial to obtain the closed string picture.

Once a closed string intepretation is available, the $1/N$ expansion
(\ref{opencon}) can be regarded
as an open string expansion, where $W_{\vec k,g}(t)$ are interpreted as
amplitudes in an open string theory at genus $g$ and with $h=| \vec k|$ holes.
According to this interpretation, the Wilson loop creates a one-cycle in the target
space where the boundaries of Riemann surfaces can end. 
The vector $\vec k$ specifies the winding numbers for the boundaries as follows:
there are $k_j$ boundaries wrapping $j$ times the one-cycle associated to the Wilson
loop. The generating
functional for connected vacuum expectation values (\ref{convev}) is interpreted as the total
free energy of an open string. The open strings that are relevant to the string
interpretation of Wilson loop amplitudes
should not be confused with the open strings that we associated to the
expansion (\ref{openf}). The open strings underlying (\ref{opencon}) should
be regarded as an open string sector in the closed string theory associated
to the resummed expansion (\ref{closedf}).

From the point of view of perturbation theory, the functions $F^{\rm p}_g(t)$ are rather formal,
and the definition
(\ref{opencl}) expresses them as a power series in $t$ whose
coefficients have to be computed order by order in perturbation theory. In some cases
the series can be exactly summed up in $h$ and the functions $F^{\rm p}_g(t)$ can then be obtained in closed
form (this is the case, for example, in some matrix models). We will see later that in the
case of Chern-Simons theory the $F^{\rm p}_g(t)$ can be also resummed to give a function of the
't Hooft coupling $t$.

Of course, the main problem of the 't Hooft program is to identify the closed string theory underlying
a gauge theory. This program has been sucessful in some cases, and string theory
descriptions have been found for two-dimensional Yang-Mills theory (Gross, 1993, and Gross and Taylor, 1993)
and for ${\cal N}=4$ Yang Mills theory in four dimensions (Maldacena, 1998). As we will see in this review,
Chern-Simons theory also admits a string theory description in terms of topological strings, which we now
introduce.

\section{Topological strings}

String theory can be regarded, at the algebraic level,
as a two-dimensional conformal field theory coupled to two-dimensional
gravity. When the conformal field theory is
in addition a topological field theory ({\it i.e.} a theory whose
correlation functions do not depend on the metric on the Riemann surface),
the resulting string theory
turns out to be very simple and in many cases can be completely solved. A string
theory which is constructed in this way is called a {\it topological string
theory}.

The starting point to obtain a topological string theory
is therefore a conformal field theory with topological invariance.
Such theories are called topological conformal field theories
and can be constructed out of
${\cal N}=2$ superconformal field theories in two dimensions by a procedure
called twisting (see Dijkgraaf {\it et al.}, 1991, for a review of these topics).
For example, one can take the
${\cal N}=2$ minimal models to obtain the so-called topological
strings in $d<1$. These models are very beautiful and interesting and are
deeply related to non-critical string theories. In this review we will consider
a more complicated class of topological string theories, where the topological
field theory is taken to be a topological sigma model with target space a
Calabi-Yau manifold. We will first review the topological sigma model and then
explain its coupling to gravity in order to obtain a topological string.
We will also introduce some ingredients of toric geometry which are needed to 
fully understand the class of models that we will consider in this review.

\subsection{Topological sigma models}

The topological sigma model was introduced and studied by Witten in
a series of papers (1988, 1990, 1991a, 1991b)
and can be constructed by twisting the ${\cal N}=2$
superconformal sigma model in two dimensions (see also Labastida and Llatas, 1991).
A detailed review of topological
sigma models and topological strings can be found in Hori {\it et al.} (2003).

The field content of the topological sigma model is the following. First, we
have a map $x: \Sigma_g \rightarrow X$ from a Riemann surface of genus
$g$ to a target space $X$, that will be a K\"ahler manifold of complex
dimension $d$. Indices in the tangent space of $X$ will be denoted by $i$,
with $i=1, \cdots, 2d$. Since
we have a complex structure, we will also have holomorphic and antiholomorphic
indices, that we will denote respectively by $I, {\overline I}$, where $I, {\overline I}
=1, \cdots, d$.
We also have Grassmann fields $\chi \in x^*(TX)$, which are
scalars on $\Sigma_g$, and a Grassmannian one form $\psi_{\alpha}$ with
values in $x^*(TX)$. This last field satisfies a selfduality condition
which implies that its only nonzero components are $\psi_{\bar z}^I
\in x^*(T^{(1,0)}X)$ and $\psi_{ z}^{\overline I}
\in x^*(T^{(0,1)}X)$, where we have picked up local
coordinates $z, \bar z$ on $\Sigma_g$. The action for the theory is:
\begin{widetext}
\be
\label{tsaction}
{\cal L}=2t\int_{\Sigma_g} d^2 z \biggl( {1 \over 2}
G_{ij} \partial_z x^i \partial_{\bar z} x^j + i G_{I \overline J}
\psi_{\bar z}^I D_{z}\chi^{\overline J} + i G_{{\overline I} J}
\psi_{z}^{\overline I} D_{\bar z}\chi^{J}- R_{I {\overline I} J {\overline J}}
\psi_{\bar z}^{I} \psi_{z}^{\overline I}\chi^J \chi^{\overline J}\biggr),
\end{equation}
\end{widetext}
where $d^2 z$ is the measure $-i dz \wedge d\bar z$, $t$ is a parameter
that plays the role of $1/\hbar$, the covariant derivative $D_{\alpha}$ is given by
\be
D_{\alpha} \chi^i =\partial_{\alpha} \chi^i + \partial_{\alpha} x^j \Gamma^i_{jk}\chi^k.
\ee
The theory also has a BRST, or topological, charge
$\CQ$ which acts on the fields according to
\begin{eqnarray}
\{ \CQ, x \} &=& \chi, \nonumber\\
\{ \CQ, \chi \} &=&0, \nonumber\\
\{ \CQ, \psi_{\bar z}^I \} &=&i\partial_{\bar z}x^I - \chi^J
\Gamma_{JK}^I \psi_{\bar z}^K, \nonumber\\
\{ \CQ, \psi_z^{\overline I} \} &=&i\partial_z x^{\overline I} -
\chi^{\overline J}
\Gamma_{{\overline J} {\overline K}}^{\overline I}
\psi_{z}^{\overline K}. \nonumber\\
\label{qtrans}
\end{eqnarray}
One can show that $\CQ^2=0$ on-shell ({\it i.e.} modulo the equations of motion).
Finally, we also have a $U(1)$ ghost number symmetry, in which
$x$, $\chi$ and $\psi$ have
ghost numbers $0$, $1$ and $-1$, respectively. Notice that the Grassmannian charge
$\CQ$ has then ghost number $1$.

The action (\ref{tsaction}) turns out to be $\CQ$-exact,
up to a topological term and terms that vanish on-shell (Witten, 1988 and 1991b):
\begin{equation}
{\cal L}=-i \{ \CQ, V\} - t \int_{\Sigma_g} x^*(J),
\label{toplag}
\end{equation}
where $J =i G_{I {\overline J}} dx^I \wedge dx^{\overline J}$ is the
K\"ahler class of $X$, $V$
(sometimes called the gauge fermion) is given by
\begin{equation}
V=t\int_{\Sigma_g} d^2z  G_{I {\overline J}} ( \psi_{\bar z}^I
\partial_z x^{\overline J} + \partial_{\bar z} x^I \psi_{z}^{\overline J})
\label{gaugef}
\end{equation}
and
\begin{equation}
\int_{\Sigma_g} x^*(J)=i \int_{\Sigma_g} dz \wedge d \bar z \, G_{I {\overline J}} \bigl( \partial_z x^I
\partial_{\bar z} x^{\overline J} -\partial_{\bar z} x^I
\partial_z x^{\overline J}\bigr).
\end{equation}
Notice that this term in (\ref{toplag}) is a topological invariant
characterizing the homotopy type of the map $x: \Sigma_g \rightarrow X$. We can
also add a coupling to a $B$-field into the action,
\be
-it\int_{\Sigma_g} x^*(B),
\end{equation}
which will replace the K\"ahler form by the complexified K\"ahler form $\omega=J+ iB$.
It is easy to covariantize (\ref{tsaction}) and (\ref{gaugef}) to introduce an
arbitrary metric $g_{\alpha \beta}$ on $\Sigma_g$. Since
the last term in (\ref{toplag}) is topological,
the energy-momentum tensor of this theory is given by:
\begin{equation}
T_{\alpha \beta}= \{\CQ, b_{\alpha \beta} \},
\label{qex}
\end{equation}
where $b_{\alpha \beta}= \delta V/\delta g^{\alpha \beta}$ and has ghost number $-1$. The fact that
the energy-momentum tensor is $\CQ$-exact means that the theory is
topological, in the sense that the partition function does not depend on the
background two-dimensional metric. This is easily proven: the partition function
is given by
\be
Z=\int {\cal D} \phi e^{-{\cal L}},
\end{equation}
where $\phi$ denotes the set of fields of the theory, and we compute it in the background of a
two-dimensional metric on the Riemann surface $g_{\alpha \beta}$. Since $T_{\alpha \beta}=
\delta {\cal L}/\delta g^{\alpha \beta}$, we find that
\be
{\delta Z \over \delta g^{\alpha \beta}}=-\langle \{\CQ, b_{\alpha \beta} \} \rangle,
\end{equation}
where the bracket denotes an unnormalized vacuum expectation value. Since $\CQ$
is a symmetry of the
theory, the above vacuum expectation value vanishes, and we find that $Z$ is metric-independent, at least formally.

The $\CQ$-exactness of the action itself has also an important consequence:
the same argument that we used above implies that the partition function
of the theory is independent of $t$. Now, since $t$ plays the role of $1/\hbar$, the limit of
$t$ large corresponds to the semiclassical approximation. Since the theory does not
depend on $t$, the
semiclassical approximation is {\it exact}. Notice that the classical configurations
for the above action are holomorphic maps  $x:
\Sigma_g \rightarrow X$. These are the instantons of the nonlinear sigma model with a
K\"ahler target, and minimize the bosonic action. The different instanton sectors are
classified topologically by the homology class
\be
\beta=x_*[(\Sigma_g)] \in H_2(X, {\bf Z}).
\ee
Sometimes it is also useful to introduce a basis $[S_i]$ of $H_2(X,{\bf Z})$, where $i=1, \cdots, b_2(X)$, in such
a way that we can expand $\beta=\sum_i n_i [S_i]$ and the instanton sectors are
labelled by $b_2(X)$ integers $n_i$.

What are the operators to consider in this theory? Since the
most interesting aspect of this model is the independence w.r.t. the metric,
we want to look for operators whose correlation functions satisfy this
condition. It is easy to see that the operators in the cohomology of $\CQ$
do the job: topological invariance requires them to be
$\CQ$-closed, and on the other hand they cannot be $\CQ$-exact, since otherwise their
correlation functions will vanish. One can also check that the $\CQ$-cohomology is
given by operators of the form
\begin{equation}
{\cal O}_{\phi}=\phi_{i_1 \cdots i_p} \chi^{i_1} \cdots \chi^{i_p},
\label{qops}
\end{equation}
where $\phi=\phi_{i_1 \cdots i_p}dx^{i_1} \cdots dx^{i_p}$ is a closed
$p$-form representing a nontrivial class in $H^p(X)$. Therefore, in this case
the $\CQ$-cohomology is in one-to-one correspondence with the de Rham cohomology of the
target manifold $X$. Also notice that the degree of the differential
form corresponds to the ghost number of the operator. Moreover, one can
derive a selection rule for correlation functions of such operators:
the vacuum expectation value $\langle {\cal O}_{\phi_1} \cdots
{\cal O}_{\phi_\ell} \rangle$ vanishes unless
\begin{equation}
\label{selrule}
\sum_{k=1}^{\ell} {\rm deg}({\cal O}_{\phi_k})=
2d(1-g) + 2\int_{\Sigma_g} x^*(c_1(X)),
\end{equation}
where ${\rm deg}({\cal O}_{\phi_k})={\rm deg}(\phi_k)$ and $c_1(X)$ is the
first Chern class of the K\"ahler manifold $X$. This selection rule
corresponds to the fact that the $U(1)$ current is anomalous, and the anomaly
is given by the r.h.s. of (\ref{selrule}), which calculates the number of
zero modes of the Dirac operator (in other words, the r.h.s.
is minus the ghost number of the vacuum). As usual in quantum field theory,
the operators with nontrivial vacuum expectation values have to soak up the
zero modes associated to the anomaly.

In what follows we will focus on Calabi-Yau threefolds, {\it i.e.} K\"ahler manifolds
of complex dimension 3, and such that $c_1(X)=0$. For these manifolds
the selection rule says that, at genus $g=0$ ({\it i.e.} when the Riemann
surface is a sphere) the correlation function of
three operators associated to 2-forms is generically nonvanishing. Since,
as we have seen, the semiclassical
approximation is exact, the correlation function can be
evaluated by counting semiclassical configurations, or in other
words by summing over worldsheet instantons. In the trivial sector
({\it i.e.} when $\beta=0$ and the image of the sphere is a point in the
target), the
correlation function is just the classical intersection number $D_1 \cap
D_2 \cap D_3$ of the three
divisors $D_i$, $i=1,2,3$, associated to the 2-forms, while the nontrivial
instanton sectors give an infinite series. The final answer looks,
schematically,
\begin{equation}
\langle {\cal O}_{\phi_1}{\cal O}_{\phi_2}
{\cal O}_{\phi_3} \rangle=(D_1 \cap D_2 \cap D_3) + \sum_{\beta}
I_{0,3, \beta}(\phi_1,\phi_2, \phi_3)
Q^{\beta}
\label{threepoint}
\end{equation}
The notation is as follows: we have introduced the complexified
K\"ahler parameters
\be
t_i=\int_{S_i} \omega,\quad i=1, \cdots, b_2(X),
\label{tipar}
\ee
where $\omega$ is the complexified K\"ahler form of $X$, and $S_i$ is a
basis of $H_2(X)$. We also define $Q_i=e^{-t_i}$, and if $\beta=\sum_i n_i [S_i]$,
then $Q^{\beta}$ denotes $\prod_i Q_i^{n_i}$. The coefficient
$I_{0,3, \beta}(\phi_1,\phi_2, \phi_3)$ ``counts'' in an appropriate sense
the number of holomorphic maps from the sphere to the Calabi-Yau that send 
the point of insertion of ${\cal O}_{\phi_i}$ to the
divisor $D_i$. It can
be shown that the coefficients $I_{0,3, \beta}(\phi_1,\phi_2, \phi_3)$
can be written as
\begin{equation}
I_{0,3, \beta}(\phi_1, \phi_2, \phi_3)=N_{0,\beta}\int_{\beta} \phi_1  \int_{\beta} \phi_2
\int_{\beta} \phi_3
\label{geno}
\end{equation}
in terms of invariants
$N_{0,\beta}$ that encode all the
information about the three-point functions (\ref{threepoint})
of the topological sigma model. The invariants $N_{0,\beta}$
are our first example of {\it Gromov-Witten invariants}. It is
convenient to put all these invariants together in a generating
functional called the {\it prepotential}:
\be
F_0(t)=\sum_{\beta} N_{0,\beta} \, Q^{\beta}.
\label{prep}
\ee

What happens if we go beyond $g=0$? For $g=1$ and $c_1(X)=0$, the
selection rule (\ref{selrule}) says that the only quantity that
may lead to a nontrivial answer is the partition function itself, while
for $g>1$ all correlation functions vanish. This
corresponds mathematically to the fact that, for a generic metric on the
Riemann surface $\Sigma_g$, there are no holomorphic maps at genus
$g>1$. In order to circumvent this problem, we have to consider the
topological string theory made out of the topological sigma model,
{\it i.e.} we have to couple the theory to
two-dimensional gravity and to consider all possible metrics on
the Riemann surface.

\subsection{Closed topological strings}

\subsubsection{Coupling to gravity}

To couple the topological sigma model to gravity, we
use the fact pointed out by Dijkgraaf {\it et al.} (1991), Witten (1995) and
Bershadsky {\it et al.} (1994) that the structure of the twisted theory is
tantalizingly close to that of the bosonic string. In the bosonic string, there is
a nilpotent BRST operator, $\CQ_{\rm BRST}$, and the energy-momentum tensor
turns out to be a $\CQ_{\rm BRST}$-commutator: $T(z)=\{\CQ_{\rm BRST}, b(z)
\}$. In addition, there is a ghost number with anomaly $3 \chi (\Sigma_g)=
6-6g$, in such a way that $\CQ_{\rm BRST}$ and $b(z)$
have ghost number $1$ and $-1$, respectively. This is precisely the
same structure that we found in (\ref{qex}), and
the composite field $b_{\alpha \beta}$ plays the role of an antighost. Therefore, one can
just follow the prescription of coupling to gravity for the bosonic string
and define a genus $g\ge 1$ free energy as follows:
\begin{equation}
F_g= \int_{{\overline M}_{g}} \langle \prod_{k=1}^{6g-6} (b, \mu_k)
\rangle,
\label{fg}
\end{equation}
where
\begin{equation}
(b, \mu_k)=\int_{\Sigma_g} d^2 z (b_{zz}(\mu_k)_{\bar z}^{~z} + b_{\bar z \bar z}
({\overline \mu}_k)_{z}^{~\bar z}),
\end{equation}
and $\mu_k$ are the usual Beltrami differentials.
The vacuum expectation value in (\ref{fg}) refers to the path integral over the
fields of the topological sigma model, and gives a differential form
on the moduli space of Riemann surfaces of genus $g$, ${\overline M}_g$,
which is then integrated over. Notice that it is
precisely when the target space is a Calabi-Yau
threefold that the anomaly (\ref{selrule})
is exactly the one of the usual bosonic string. In that
sense, one can say that topological strings whose target is a Calabi-Yau
threefold are critical.

It turns out that the free energies $F_g$, $g\ge 1$, can be also evaluated
as a sum over instanton sectors, like in the
topological sigma model. Therefore they have
the structure
\begin{equation}
F_g(t)= \sum_{\beta}N_{g,\beta}\, Q^{\beta},
\label{gwgenf}
\end{equation}
where $N_{g, \beta}$ ``count" in an appropriate sense the number
of curves of genus $g$ and in the two-homology class $\beta$.
We will refer to $N_{g, \beta}$ as the Gromov-Witten invariant of the
Calabi-Yau $X$ at genus $g$ and in the class $\beta$. They generalize the
Gromov-Witten invariants at genus $0$ that were introduced in (\ref{geno}).

\subsubsection{Mathematical description}

The Gromov-Witten invariants that we introduced in (\ref{gwgenf})
can be defined in a rigorous mathematical way, and
have played an important role in algebraic geometry and symplectic geometry. We
will now give a short summary of the main mathematical ideas involved in
Gromov-Witten theory.

The coupling of the model to gravity involves the moduli
space of Riemann surfaces ${\overline M}_g$, as we have just seen.
In order to construct the Gromov-Witten invariants in full generality
we also need the moduli space of possible metrics (or equivalently, complex structures)
on a Riemann surface with punctures, which
is the famous Deligne-Mumford space
${\overline M}_{g,n}$ of $n$-pointed stable curves (the definition of what
stable means can be found for example in Harris and Morrison, 1998). The relevant moduli space in
the theory of topological strings ${\overline
M}_{g,n}(X,\beta)$ is a generalization of ${\overline M}_{g,n}$, and depends on
a choice of a two-homology class $\beta$ in $X$. Very roughly, a
point in ${\overline M}_{g,n}(X,\beta)$ can be written as
$(f,\Sigma_g, p_1, \cdots, p_n)$ and is given by (a) a
point in ${\overline M}_{g,n}$, {\it i.e.} a
Riemann surface with $n$ punctures, $(\Sigma_g, p_1,
\cdots, p_n)$, together with a choice of complex structure on $\Sigma_g$, and
(b) a map $f:\Sigma_g \rightarrow X$ which is holomorphic
with respect to this choice of complex structure and such that
$f_*[\Sigma_g]=\beta$. The set of all such points forms a good moduli
space provided a
certain number of conditions are satisfied (see for example Cox and Katz (1999) and Hori {\it
et al.} (2003) for
a detailed discussion of these issues). ${\overline
M}_{g,n}(X,\beta)$ is the basic moduli space
we will need in the theory of topological strings. Its complex virtual dimension is
given by:
\begin{equation}
(1-g)(d -3) + n + \int_{\Sigma_g} f^*(c_1(X)),
\end{equation}
which is given by the r.h.s. of (\ref{selrule}) plus $3g-3+n$, which
is the dimension of ${\overline M}_{g,n}$ and takes into account the
extra moduli that come from the coupling to two-dimensional gravity.
We also have two natural maps
\begin{eqnarray}
\pi_1:  {\overline
M}_{g,n}(X,\beta)& \longrightarrow & X^n,\nonumber\\
\pi_2:    {\overline
M}_{g,n}(X,\beta) &\longrightarrow &  {\overline M}_{g,n}.
\end{eqnarray}
The first map is easy to define: given a point
$(f,\Sigma_g, p_1, \cdots, p_n)$ in ${\overline
M}_{g,n}(X,\beta)$, we just compute $(f(p_1), \cdots, f(p_n))$. The second
map essentially sends $(f,\Sigma_g, p_1, \cdots, p_n)$ to
$(\Sigma_g, p_1, \cdots, p_n)$, {\it i.e.} forgets
the information about the map and keeps the information about the
punctured curve. We can now formally define the
Gromov-Witten invariant $I_{g,n,\beta}$ as follows. Let us consider
cohomology classes $\phi_1, \cdots, \phi_n$ in $H^*(X)$. If we pullback
their tensor product to $H^*({\overline
M}_{g,n}(X,\beta))$ via $\pi_1$, we get a differential form on the moduli
space of maps that we can integrate (as long as there is a well-defined
fundamental class for this space):
\begin{equation}
I_{g,n,\beta}(\phi_1, \cdots, \phi_n)=\int_{{\overline
M}_{g,n}(X,\beta)} \pi_1^* (\phi_1 \otimes \cdots \otimes \phi_n).
\label{gwinv}
\end{equation}
The Gromov-Witten
invariant $I_{g,n,\beta}(\phi_1, \cdots, \phi_n)$ vanishes unless the
degree of the form equals the dimension of the moduli space. Therefore, we
have the following constraint:
\begin{equation}
{1 \over 2}\sum_{i=1}^n {\rm deg}(\phi_i)=(1-g)(d -3) + n +
\int_{\Sigma_g} f^*(c_1(X)).
\label{degcons}
\end{equation}
Notice that Calabi-Yau threefolds play a special role in the theory,
since for those targets the virtual dimension only depends on the number of
punctures, and therefore the above condition is always satisfied if the
forms $\phi_i$ have degree 2. These invariants generalize the invariants
obtained from topological sigma models. In particular, $I_{0,3, \beta}$ are
the invariants involved in the evaluation of correlation functions of the
topological sigma model with a Calabi-Yau threefold as its target in
(\ref{threepoint}). When $n=0$, one gets an
invariant $I_{g,0,\beta}$ which does not require any
insertions. This is precisely the Gromov-Witten invariant
$N_{g,\beta}$ that appears in (\ref{gwgenf}). Notice that these invariants
are in general {\it rational}, due to the orbifold character of the moduli
spaces involved.

By using the Gysin map $\pi_{2!}$, one can reduce any integral of the
form (\ref{gwinv}) to an integral over
the moduli space of curves ${\overline M}_{g,n}$. The resulting integrals involve
two types of differential forms. The first type of forms are the Mumford
classes $\psi_i$, $i=1, \cdots, n$, which are constructed as follows.
We first define the line bundle ${\mathcal L}_i$ over ${\overline M}_{g,n}$ to be the line
bundle whose fiber over each curve $\Sigma_g$ is the cotangent space
of $\Sigma_g$ at $p_i$ (where $p_i$ is the $i$-th marked point). We then have,
\begin{equation}
\label{psiclas}
\psi_i =c_1 ({\mathcal L}_i),\,\,\,\ i=1, \cdots, n.
\end{equation}
The second type of differential forms are the Hodge classes $\lambda_j$, $j=1, \cdots, g$,
which are defined as follows. On ${\overline M}_{g,n}$ there is a complex vector bundle $\DE$
of rank $g$, called the Hodge bundle, whose fiber at a point $\Sigma_g$ is
$H^0 (\Sigma_g, K_{{\Sigma}_g})$ ({\it i.e.} the space of holomorphic sections of the canonical line
bundle $K_{{\Sigma}_g}$ of $\Sigma_g$). The Hodge classes are simply the Chern
classes of this bundle,
\be
\lambda_j=c_j(\DE).
\label{hodgeclass}
\ee
The integrals of the $\psi$ classes can be obtained by the
results of Witten (1991a) and Kontsevich (1992),
while the integrals involving
$\psi$ and $\lambda$ classes (the so-called Hodge integrals) can be in
principle computed by reducing them to pure $\psi$ integrals (Faber, 1999).
Explicit formulae for some Hodge integrals can be found for
example in Getzler and Pandharipande (1998) and Faber and Pandharipande (2000). 
As we will see, one of the outcomes of the string/gauge correspondence for 
Chern-Simons theory is an explicit formula for a wide class of Hodge integrals.

\subsubsection{Integrality properties}

The free energies $F_g$ of topological string theory,
which contain information about the Gromov-Witten invariants
of the Calabi-Yau manifold $X$, turn out to play an important role in
type IIA string theory: they
capture certain couplings in the four-dimensional ${\cal N}=2$
supergravity which is obtained when type IIA theory is compactified
on $X$. For example, the prepotential $F_0$ encodes the information about
the effective action for vector multiplets up to two derivatives.
As shown by Bershadsky {\it et al.} (1994) and Antoniadis
{\it et al.} (1994), the higher genus amplitudes $F_g$ with $g \ge 1$ can be
also interpreted as couplings in the 4d supergravity theory, involving the curvature
and the graviphoton field strength.

This connection between topological strings and usual type II
superstrings has been a source of insights for both models, and in
particular has indicated a hidden integrality structure in
the Gromov-Witten invariants $N_{g, \beta}$. In order to
make manifest this structure it is
useful to introduce a generating functional for the all-genus
free energy:
\begin{equation}
F(g_s,t)=\sum_{g=0}^{\infty} F_g(t) g_s^{2g-2}.
\label{freen}
\end{equation}
The parameter $g_s$ can be regarded as a formal variable, but in the context
of type II strings it is nothing but the string coupling constant. Gopakumar and Vafa (1998b)
showed that the generating
functional (\ref{freen}) can be written as a generalized index that
counts BPS states in the type IIA superstring theory compactified
on $X$, and this leads to the following structure result for
$F(g_s, t)$:
\begin{equation}
F(g_s,t)=\sum_{g=0}^{\infty} \sum_{\beta} \sum_{d=1}^{\infty}
n^g_{\beta} {1\over d}\biggl( 2 \sin {d g_s \over 2} \biggr)^{2g-2} Q^{d\beta},
\label{gvseries}
\end{equation}
where $n^g_\beta$, known as Gopakumar-Vafa invariants, are {\it integer}
numbers. Therefore,
Gromov-Witten invariants of closed strings, which are in general
rational, can be written in terms
of integer invariants. In fact, by knowing the Gromov-Witten
invariants we can explicitly compute the Gopakumar-Vafa invariants from
(\ref{gvseries}). The Gopakumar-Vafa invariants can be also computed
in some circumstances directly in terms of the geometry of embedded curves,
and in many cases their computation only involves elementary algebraic
geometry (Katz, Klemm, and Vafa, 1999). However, a rigorous mathematical
definition of the invariants is not known yet.

There is also a contribution of constant maps to $F_g$ for $g\ge 2$
which has not been included in
(\ref{gvseries}) and is given by $N_{g,0}$. It was shown by Bershadsky {\it et al.}
(1994) (see also Getzler and Pandharipande, 1998) that
this contribution can be written as a Hodge integral
\be
N_{g,0}=(-1)^g {\chi (X) \over 2} \int_{{\overline M}_g} c_{g-1}^3(\DE), \,\,\,\,\,\,\,\,
g \ge 2,
\label{hint}
\ee
where $\chi(X)$ is
the Euler characteristic of the Calabi-Yau manifold $X$. The above integral
can be evaluated explicitly to give (Faber and Pandharipande, 2000)
\be
N_{g,0}=
{(-1)^g \chi(X) |B_{2g} B_{2g-2}| \over
4g (2g-2)(2g-2)!}.
\end{equation}
This can be also deduced from the physical picture of Gopakumar and Vafa (1998b) and
from type IIA/heterotic string duality (Mari\~no and Moore, 1999).

It is easy to show that the Gopakumar-Vafa formula (\ref{gvseries})
predicts the following expression for $F_g(t)$:
\ben
F_g (t)& =&
\sum_{\beta} \biggl( { |B_{2g}| n_\beta^0 \over
2g (2g-2)!} + {2 (-1)^g n_\beta^2 \over (2g-2)!}
\pm \cdots  \nonumber \\
& & - {g-2 \over 12} n^{g-1}_\beta + n_\beta^g\biggr) {\rm Li}_{3-2g}(Q^\beta),
\label{multibuble}
\een
and ${\rm Li}_j$ is the polylogarithm of index $j$ defined by
\begin{equation}
{\rm Li}_j (x)= \sum_{n=1}^{\infty} {x^n \over n^j}.
\end{equation}
The appearance of the polylogarithm of order $3-2g$ in $F_g$ was
first predicted from type IIA/heterotic string duality by Mari\~no and Moore (1999).

\subsection{Open topological strings}

One can extend many of the previous results to open topological strings.
The natural starting point
is a topological sigma model in which the worldsheet is now a Riemann
surface $\Sigma_{g,h}$ of genus $g$ with $h$ holes. Such models were analyzed in detail by
Witten (1995). The main issue is of course to specify boundary conditions for
the maps $f: \Sigma_{g,h} \rightarrow X$. It turns out that
the relevant boundary conditions are Dirichlet and are specified by
Lagrangian submanifolds of the Calabi-Yau $X$. A Lagrangian
submanifold $\CL$ is a cycle where the K\"ahler form vanishes:
\be
\label{lag}
J|_{\CL} =0.
\end{equation}
If we denote by $C_i$, $i=1,
\cdots, h$ the boundaries of $\Sigma_{g,h}$ we have to pick a
Lagrangian submanifold $\CL$, and consider holomorphic maps such that
\begin{equation}
f(C_i)\subset \CL.
\label{bound}
\end{equation}
These boundary
conditions are a consequence of requiring $\CQ$-invariance at the
boundary. One also has boundary conditions on the Grassmann fields of the
topological sigma model, which require that $\chi$ and $\psi$ at the boundary $C_i$
take values on $f^*(T\CL)$.

We can also couple the theory to
Chan-Paton degrees of freedom on the boundaries,
giving rise to a $U(N)$ gauge symmetry.
The model can then be interpreted as a topological open string theory in the
presence of $N$ {\it topological D-branes} wrapping the Lagrangian submanifold
$\CL$. The Chan-Paton factors give rise to a boundary term in the
presence of a gauge connection. If $A$ is a $U(N)$ connection
on $\CL$, then the
path integral has to be modified by inserting
\be
\prod_i{\rm Tr} \, {\rm P}\, \exp \oint_{C_i} f^*(A),
\label{chancoup}
\end{equation}
where we pullback the connection to $C_i$ through the map $f$, restricted to the
boundary. In contrast to physical D-branes in
Calabi-Yau manifolds,
which wrap special Lagrangian submanifolds (Becker {\it et al.}, 1995;
Ooguri {\it et al.}, 1996), in the
topological framework the conditions are relaxed to just Lagrangian.

Once boundary conditions have been specified, one can define
the free energy of the topological string theory by summing
over topological sectors, similarly to what we did in the closed case.
In order to specify the
topological sector of the map, we have to give two different kinds of data:
the boundary part and the bulk part. For the bulk part, the topological
sector is labelled by relative homology classes, since we are requiring the
boundaries of $f_*[\Sigma_{g,h}]$ to end on $\CL$. Therefore, we will
set
\begin{equation}
f_*[\Sigma_{g,h}]=\beta \in H_2(X, \CL)
\label{bulkpart}
\end{equation}
To specify the topological sector of the boundary, we will assume that
$b_1 (\CL)=1$, so that $H_1 (\CL)$ is generated by a nontrivial
one cycle $\gamma$. We then have
\begin{equation}
f_*[C_i]=w_i \gamma, \,\,\,\,\, w_i \in {\bf Z},\,\,\,\,
i=1, \cdots, h,
\label{wind}
\end{equation}
in other words, $w_i$ is the winding number associated to the map $f$
restricted to $C_i$. We will collect these integers into a single
$h$-uple denoted by $w=(w_1, \cdots, w_h)$.

There are various generating functionals that we can consider, depending on
the topological data that we want to keep fixed. It is very useful to fix
$g$ and the winding numbers, and sum over all bulk classes. This produces
the following generating
functional of open Gromov-Witten
invariants:
\begin{equation}
F_{w,g} (t) =\sum_{\beta} F_{w,g,\beta}\,Q^{\beta}.
\end{equation}
In this equation, the sum is over relative homology classes $\beta \in H_2(X,\CL)$, and
the notation is as in (\ref{threepoint}). The quantities $F_{w,g, \beta}$
are open Gromov-Witten invariants.
They ``count" in an appropriate sense the number of holomorphically
embedded Riemann surfaces of genus $g$ in $X$ with Lagrangian boundary
conditions specified by $\CL$, and in the class represented
by $\beta, w$. They are in general rational numbers.

In order to consider all topological sectors, we have to introduce a
matrix $V$ which makes possible to take into account
different sets of winding numbers $w$. The total free energy is defined by
\begin{widetext}
\be
F(V)=  \sum_{g=0}^{\infty} \sum_{h=1}^{\infty}
\sum_{w_1, \cdots, w_h} {i^h \over h!}
g_s^{2g-2+h} F_{w,g} (t)
{\rm Tr}\,V^{w_1} \cdots {\rm Tr}\, V^{w_h},
\label{totalfreeopen}
\ee
\end{widetext}
where $g_s$ is the string coupling constant.
The factor $i^h$ is introduced for convenience, while
the factor $h!$ is a symmetry factor which takes into
account that the holes are indistinguishable.

In the case of open topological strings one can also
write the open Gromov-Witten invariants in terms of a new set of integer
invariants that we will denote by $n_{w,g,\beta}$. The integrality
structure of open Gromov-Witten invariants
was derived by Ooguri and Vafa (2000) and by Labastida, Mari\~no, and
Vafa (2000) following arguments similar to those of
Gopakumar and Vafa (1998b). According to this structure,
the free energy of open topological string theory in the sector
labelled by $w$ can be written in terms of the integer invariants
$n_{w, g, \beta}$ as follows:
\begin{widetext}
\be
\sum_{g=0}^{\infty}
g_s^{2g-2 +h}
F_{w,g}(t) = {1\over \prod_i w_i }
\sum_{g=0}^{\infty}\sum_{\beta}\sum_{d|w}  (-1)^{h+g}\,
n_{w/d, g, \beta}\, d^{h-1}
\biggl( 2\sin {d g_s \over 2} \biggr)^{2g-2}
\prod_{i=1}^h \biggl( 2\sin {w_i g_s \over 2} \biggr) Q^{d\beta}.
\label{multopen}
\ee
\end{widetext}
Notice there is one such identity for each $w$.
In this expression, the sum is over all integers $d$ which satisfy that
$d|w_i$ for all $i=1, \cdots, h$.
When this is the case, we define the $h$-uple $w/d$
whose $i$-th component is $w_i/d$. The expression (\ref{multopen})
can be expanded to give formulae for the
different genera. For example, at $g=0$ one simply finds
\be
\label{multiopen}
 F_{w, g=0, \beta}= (-1)^{h}
\sum_{d|w} d^{h-3} n_{w/d , 0, \beta/d}.
\ee
where the integer $d$ has to divide the vector $w$ (in the
sense explained above) and it is understood that $n_{w_d,0,\beta/d}$
is zero if $\beta/d$ is not a relative homology class. Formulae for higher genera
can be easily worked out from (\ref{multopen}), see Mari\~no and Vafa (2002) for examples.

When all the winding numbers $w_i$ are positive, one can label $w$
in terms of a vector $\vec k$. Given an $h$-uple $w=(w_1, \cdots, w_h)$,
we define a vector $\vec k$ as follows: the $i$-th
entry of $\vec k$ is the number of $w_j$'s
which take the value $i$. For example, if $w=(1,1,2)$,
the corresponding $\vec k$ is $\vec k =(2,1,0,\cdots)$.
In terms of $\vec k$, the number of holes and the total winding number are
given by
\begin{equation}
h=|\vec k|,\,\,\,\ \ell =\sum_i w_i=\sum_j  j k_j.
\end{equation}
Note that a given $\vec k$ will correspond to many $w$'s
which differ by a permutation of their entries. In fact
there are $h!/\prod_j k_j!$ $h$-uples $w$ which
give the same vector $\vec k$ (and the same amplitude). We can
then write the total free energy for positive winding numbers as
\begin{equation}
F(V)=\sum_{g=0}^{\infty} \sum_{\vec k} {i^{|\vec k|} \over \prod_j k_j!}
g_s^{2g-2+h} F_{\vec k, g} (t) \Upsilon_{\vec k} (V),
\label{freevk}
\end{equation}
where $\Upsilon_{\vec k} (V)$ was introduced in (\ref{ups}). 

We have considered for
simplicity the case in which the boundary conditions are specified by a single
Lagrangian submanifold with a single nontrivial one-cycle. In case there are more
one-cycles in the geometry, say $L$, providing possible boundary conditions for the open strings,
the above formalism has to be generalized in an obvious way: one needs to
specify $L$ sets of winding numbers $w^{(\alpha)}$, and the generating
functional (\ref{freevk}) depends on $L$ different
matrices $V_{\alpha}$, $\alpha=1, \cdots, L$. It is useful to write the free energy
(\ref{freevk}) as
\be
F(V)=\sum_R F_R(g_s, t) {\rm Tr}_R\, V
\label{fvr}
\ee
by using Frobenius formula (\ref{frob}). The total partition function $Z=e^F$
can then be written as
\be
Z(V)=\sum_R Z_R (g_s, t) {\rm Tr}_R \, V
\label{zmanyv}
\ee
by simply expanding (\ref{fvr}) as a formal power series in $V$.
One has for example $F_{\tableau{1}}=Z_{\tableau{1}}$, $F_{\tableau{2}}=
Z_{\tableau{2}}-Z_{\tableau{1}}^2/2$, and so on. When there are $L$ one-cycles in the 
target geometry providing boundary conditions, 
the total partition function has the structure
\be
Z(V_i)=\sum_{R_1, \cdots, R_L}  Z_{R_1 \cdots R_L} (g_s, t) \prod_{\alpha=1}^L{\rm Tr}_{R_{\alpha}} \, 
V_{\alpha}.
\label{zr}
\ee

It turns out that the integer invariants $n_{w, g,
\beta}$ appearing in (\ref{multopen}) are not the most fundamental ones. As we
have seen, if all the winding
numbers are positive we can represent $w$ by a vector $\vec
k=(k_1, k_2, \cdots)$. As we explained in II.F, such a vector can
be interpreted as a label for a conjugacy class $C(\vec k)$ of the
symmetric group $S_{\ell}$, where $\ell$ is the total
winding number. The invariant $n_{w, g,
\beta}$ will be denoted in this case as $n_{\vec k, g, \beta}$. It turns out that
this invariant can be written as (Labastida, Mari\~no, and
Vafa, 2000)
\be n_{\vec k, g, \beta}=\sum_R
\chi_R (C(\vec k)) N_{R, g, \beta},
\label{openBPS} \ee
where $ N_{R,
g, \beta}$ are integer numbers labelled by representations of the
symmetric group, {\it i.e.} by Young tableaux, and $\chi_R$ is the
character of $S_{\ell}$ in the representation $R$. The above
relation is invertible, since by orthonormality of the characters
one has
\be N_{R, g, \beta}=\sum_{\vec k} { \chi_R (C(\vec k)) \over
z_{\vec k}} n_{\vec k, g, \beta}, \ee
where $z_{\vec k}$ is given in (\ref{zk}).
Notice that integrality of $N_{R,g,\beta}$ implies integrality of $n_{\vec k,
g, \beta}$, but not the other way around. In that sense, the integer invariants $N_{R,
g, \beta}$ are the most fundamental ones. When there are both positive and negative
winding numbers, we can introduce two sets of vectors $\vec k^{(1)}$, $\vec k^{(2)}$
associated to the positive and the negative winding numbers, respectively, and following
the same steps we can define BPS invariants $n_{\vec k^{(1)}, \vec k^{(2)},g,\beta}$
and $N_{R_1,R_2,g,\beta}$.

In contrast to conventional Gromov-Witten invariants,
a rigorous theory of open Gromov-Witten invariants is not
yet available. However, localization techniques make possible to compute them in
some situations (Katz and Liu, 2002; Li and Song, 2002; Graber and Zaslow, 2002; Mayr, 2002).

\subsection{Some toric geometry}

So far we have considered topological string theory
on general Calabi-Yau threefolds. We will now restrict ourselves to a
particular class of geometries, namely noncompact, toric Calabi-Yau threefolds.
These are threefolds that have the structure of a fibration with
torus fibers. In particular, the manifolds we will be interested in have the
structure of a fibration of $\IR^3$ by ${\bf T}^2 \times \IR$. It turns out that
the geometry of these threefolds can be packaged in a two-dimensional graph which
encodes the information about the degeneration locus of the fibration. We will 
often call these graphs the {\it toric diagrams} of the 
corresponding Calabi-Yau manifolds. Instead of
relying on general ideas of toric geometry (which can be found for example in Cox and
Katz, 1999, and in Hori {\it et al.}, 2003), we will use
the approach developed by Leung and Vafa (1998), Aganagic and Vafa (2001), and
specially Aganagic, Klemm, Mari\~no, and Vafa (2003).

\subsubsection{${\bf C}^3$}

In the approach to toric geometry developed by Aganagic, Klemm, Mari\~no, and Vafa (2003),
noncompact toric Calabi-Yau threefolds are constructed out of an elementary building block,
namely ${\bf C}^3$. We will now exhibit its structure as a ${\bf T}^2 \times \IR$
fibration over $\IR^3$, and we will
encode this information in a simple trivalent, planar graph.

Let $z_i$ be complex coordinates on ${\bf C}^3$, $i=1,2,3$. We introduce
three functions or Hamiltonians
\ben
\label{hamil}
r_{\alpha}(z)&=&|z_1|^2-|z_3|^2, \nonumber\\
r_{\beta} (z)&=&|z_2|^2-|z_3|^2, \nonumber\\
r_{\gamma} (z)
&=&{\rm Im}(z_1 z_2 z_3).
\end{eqnarray}
These Hamiltonians generate three flows on ${\bf C}^3$ via the standard
symplectic form $\omega = \sum_i dz_i \wedge d\overline z_i$
on ${\bf C}^3$ and the Poisson brackets
\be
\partial_{\upsilon} z_i = \{
r_{\upsilon},
z_i\}_{\omega},\quad \upsilon=\alpha,\beta,\gamma.
\end{equation}
This gives the fibration structure that we were looking for: the base of the
fibration, $\IR^3$, is parameterized by the Hamiltonians (\ref{hamil}),
while the fiber ${\bf T}^2\times \IR$ is parameterized by the flows associated
to the Hamiltonians. In particular,
the ${\bf T}^2$ fiber is generated by the circle actions
\be\label{caction}
  e^{i \alpha r_\alpha + i \beta r_\beta}:\;\;
(z_1,z_2,z_3)\;\rightarrow \; (e^{i\alpha} z_1,e^{i\beta} z_2,
e^{-i(\alpha+\beta)}z_3),\end{equation}
while $r_\gamma$ generates the real line $\IR$.
We will call the cycle generated by $r_{\alpha}$
the $(0,1)$ cycle, and the cycle generated by $r_{\beta}$ the $(1,0)$ cycle.

Notice that the $(0,1)$ cycle
degenerates over the subspace of
${\bf C}^3$ described by $z_1=0=z_3$, which
is the subspace of the base $\IR^3$ given by $r_\alpha =r_\gamma=0$, $r_\beta\geq 0$.
Similarly, over $z_2=0=z_3$ the
$(1,0)$-cycle degenerates over the subspace $r_\beta = r_\gamma =0$ and
$r_\alpha\geq 0$. Finally, the one-cycle parameterized by
$\alpha+\beta$ degenerates over $z_1=0=z_2$, where
$r_\alpha-r_\beta =0 =r_\gamma$ and $r_\alpha\leq 0$.
\begin{figure}
\scalebox{.7}{\includegraphics{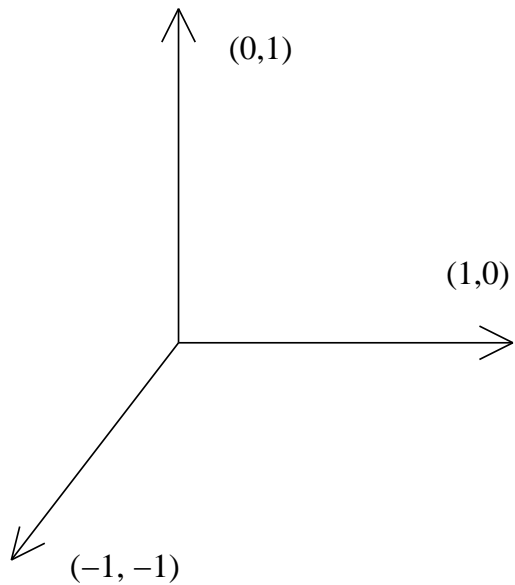}}
\caption{\label{vgraph}
This graph represents the degeneration locus
of the ${\bf T}^2\times \IR$ fibration
of ${\bf C}^3$ in the base $\IR^3$ parameterized 
by $(r_\alpha,r_\beta,r_\gamma)$.}\end{figure}

We will represent the ${\bf C}^3$ geometry by a graph which encodes the
degeneration loci in the $\IR^3$ base. In fact, it is useful to have a planar
graph by taking $r_{\gamma}=0$ and drawing the lines in the
$r_{\alpha} -r_{\beta}$ plane.
The degeneration loci will then be straight lines described
by the equation $p r_\alpha+ q r_\beta = {\rm const}$. Over this line
the $(-q,p)$ cycle of the ${\bf T}^2$ degenerates. Therefore
we correlate the degenerating cycles unambiguously with the lines in the graph (up
to $(q,p)\rightarrow (-q,-p)$). This yields the
graph in Fig. \ref{vgraph}, drawn in the $r_\gamma=0$ plane.

There is a symmetry in the
${\bf C}^3$ geometry that makes possible to represent it by different
toric graphs. These graphs are characterized by
three vectors $v_i$ that are obtained from the ones in Fig. \ref{vgraph} by
an ${\rm Sl}(2, {\bf Z})$ transformation. The vectors have to satisfy
\be
\label{sumvzero}
\sum_i v_i=0.
\ee
The ${\rm Sl}(2, {\bf Z})$ symmetry is inherited from the
${\rm Sl}(2, {\bf Z})$ symmetry of ${\bf T}^2$ that appeared in II.C
in a very different context. In the above discussion the
generators $H_1({\bf T}^2)$ have been chosen to be
the one-cycles associated to $r_\alpha$ and $r_\beta$, but there are
other choices that differ from this one by an ${\rm Sl}(2,{\bf Z})$
transformation on the ${\bf T}^2$. For example, we can choose
$r_\alpha$ to generate a $(p,q)$ one-cycle and $r_\beta$ a $(t,s)$
one-cycle, provided that $ps-qt=1$. These different choices
give different trivalent graphs. As we will see in the examples below,
the construction of general toric geometries requires in fact these more
general graphs representing ${\bf C}^3$ .

\subsubsection{More general geometries}

The non compact, toric Calabi-Yau threefolds
that we will study can be described as
symplectic quotients. Let us consider the complex
linear space ${\bf C}^{N+3}$,
described by $N+3$ coordinates $z_1, \cdots, z_{N+3}$, and let us
introduce $N$ real equations of the form
\be
\label{Dt}
\mu_A=\sum_{j=1}^{N+3} Q^j_A |z_j|^2 = t_A, \qquad A=1, \cdots, N.
\end{equation}
In this equation, $Q^j_A$ are integer numbers satisfying
\be
\label{flat}
\sum_{j=1}^{N+3} Q^j_A=0.
\end{equation}
Furthermore, we consider the action of the group
$G_N=U(1)^N$ on the $z's$ where the
$A$-th $U(1)$ acts on $z_j$ by
$$z_j \, \rightarrow \,\exp(i \; Q^j_A\; \alpha_A) z_j.$$
The space defined by the equations (\ref{Dt}), quotiented by the
group action $G_N$,
\be
X= \bigcap_{A=1}^N  \mu_A^{-1}(t_A)/G_N
\label{xquot}
\ee
turns out to be a Calabi-Yau manifold (it can be seen that the condition
(\ref{flat}) is equivalent to the Calabi-Yau condition). The $N$ parameters
$t_A$ are K\"ahler moduli of the
Calabi-Yau. This mathematical description of $X$ appears in the
study of two-dimensional
linear sigma model with ${\cal N}=(2,2)$ supersymmetry (Witten, 1993).
The theory has $N+3$ chiral fields, whose lowest components
are the $z$'s and are charged under
$N$ vector multiplets with charges $Q^j_A$. The
equations (\ref{Dt}) are the D-term equations,
and after dividing by the $U(1)^N$ gauge group we obtain the Higgs
branch of the theory.

The Calabi-Yau manifold $X$ defined in (\ref{xquot})
can be described by ${\bf C}^3$ geometries glued together in an
appropriate way. Since each of these ${\bf C}^3$'s is represented by
the trivalent vertex depicted in Fig. \ref{vgraph}, we will be able to encode the
geometry of (\ref{xquot}) into a trivalent graph.
In order to provide this description, we must first
find a decomposition of the set of all coordinates
$\{z_j\}_{j=1}^{N+3}$ into triplets $U_{a} = (z_{i_a}, z_{j_a},z_{k_a})$ that
correspond to the decomposition of $X$ into ${\bf C}^3$ patches. We pick one
of the patches and we associate to it two Hamiltonians $r_{\alpha}$, $r_{\beta}$
as we did for ${\bf C}^3$ before. These two coordinates will be global
coordinates in the
base $\IR^3$, therefore they will generate a globally
defined ${\bf T}^2$ fiber. The third coordinate in the base
is $r_{\gamma} = {\rm Im}(\prod_{j=1}^{N+3}z_j)$,
which is manifestly gauge invariant and moreover, patch by patch,
can be identified with the coordinate used in the ${\bf C}^3$
example above. The equation (\ref{Dt}) can then be used to find the action of
$r_{\alpha, \beta}$ on the other patches.

We will now exemplify this procedure with two important
examples: the resolved conifold and the local $\IP^2$ geometry.

\subsubsection{The resolved conifold}

\begin{figure}
\scalebox{.65}{\includegraphics{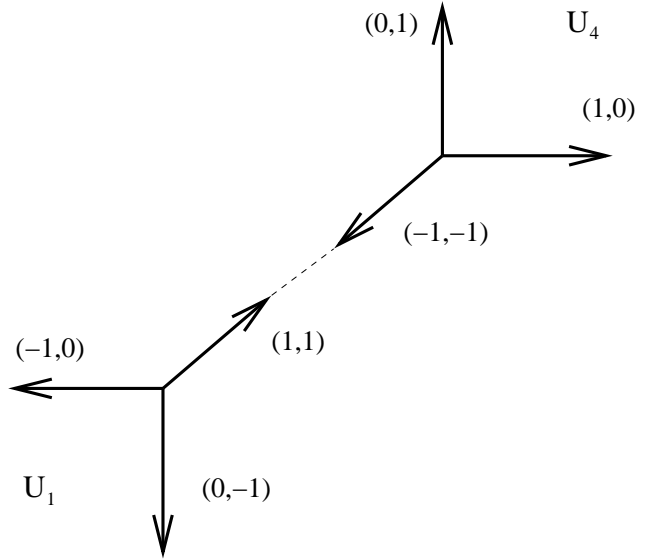}}
\caption{\label{res}
The graph associated to the resolved conifold ${\cal O}(-1)
\oplus {\cal O}(-1) \rightarrow \IP^1$. This manifold is made out of two 
${\bf C}^3$ patches glued through a common edge.}
\end{figure}
The simplest Calabi-Yau manifold is probably the so-called
resolved conifold, which is the
total space of the bundle
${\cal O}(-1) \oplus {\cal O}(-1) \rightarrow \IP^1$. This
manifold has a description of the form (\ref{xquot}), with $N=1$.
There is only one constraint given by
\be
|z_1|^2 + |z_4|^2 -|z_2|^2- |z_3|^2=t
\label{toricres}
\end{equation}
and the $U(1)$ group acts as
\be
\label{uquot}
z_1, z_2, z_3, z_4 \rightarrow e^{i \alpha} z_1, e^{-i \alpha}
z_2, e^{-i \alpha} z_3,
 e^{i \alpha} z_4.
\end{equation}
 Notice that, for $z_2=z_3=0$,
(\ref{toricres}) describes a $\IP^1$ whose area is proportional to
$t$. Therefore, $(z_1, z_4)$ can be taken as homogeneous
coordinates of the $\IP^1$ which is the basis of the fibration,
while $z_2, z_3$ can be regarded as coordinates for the fibers.

Let us now give a description in terms of ${\bf C}^3 $ patches glued together.
The first patch will be defined by $z_4 \not= 0$. Using
(\ref{toricres}) we can solve for the modulus of $z_4$ in terms of the other coordinates,
and using the $U(1)$ action we can gauge away its phase. Therefore, the patch
will be parameterized by $U_4=(z_1, z_2, z_3)$. The Hamiltonians will be in this
case
 \ben
\label{hamilres}
r_{\alpha}(z)&=&|z_2|^2-|z_1|^2, \nonumber\\
r_{\beta} (z)&=&|z_3|^2-|z_1|^2,
\end{eqnarray}
which generate the actions
\be\label{actionres}
 e^{i \alpha r_\alpha + i \beta r_\beta}:\;\;
(z_1,z_2,z_3)\;\rightarrow \; (e^{-i(\alpha+ \beta)} z_1,e^{i\alpha} z_2,
e^{i\beta}z_3).\end{equation}
This patch will be represented by the same graph that we found for ${\bf C}^3$.
The other patch will be defined by $z_1 \not =0$, therefore we can write it
as $U_1=(z_4, z_2, z_3)$. However, in this patch $z_1$ is no longer a natural coordinate,
but we can use (\ref{toricres}) to rewrite the Hamiltonians as
\ben
r_{\alpha}(z)&=&|z_4|^2-|z_3|^2 -t , \nonumber\\
r_{\beta} (z)&=&|z_4|^2-|z_2|^2 -t,
\end{eqnarray}
generating the action
\be
 e^{i \alpha r_\alpha + i \beta r_\beta}:\;\;
(z_4,z_2,z_3)\;\rightarrow \; (e^{i(\alpha+ \beta)} z_4,e^{-i\beta} z_2,
e^{-i\alpha}z_3).\end{equation}
The degeneration loci in this patch are the following: i) $z_4=0=z_2$, corresponding to
the line $r_{\beta}=-t$ where a $(-1,0)$ cycle degenerates; ii) $z_4=0=z_3$, corresponding to
the line $r_{\alpha}=-t$, where a $(0,1)$ cycle degenerates; iii) finally, $z_2=0=z_3$, where
$r_{\alpha}-r_{\beta}=0$, and a cycle $(1,1)$ degenerates. This patch is identical to the first one,
and they are joined together through the common edge where $z_2=0=z_3$. The full construction
is represented in Fig. \ref{res}. Notice that the common edge of the graphs represents
the $\IP^1$ of the resolved conifold: along this edge, one of the ${\bf S}^1$s of ${\bf T}^2$
has degenerated, while the other only degenerates at the endpoints.
An ${\bf S}^1$ fibration of an interval which degenerates at its endpoints is nothing
but a two-sphere. The length of the edge is $t$, the K\"ahler parameter associated to
the $\IP^1$.

\subsubsection{${\cal O}(-3) \rightarrow \IP^2$}

Let us now consider a more complicated example, namely
the non-compact Calabi-Yau manifold ${\cal O}(-3)
\rightarrow \IP^2$. This is the total space of $\IP^2$ together with its
anticanonical bundle, and it is often called local $\IP^2$.
We can describe it again as in (\ref{xquot}) with
$N=1$. There are
four complex variables, $z_0, \cdots, z_3$, and the constraint (\ref{Dt}) reads now
\be
\label{smb}
|z_1|^2 + |z_2|^2 +|z_3|^2 -3 |z_0|^2=t.
\end{equation}
The $U(1)$ action on the $z$s is
\be
z_0, z_1, z_2, z_3 \rightarrow e^{-3i \alpha} z_0, e^{i \alpha}
z_1, e^{i \alpha} z_2,
 e^{i \alpha} z_3.
\end{equation}
Notice that $z_{1,2,3}$ describe the basis $\IP^2$, while $z_0$ parameterizes
the complex direction of the fiber.

Let us now give a description in terms of glued ${\bf C}^3$ patches.
There are three patches $U_{i}$ defined by $z_{i}\neq 0$,
for $i=1,2,3$, since at least one of these three coordinates must be
non-zero in $X$. All of these three patches look like ${\bf C}^3$.
For example,
for $z_3\neq 0$, we can ``solve'' again for $z_3$ in terms of the other
three unconstrained coordinates which then parameterize ${\bf C}^3$:
$U_3 = (z_0,z_1,z_2)$. Similar statement holds for the other two
patches.
\begin{figure}
\scalebox{.6}{\includegraphics{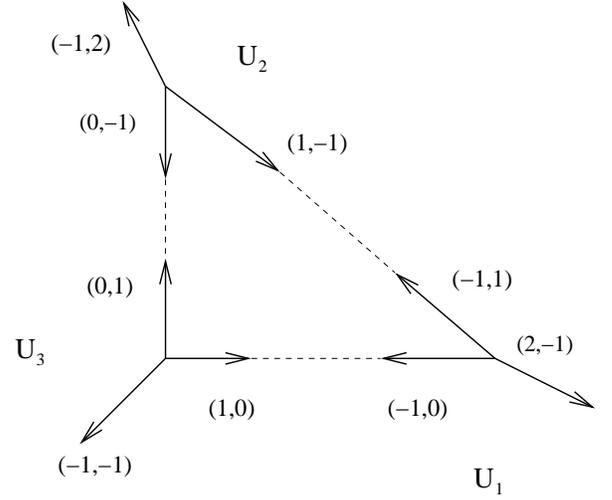}}
\caption{\label{figptwo}
The graph of ${\cal
O}(-3)\rightarrow \IP^2$. This manifold is built out of three
${\bf C}^3$ patches.}\end{figure}
Let us now construct the corresponding degeneration graph.
In the $U_3= (z_0,z_1,z_2)$ patch we take as our Hamiltonians
\ben
r_\alpha &=& |z_1|^2-|z_0|^2, \nonumber\\
r_\beta &=& |z_2|^2-|z_0|^2.
\end{eqnarray}
 The graph of the degenerate fibers in the
$r_\alpha-r_\beta$ plane is the same
as in the ${\bf C}^3$ example, Fig. \ref{vgraph}.
The third direction
in the base, $r_\gamma$ is now given by the gauge invariant product
$r_\gamma= {\rm Im}(z_0 z_1 z_2 z_3)$.
The same two Hamiltonians $r_{\alpha,\beta}$  generate the action
in the $U_{2}=(z_0,z_1,z_3)$ patch, and we use the constraint (\ref{smb})
to rewrite them as follows: since both $z_{0}$ and
$z_1$ are coordinates of this patch $r_\alpha$ does not
change. On the other hand,
$r_\beta$ must be rewritten since $z_2$ is not a natural coordinate here. We then
find:
\ben
r_\alpha &=& |z_1|^2-|z_0|^2, \nonumber\\
r_{\beta} &=& t+ 2 |z_0|^2 - |z_1|^2 - |z_3|^2,
\end{eqnarray}
hence
$$e^{(i\alpha r_\alpha +i \beta r_\beta)}: (z_0,z_1,z_3)\rightarrow
(e^{i(-\alpha+2\beta)} z_0, e^{i(\alpha-\beta)} z_1, e^{-i\beta}z_3),$$
We see from the above that
the fibers degenerate over three lines: i) $r_\alpha+r_\beta=t,$
corresponding to $z_0=0=z_3$, and where a $(-1,1)$
cycle degenerates;
ii) there is a line over which a $(-1,2)$ cycle degenerates
where $z_1=0=z_3$, $2r_{\alpha}+r_{\beta}=t$,
and
finally, iii) there is a line over which $r_\alpha=0$, and a
$(0,1)$-cycle degenerates. The $U_1$ patch is similar, and
we end up with the graph for ${\cal O}(-3)\rightarrow \IP^2$ shown in
Fig. \ref{figptwo}.

\subsubsection{Lagrangian submanifolds}

In order to consider open string amplitudes in the above Calabi-Yau geometries,
we have to construct Lagrangian submanifolds providing
boundary conditions, as we explained in IV.C.
Let us start by considering the ${\bf C}^3$ geometry discussed above. In this case,
one can easily construct Lagrangian submanifolds following the work of Harvey and Lawson
(1982). In terms of the Hamiltonians
in (\ref{hamil}), we have three types of them:
\ben
L_1:& &  \,\,\,\,\,\, r_{\alpha}=0, \qquad r_{\beta}=r_1, \qquad r_{\gamma}\ge 0.
\nonumber\\
L_2:& &  \,\,\,\,\,\, r_{\alpha}=r_2, \qquad r_{\beta}=0, \qquad r_{\gamma}\ge 0.
\nonumber\\
L_3:& &  \,\,\,\,\,\, r_{\alpha}=r_{\beta}=r_3, \qquad r_{\gamma}\ge 0,
\label{lagcthree}
\een
where $r_i$, $i=1,2,3$ are constants. It is not difficult to check that the above submanifolds are 
indeed Lagrangian (they
turn out to be Special Lagrangian as well). In terms of the graph description we
developed above, they correspond
to points in the edges of the planar graph spanned by
$(r_{\alpha}, r_{\beta})$, and they project to semi-infinite straight lines on the
basis of the fibration ${\bf R}^3$ parameterized by $r_{\gamma}\ge 0$. Since they are located at the
edges, where one of the circles of the fibration degenerates, they have
the topology of ${\bf C} \times {\bf S}^1$.

It is easy to generalize the construction to other toric geometries, like the resolved
conifold or local $\IP^2$: Lagrangian submanifolds with the topology of
${\bf C} \times {\bf S}^1$ are just given by points on the edges of the planar graphs.
Such Lagrangian submanifolds were first
considered in the context of open topological string theory by Aganagic and Vafa (2000),
and further studied by Aganagic, Klemm, and Vafa (2002).

\subsection{Examples of closed string amplitudes}

Gromov-Witten invariants of Calabi-Yau threefolds
can be computed in a variety of ways. A powerful technique which can be
made mathematically rigorous is the localization
technique pioneered by Kontsevich (1995). For compact Calabi-Yau
manifolds, only $N_{g=0,\beta}$ have been computed rigorously.
For non-compact, toric Calabi-Yau manifolds
one can compute $N_{g,\beta}$ for arbitrary genus using
these localization techniques. We will now present some results
for the topological string amplitudes $F_g$ of the geometries we described
above.

The resolved conifold ${\cal O}(-1) \oplus {\cal O}(-1)
\rightarrow \IP^1$ has one single K\"ahler parameter $t$
corresponding to the $\IP^1$ in the
base, and its total free energy is given by
\be
F(g_s,t)=\sum_{d=1}^{\infty} {1\over d \Bigl(2 \sin {d g_s \over
2} \Bigr)^{2}} Q^{d}, \label{resf}
\end{equation}
where $Q=e^{-t}$. We see that the only nonzero Gopakumar-Vafa invariant is
$n_1^0=1$. On the other hand, this model already has an infinite
number of nontrivial $N_{g,\beta}$ invariants, which can be obtained
by expanding the above expression in powers of $g_s$. The above closed expression
was obtained in Gromov-Witten theory by Faber and Pandharipande (2000).

The space ${\cal O}(-3)
\rightarrow \IP^2$ has also one single K\"ahler parameter, corresponding to the hyperplane
class of $\IP^2$.
By using the localization
techniques of Kontsevich, adapted to the noncompact case,
one finds (Chiang {\it et al.}, 1999; Klemm and Zaslow, 2001)
\ben
\label{localp2fs}
F_0 (t)&=& -{t^3 \over 18} + 3 \,Q -{45 \,Q^2 \over 8} + {244 \,Q^3 \over 9} -
{12333 \, Q^4 \over 64}  \cdots \nonumber\\
F_1 (t)&=& -{t \over 12} + {Q \over 4} -{3 \, Q^2 \over 8} -{23 \, Q^3
\over 3} + {3437 \, Q^4 \over 16} \cdots\nonumber\\
F_2 (t)&=& { \chi (X) \over 5720} + {Q \over 80} + {3 \, Q^3 \over 20}
+{514 \, Q^4 \over 5} \cdots,
\end{eqnarray}
and so on. In (\ref{localp2fs}), $t$ is the K\"ahler class of the manifold, 
$Q=e^{-t}$, and $\chi(X)=2$ is the Euler characteristic of the local $\IP^2$. The first
term in $F_0$ is proportional to the intersection number $H^3$
of the hyperplane class, while the first term in $F_1$ is
proportional to the intersection number between $H$ and $c_2(X)$. The first term in $F_2$
is the contribution of constant maps.

As we explained above, we can express the closed string amplitudes
in terms of Gopakumar-Vafa invariants. Let us introduce a generating
functional for integer invariants as follows:
\begin{equation}
f (z, Q)=\sum_{g,\beta} n_{\beta}^g z^g Q^{\beta},
\end{equation}
where $z$ is a formal parameter.
For local $\IP^2$ we find
\ben
\label{pgv}
f (z,Q)& = & 3 \, Q  - 6\, Q^2 + (27 -10\, z)\, Q^3  \nonumber  \\
&-&(192 - 231\, z  + 102 \, z^2
- 15 \, z^3) \, Q^4 \nonumber \\ &+& {\cal O}(Q^5).
\een

It should be mentioned that there is of course a very powerful method to
compute the amplitude $F_g$, namely mirror symmetry. In the mirror symmetric
computation, the $F_g$ amplitudes are deeply related to the variation of
complex structures on the
Calabi-Yau manifold (Kodaira-Spencer theory) and can be computed through
the holomorphic anomaly equations of Bershadsky {\it et al.} (1993 and 1994).
Gromov-Witten invariants of non-compact, toric Calabi-Yau threefolds
have been computed with mirror symmetry by Chiang {\it et al.} (1999),
Klemm and Zaslow (2001) and Katz, Klemm, and Vafa (1999).

\section{Chern-Simons theory as a string theory}

In this section we show that the 't Hooft program to interpret
the $1/N$ expansion of a gauge theory in terms of a string theory can be
realized in detail in the case of Chern-Simons theory on the three-sphere.

\subsection{Topological open strings on $T^*M$}

In order to give a string theory interpretation of Chern-Simons theory
on ${\bf S}^3$, a
good starting point is to give an open string interpretation of the $1/N$
expansion of the free energy (\ref{opencl}). This was done by
Witten (1995) in a remarkable paper, and we will follow his analysis
very closely.

Let $M$ be an arbitrary (real) three-dimensional manifold, and consider
the six-dimensional space given by $M$ together
by its cotangent bundle, $T^*M$. This space is a symplectic
manifold. If we pick local coordinates $q_a$ on $M$,
$a=1,2,3$, and local coordinates for the fiber $p_a$, the symplectic form can be
written as
\be
J=\sum_{a=1}^3 dp_a \wedge dq_a.
\end{equation}
One can find a complex structure on $T^*M$ such that $J$ is a K\"ahler form, so $T^*M$ can
be regarded as a K\"ahler manifold. Since the curvature of the cotangent bundle cancels
exactly the curvature of $M$, it is Ricci-flat, therefore it is a Calabi-Yau manifold. In fact
$T^* {\bf S}^3$ is a well-known Calabi-Yau, namely the {\it deformed conifold}.
The deformed conifold is usually described by
the algebraic equation
\be
\sum_{\mu=1}^4 \eta_{\mu}^2 =a.
\label{defconifold}
\end{equation}
To see that this describes $T^* {\bf S}^3$, let us
write $\eta_{\mu}= x_{\mu} + i v_{\mu}$, where $x_{\mu}$,
$v_{\mu}$ are real coordinates, and let us take $a$ to be real.
We find the two equations
\ben
\sum_{\mu=1}^4 ( x_{\mu}^2 -v_{\mu}^2) & =& a, \nonumber\\
\sum_{\mu=1}^4 x_{\mu} v_{\mu}&=&0.
\end{eqnarray}
The first equation indicates that the
locus $v_{\mu}=0$, $\mu=1, \cdots, 4$, describes a
sphere ${\bf S}^3$ of radius $R^2=a$, and the second equation shows that
the $v_{\mu}$ are coordinates for the cotangent space. Therefore,
(\ref{defconifold}) is nothing but $T^*{\bf S}^3$.

It is obvious that $M$ is a Lagrangian submanifold in $T^*M$, since $J$ vanishes
along $p_a={\rm const}.$ Since we have a Calabi-Yau manifold together with a Lagrangian
submanifold in it, we can consider a
system of $N$ topological D-branes wrapping $M$, thus providing Dirichlet
boundary conditions for topological open strings on $T^*M$. Our goal now
is to obtain a spacetime
action describing the dynamics of these topological
D-branes, and as we will see this action is nothing but Chern-Simons theory on $M$. 
This will prove the sought-for realization of Chern-Simons theory in terms of 
open strings.

\subsection{Open string field theory}

In order to obtain the spacetime description of open strings
on $T^*M$ we will use string field theory. 
We briefly summarize here some basic ingredients of the cubic string field theory
introduced by Witten (1986) to describe the spacetime dynamics of open
bosonic strings, since we will use the same model to describe topological strings.

In bosonic open string
field theory, we consider the worldsheet of the string to be an infinite strip parameterized
by a spatial coordinate $0 \le \sigma \le \pi$ and a time coordinate
$-\infty< \tau <\infty$,
and we pick the flat metric $ds^2 =d\sigma^2 + d\tau^2$. We then
consider maps $x: I \rightarrow X$, with $I=[0, \pi]$ and $X$ the target of the string. The
 string field is a functional of open string configurations $\Psi [x(\sigma))]$, with
ghost number one (although we will not indicate it explicitly, this 
string functional depends as well on the ghost fields). Witten (1986) defines two operations on the space of string functionals.
The first one is the {\it integration}, which 
is defined formally by folding the string around its midpoint and gluing
the two halves:
\be
\int \Psi =\int {\cal D} x(\sigma) \prod_{0\le \sigma \le \pi/2}
\delta[ x(\sigma) -x(\pi -\sigma)] \Psi [x(\sigma)].
\end{equation}
The integration has ghost number $-3$, which is the ghost number of the vacuum. This
corresponds to the usual fact that in open string theory on the disc one has to
soak up three zero modes. One also defines an associative, noncommutative {\it star product}
$\star$ of string functionals through
the following equation:
\begin{widetext}
\be
 \int \Psi_1 \star \cdots \star \Psi_N =  \int \prod_{i=1}^N {\cal D} x_i (\sigma)
\prod_{i=1}^N \prod_{0\le \sigma \le \pi/2}
\delta[ x_i(\sigma) -x_{i+1}(\pi -\sigma)] \Psi_i [x_i(\sigma)],
\ee
\end{widetext}
where $x_{N+1}\equiv x_1$. The star product simply glues the
string together by folding them around their midpoints, and gluing
the first half of one with the second half of the following (see for example
the review of Taylor and Zwiebach (2003) for more details), and it doesn't change the
ghost number.
In terms of these geometric operations, the string
field action is given by
\be
S={1 \over g_s}
\int \biggl( {1 \over 2} \Psi \star Q_{\rm BRST} \Psi + {1 \over 3} \Psi \star \Psi
\star \Psi \biggr).
\label{cubicsft}
\end{equation}
Notice that the integrand has ghost number $3$, while the
integration has ghost number $-3$, so that the action (\ref{cubicsft}) has zero
ghost number. If we
add Chan-Paton factors, the string field is promoted to a $U(N)$ matrix of
string fields, and the integration in (\ref{cubicsft}) includes a trace
${\rm Tr}$. The action (\ref{cubicsft}) has all
the information about the spacetime dynamics of open bosonic strings, with
or without D-branes. In
particular, one can derive the Born-Infeld action describing the dynamics
of D-branes from the above action (Taylor, 2000).

We will not need all the technology of string field
theory in order to understand open topological strings. The only piece of
relevant information is the following: the string functional is a function
of the zero mode of the string (which corresponds to the position of the string
midpoint), and of the higher oscillators. If we decouple all
the oscillators, the string functional becomes an ordinary function
of spacetime, the $\star$ product becomes the usual product of functions, and
the integral is the usual integration of functions. The decoupling of the
oscillators is in fact the pointlike limit of string theory. As we will see, this
is the relevant limit for topological open strings on $T^*M$.

\subsection{Chern-Simons theory as an open string theory}

We can now exploit
again the analogy between open topological strings and the open bosonic
string that we used to define the coupling of topological sigma
models to gravity ({\it i.e.}, that both have a nilpotent
BRST operator and an energy-momentum tensor that is $\CQ_{\rm
BRST}$-exact). Since both theories have a similar structure,
the spacetime dynamics of topological D-branes in
$T^*M$ is governed as well by (\ref{cubicsft}), where $\CQ_{\rm BRST}$ is given in
this case by the
topological charge defined in (\ref{qtrans}), and where the star product and
the integration operation are as in the bosonic string. The construction of the
cubic string field theory also requires the existence of a ghost number symmetry, which
is also present in the topological sigma model, as
we discussed in IV.A. It is convenient to consider the ghost number charge 
shifted by $-d/2$ with
respect to the assignment presented in IV.A (here, $d$ is the dimension of the
target). The shifted ghost number is actually the axial charge of the original ${\cal N}=2$ 
superconformal theory in the Ramond sector. When $d=3$ this corresponds to the normalization used by Witten (1986)
in which the ghost vacuum of the $bc$ system
is assigned the ghost number $-1/2$.

In order to provide the string field theory description of open topological strings on
$T^*M$, we have to determine the precise content of the string field, the $\star$ algebra and
the integration of string functionals for this particular model. As in the conventional
string field theory of the bosonic string, we have to consider the Hamiltonian description
of topological open strings. We then take $\Sigma$ to be an infinite strip and
consider maps $x: I \rightarrow T^*M$, with $I=[0, \pi]$, such that $\partial I$ is mapped to
$M$. The Grassmann field $\psi$, being a one-form on $\Sigma$, can be split as
$\psi=\psi_{\sigma}d\sigma + \psi_{\tau} d\tau$, but due to the
self-duality condition only one of them, say $\psi_{\tau}$, is independent.
The canonical commutation relations can be read out of the Lagrangian (\ref{tsaction}):
\ben
\Bigl[ {d x^i \over d\tau}(\sigma), x^j (\sigma')\Bigr]&=&
-{i \over t}G^{ij} \delta (\sigma -\sigma'), \nonumber\\
\{ \psi_{\tau}(\sigma), \chi (\sigma')\}&=&{1 \over t} \delta (\sigma -\sigma').
\end{eqnarray}
The Hilbert space is made out of functionals $\Psi[x(\sigma),\cdots]$, where $x$ is a map
from the interval as we have just described, and the $\cdots$ refer to the Grassmann
fields (which play here the r\^ole of ghost fields). The Hamiltonian is obtained, as usual in string theory,
by
\be
L_0 = \int_0^{\pi} d\sigma T_{00}.
\end{equation}
The bosonic piece of $T_{00}$ is just
\be
t G_{ij} \Bigl( {d x^i \over d \sigma}{d x^j \over d \sigma} +
{d x^i \over d \tau}{d x^j \over d \tau}\Bigr),
\end{equation}
and using the canonical commutation relations we find:
\be
L_0 = \int_0^{\pi} d\sigma \biggl( -{1 \over t} G^{ij} {\delta^2
\over \delta x^i(\sigma) \delta x^j(\sigma)}+
t G_{ij} {d x^i \over d \sigma}{d x^j \over d \sigma} \biggr).
\end{equation}
We then see that string functionals with $dx^i/d\sigma\not=0$ cannot
contribute: since the physics is $t$-independent, we can take
$t\rightarrow \infty$, where they get infinitely massive and decouple from the spectrum.
Therefore, the map $x: I \rightarrow T^*M$ has to be constant and in particular it must be a point
in $M$. A similar analysis holds for the Grassmann fields as
well, and the conclusion is that the string functionals are functions of the commuting and
the anticommuting zero modes. Denoting them by $q^a$, $\chi^a$, the string functional
reduces to
\be
\Psi=A^{(0)}(q) + \sum_{p=1}^3 \chi^{a_1} \cdots \chi^{a_p} A^{(p)}_{a_1 \cdots a_p}.
\label{stringf}
\end{equation}
These functionals can be interpreted as differential forms on $M$. A differential form
of degree $p$ will have ghost number $p$, or equivalently, shifted ghost number charge $p-3/2$. If we have
$N$ D-branes wrapping $M$, the above differential forms take values in the adjoint
representation of the gauge group ({\it i.e.} they are valued in the $U(N)$ Lie algebra).
On these functionals, the $\CQ$ symmetry acts as the exterior differential, and
$\{ \CQ,\Psi \}=0$ if the differential forms are closed. Of course
in string field theory we do not restrict ourselves to functionals in the $\CQ$ cohomology.
We rather compute the string field action for arbitrary functionals, and then the condition
of being in the $\CQ$-cohomology arises as a linearized equation of motion.

We are now ready to write the string field action for topological open strings
on $T^*M$ with Lagrangian
boundary conditions specified by $M$. We have seen that the relevant
string functionals are of the
form (\ref{stringf}). Since in string field theory the string field has ghost number one
(equivalently, $U(1)_R$ charge $-1/2$), we
see that
\be
\Psi = \chi^a A_a(q),
\end{equation}
where $A_a(q)$ is a Hermitian matrix. In other words, the string field is just
a $U(N)$ gauge connection on $M$. Since the string field only depends on commuting
and anticommuting zero modes, the integration of string functionals becomes
ordinary integration of forms on $M$, and the star product becomes the
usual wedge products of forms. We then have the following dictionary:
\be
\begin{array}{ccc}
 \Psi \rightarrow A, & \,  & \CQ_{\rm BRST}\rightarrow d\\
\,  & \,  &\, \\
\star \rightarrow \wedge, & \,  & \int \rightarrow \int_M.
\end{array}
\end{equation}
The string field action (\ref{cubicsft}) is then the usual
Chern-Simons action for $A$, and by comparing
with (\ref{csact}) we have the following relation
between the string coupling constant
and the Chern-Simons coupling:
\be
g_s ={2 \pi \over k+N},
\end{equation}
after taking into account the shift $k \rightarrow k+N$.

This result is certainly remarkable. In the usual
open bosonic string, the string field
involves an infinite tower of string excitations. For the open topological
string, the topological character of the model implies that all excitations
decouple, except for the lowest
lying one. In other words, the usual
reduction to a finite number of degrees of freedom that occurs in
topological theories downsizes the string field to a single excitation.
In physical terms, what is happening is that string theory reduces in this
context to its pointlike limit, since the only relevant degree of freedom of the
string is its zero mode, which describes the motion of a pointlike particle. As
expected, the dynamics reduces then to a usual quantum field theory.

However, as explained by Witten (1995), since open topological string theory is a
theory that describes open string instantons with Lagrangian boundary conditions, we should
expect to have corrections to the above result due to nontrivial worldsheet instantons.
It is easy to see that instantons $x: \Sigma \rightarrow T^*M$ such that
$x(\partial \Sigma) \subset M$
are necessarily constant. Notice first that $J=d\rho$, where
\be
\rho= \sum_{a=1}^3 p_a dq_a,
\end{equation}
and $p_a$ vanishes on $M$. Since $x$ is a holomorphic map, the instanton action
equals the topological piece $-\int_{\Sigma} x^* (J)$. This can be evaluated to be
\be
\int_{\Sigma} x^*(J)= \int_{\partial \Sigma} x^*(\rho)=0
\end{equation}
since $x(\partial \Sigma) \subset M$. Holomorphic maps
with the above boundary conditions are necessarily constant, and there are
no worldsheet instantons in the geometry. Therefore, there are
no instanton corrections to the Chern-Simons
action that we derived above.

One of the
immediate consequences of the Chern-Simons spacetime description of open topological
strings on $T^*M$ is that the coefficient
$F_{g,h}$ in the perturbative expansion (\ref{openf}) of Chern-Simons
theory on $M$ is given by the
free energy of the topological string theory at genus $g$ and $h$ holes.
What is then the interpretation of the fatgraph associated to $F_{g,h}$ from the
point of view of the topological string theory on $T^*M$? Even though there are no
``honest" worldsheet instantons in this geometry,
there are degenerate instantons of zero area
in which the Riemann surface degenerates to a graph in $M$. It is well-known that the
moduli space of open Riemann surfaces contains this type of configurations. In the case
at hand, the fatgraphs appearing in the $1/N$ expansion of Chern-Simons theory on $M$
are precisely the graphs that describe the degenerate instantons of the geometry.
This model gives then a very concrete realization of the string picture of the $1/N$
expansion discussed in III.

\subsection{More general Calabi-Yau manifolds}

In the previous section we have presented an explicit description of open topological
strings on $T^*M$, following Witten (1995). What happens if the target is a more general
Calabi-Yau manifold?

\begin{figure}[!ht]
\scalebox{.6}{\includegraphics{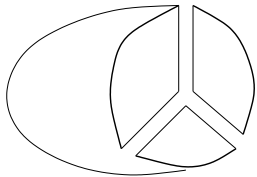}}
\caption{This figure shows a partially degenerate worldsheet instanton of genus $g=0$ and
with $h=3$ ending on an unknot. The instanton is made out of a honest
holomorphic disk and the degenerate piece, which is a fatgraph.}
\label{degen}
\end{figure}

Let us consider a Calabi-Yau
manifold $X$ together with some Lagrangian submanifolds $M_i \subset X$,
with $N_i$ D-branes wrapped over $M_i$.
In this case the spacetime description of topological open strings will have
two contributions. First of all, we have the contributions
of degenerate holomorphic curves. These are captured
by Chern-Simons theories on the manifolds $M_i$, following the
same mechanism that we described for $T^*M$. However, as
pointed out by Witten (1995), for a general Calabi-Yau $X$ we may also have
honest open string instantons contributing to the spacetime description, which
will be embedded holomorphic Riemann surfaces with boundaries ending on the Lagrangian
submanifolds $M_i$. An open string instanton $\beta$ will intersect
the $M_i$ along one-dimensional curves ${\cal K}_i(\beta)$, which
are in general knots inside $M_i$. We know from (\ref{chancoup}) that
the boundary of such an instanton will give a Wilson loop insertion
in the spacetime action of the form $\prod_i {\rm Tr}U_{{\cal K}_i(\beta)}$, where
$U_{{\cal K}_i(\beta)}$ is the holonomy of the Chern-Simons
connection on $M_i$ along the knot ${\cal K}_i(\beta)$. In addition,
this instanton will be weighted by its area (which corresponds to the
closed string background).
We can then take into account the contributions of all instantons
by including the corresponding Chern-Simons theories
$S_{\rm CS}(A_i)$, which account for the degenerate instantons,
coupled in an appropriate way with the ``honest" holomorphic
instantons. The spacetime action will then have the form
\be
\label{impo}
S(A_i)= \sum_i S_{\rm CS}(A_i)+\sum_{\beta} e^{-\int_{\beta}\omega}
\prod_i {\rm Tr}U_{{\cal K}_i(\beta)}
\end{equation}
where $\omega$ is the complexified K\"ahler form and 
the second sum is over ``honest'' holomorphic instantons $\beta$.
Notice that all the Chern-Simons theories $S_{\rm CS}(A_i)$
have the same coupling constant, equal to the string coupling constant.
More precisely,
\be
\label{equalg}
{2\pi \over k_i+N_i}=g_s.\end{equation}
In the action (\ref{impo}), the honest holomorphic instantons are
put ``by hand'' and in principle one has to solve a nontrivial enumerative problem
to find them. Once they are included in the action, the path integral over
the Chern-Simons connections will join degenerate instantons
to these honest worldsheet
instantons: if we have a honest worldsheet instanton ending on a knot ${\cal K}$,
it will give rise to a Wilson loop operator in (\ref{impo}), and the
$1/N$ evaluation of the vacuum expectation value will generate all possible
fatgraphs $\Gamma$ joined to the knot ${\cal K}$, producing in this way
partially degenerate worldsheet instantons (the fatgraphs are
interpreted, as before, as degenerate instantons). An example of this
situation is depicted in Fig. \ref{degen}. This more complicated scenario
was explored by Aganagic and Vafa (2001), Diaconescu, Florea, and Grassi (2003a, 2003b),
and Aganagic, Mari\~no, and Vafa (2004). We will give examples of
(\ref{impo}) in VI.

\subsection{The conifold transition and the large $N$ duality}

We know now that Chern-Simons theory on ${\bf S}^3$ is a
topological open string theory on $T^*{\bf S}^3$. The next step is to see
if there is a {\it closed} string theory leading to the resummation
(\ref{closedf}). As shown by Gopakumar and Vafa in an important paper (1999),
the answer is yes.

One way to motivate their result is as follows: since the holes of
the Riemann surfaces are due to the
presence of D-branes, ``filling the holes'' to get the closed strings means
getting rid of the D-branes. But this is precisely what happens in another
large $N$ duality, namely the AdS/CFT correspondence (Maldacena, 1998),
where type IIB theory in flat space in the presence of D-branes is
conjectured to be equivalent to type IIB theory in ${\rm AdS}_5 \times {\bf
S}^5$ with no D-branes, and where the radius of the ${\bf S}^5$ is related
to the number of D-branes. The reason for that is that, at large $N$, the
presence of the D-branes can be traded by a deformation of the background
geometry. In other words, we can make the branes disappear if we change the
background geometry at the same time: as emphasized by Gopakumar and Vafa,
large $N$ dualities relating open and closed strings should be associated
to transitions in the geometry. This reasoning suggests to look for a
transition involving the background
$T^* {\bf S}^3$. It turns out that such a transition is well-known in the
physical and the mathematical literature, and it is called the conifold
transition (see for example Candelas and de la Ossa, 1990). Let us explain this in detail.

The algebraic equation describing the deformed conifold is
(\ref{defconifold}).
It is useful to rewrite this equation as follows.
Introduce the following complex coordinates:
\be
\begin{array}{ccc}
\label{newcoords}
x=\eta_1 + i \eta_2, & & v=i(\eta_3 -i \eta_4), \\
u=i(\eta_3 + i \eta_4), & & y= \eta_1 -i \eta_2.
\end{array}
\end{equation}
The deformed conifold can be now written as
\be
xy=uv+a.
\label{defalt}
\end{equation}
Notice that in this parameterization the geometry has a ${\bf T}^2$ fibration
\be
x,y,u,v \rightarrow xe^{-i\alpha},ye^{i\alpha},ue^{-i\beta},ve^{i\beta}
\label{torusact}
\end{equation}
where the $\alpha$ and $\beta$ actions above can be taken to generate
the $(0,1)$ and $(1,0)$ cycles of the ${\bf T}^2$, respectively. The ${\bf T}^2$
fiber can degenerate to ${\bf S}^1$ by collapsing
one of its one-cycles. In (\ref{torusact}), for example,
the $U(1)_{\alpha}$ action fixes $x=0=y$ and therefore fails to
generate a circle there.
In the total space, the locus where this happens, i.e. the $x=0=y$
subspace of $X$, is a cylinder $uv=-a$. Similarly, the locus where the
other circle collapses, $u=0=v$, gives another cylinder $xy=a$.
Therefore, we can regard the whole
geometry as a ${\bf T}^2 \times\IR$ fibration over $\IR^3$: if we define
$z=uv$, the $\IR^3$ of the base is given by ${\rm Re}(z)$ and the axes
of the two cylinders. The fiber is given by the circles of the two
cylinders, and by ${\rm Im}(z)$. The $U(1)_{\alpha}$
fibration degenerates at $z=-a$, while the $U(1)_{\beta}$ fibration
degenerates at $z=0$. This is the same kind of fibration
structure that we found when discussing the geometries of the form (\ref{xquot}).
\begin{figure}
\scalebox{.6}{\includegraphics{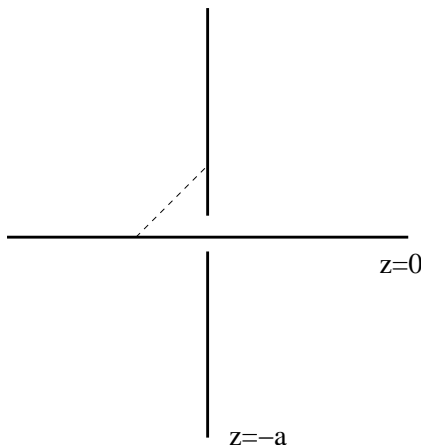}}
\caption{\label{deformedf}
This figure represents $T^*{\bf S}^3$, regarded as a ${\bf T}^2
\times \IR$ fibration of $\IR^3$. Two of the directions represent the axes
of the two cylinders, and the third direction represents the real axis of
the $z$-plane. The dashed line represents the ${\bf S}^3$ cycle.}
\end{figure}

\begin{figure*}
\scalebox{.4}{\includegraphics{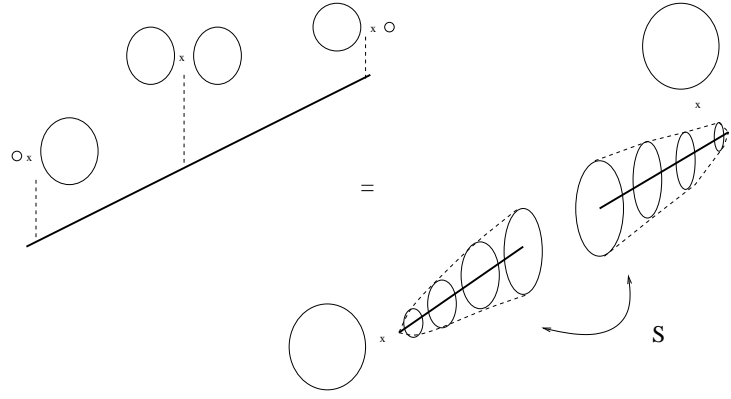}}
\caption{\label{torusfiber}
On the left hand side, we represent ${\bf S}^3$ as a ${\bf T}^2$ fibration
over the interval. One of the
circles of the torus degenerates over one endpoint, while the other circle degenerates over the
opposite endpoint. Each of the degenerating circles fibers over half the interval to produce a
disc $D$, and on the right hand side the three-sphere is equivalently realized as two ${\bf S}^1 \times D$
glued through an $S$ transformation.}\end{figure*}

As we did in section IV, it is very useful to represent the above geometry by
depicting the singular loci of the torus action in the base $\IR^3$.
The loci where the cycles of the torus collapse, which are cylinders,
project to lines in the base space. This is
shown in Fig. \ref{deformedf}. Notice that the
${\bf S}^3$ of the deformed conifold geometry is realized in this picture
as a ${\bf T}^2$ fibration
over an interval $I$. This interval is represented in Fig. \ref{deformedf} by a dashed line in
the $z$-plane between $z=-a$ (where the $(0,1)$ cycle collapses) and $z=0$
(where the collapsing cycle is the $(1,0)$). The geometric description
of ${\bf S}^3$ that is obtained in this way is in fact equivalent to the description
given in II in terms of a Heegaard splitting along solid tori. To see this,
let us cut the three-sphere in two pieces
by cutting the interval $I$ in
two smaller intervals $I_{1,2}$ through its midpoint. Each of the halves is a
fibration of ${\bf T}^2 =
{\bf S}^1 \times {\bf S}^1_c$ over an interval $I_i$, where ${\bf S}^1_c$ denotes the
collapsing
cycle. Of course the nontrivial part of the
fibration refers to the collapsing cycle, so we can see each of the halves as
${\bf S}^1$ times the
fibration of the collapsing cycle over $I_i$, which is nothing but a disk. In other words, we
are constructing the three-sphere by gluing two manifolds of the form ${\bf S}^1 \times D$.
These are of course two solid tori, which are glued after exchanging the two cycles, {\it i.e.}
after performing an $S$ transformation. This is shown in Fig. \ref{torusfiber}.

\begin{figure*}
\scalebox{.6}{\includegraphics{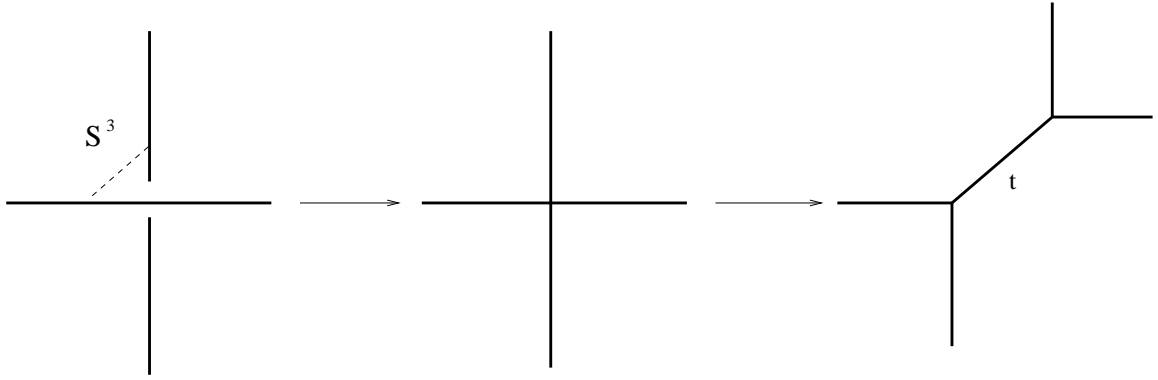}}
\caption{\label{transition}
A pictorial representation of the conifold transition.
The figure on the left represents the deformed conifold with an ${\bf S}^3$, the figure
on the center is the conifold singularity, and the figure on the right
is the resolved conifold. }
\end{figure*}

The conifold singularity appears when $a=0$ in (\ref{defalt})
and the three-sphere collapses. This is
described by the equation
\be
xy=uv.
\label{conifold}
\end{equation}
In algebraic geometry, singularities can be avoided in two ways, in general. The
first way is to deform the complex geometry, and in our case this leads to
the deformed conifold (\ref{defconifold}). The other way is to resolve
the singularity, for example by performing a blow up, and this leads to the
resolved conifold geometry (see for example Candelas and de la Ossa, 1990). The resolution of
the geometry can be explained as follows. The equation (\ref{conifold}) can be solved by
\be
x=\lambda v, \,\,\,\,\,\, u=\lambda y
\label{rescon}
\end{equation}
where $\lambda$ is an inhomogeneous coordinate in $\IP^1$. (\ref{rescon}) can be
interpreted as defining the bundle ${\cal O}(-1) \oplus {\cal O}(-1)
\rightarrow \IP^1$. To make
contact with the toric description
given in (\ref{toricres}), we put $x=z_1 z_3$, $y=z_2 z_4$, $u=z_1 z_2$ and
$v=z_3 z_4$. We then see that $\lambda=z_1/z_4$ is the inhomogenous coordinate
for the $\IP^1$ described in (\ref{toricres}) by $|z_1|^2 + |z_4|^2=t$.
We therefore have a {\it conifold transition} in
which the three-sphere of the deformed conifold
shrinks to zero size as $a$ goes to zero, and then a two-sphere of size $t$
grows giving the resolved conifold. In terms of the coordinates $z_1,
\cdots, z_4$, the ${\bf T}^2$ action (\ref{torusact}) becomes
\be
\label{inht}
z_1, z_2, z_3, z_4 \rightarrow e^{-i(\alpha +\beta)} z_1,
e^{i\alpha } z_2,e^{i\beta } z_3, z_4.
\end{equation}
This ${\bf T}^2$ fibration is precisely (\ref{actionres}). Notice that the
singular loci of fibration of the resolved conifold
which is encoded in the trivalent graph of
Fig. \ref{res} is inherited from the singular loci depicted in Fig. \ref{deformedf}.
The transition from the deformed to the resolved conifold can then be represented
pictorially as in Fig. \ref{transition}.

Now we are ready to state the conjecture of Gopakumar and Vafa (1999).
We know that Chern-Simons theory is an open topological string on the
deformed conifold geometry with $N$ topological D-branes wrapping the
three-sphere. The conjecture is that at large $N$ the D-branes
induce a conifold transition in the background geometry, so that we end up with the resolved
conifold and no D-branes. But in the absence of D-branes that enforce boundary
conditions we just have a theory of closed
topological strings. Therefore, {\it Chern-Simons theory on ${\bf S}^3$
is equivalent to closed topological string theory on the resolved conifold.} As we will see,
the relation between the parameters is the following: the closed string coupling constant
is the open string coupling constant, or equivalently the Chern-Simons effective coupling constant
as in (\ref{oscc}). The size of the $\IP^1$ in the resolved geometry is given by the 't Hooft coupling
of Chern-Simons theory,
\be
t=ig_s N = xN.
\label{thooft}
\end{equation}

This conjecture has been proved by embedding the
duality in type II superstring
theory (Vafa, 2001a) and lifting it to
M-theory (Acharya, 2000; Atiyah, Maldacena, and Vafa, 2001). There is also
a worldsheet derivation of the
duality due to Ooguri and Vafa (2002). In the next subsection, we will give
evidence for the conjecture at the level of the free energy.

\subsection{A test of the duality: the partition function}

A nontrivial test of the duality advocated by Gopakumar and Vafa is to
verify that the free energy of $U(N)$ Chern-Simons theory on the three-sphere agrees
with the free energy of closed topological strings on the resolved
conifold.
So far, what we have uncovered is the open string expansion of Chern-Simons
theory, which is (order by order in $x$) determined by the perturbative
expansion. In order to find a closed string interpretation, we have to sum
over the holes, as in (\ref{opencl}). The `t Hooft parameter $t$ is given by
(\ref{thooft})
and
\be
F^{\rm p}_g(t) = \sum_{h=1}^{\infty}F_{g,h}^{\rm p} (-it)^h.
\end{equation}
Let us first focus on $g\ge 2$.
To perform the sum explicitly, we write again the $\zeta$ function
as $\zeta(2g-2+ 2p)=\sum_{n=1}^{\infty} n^{2-2g-2p}$, and use the
binomial series,
\be
{1 \over (1-z)^q}=\sum_{n=0}^{\infty} {q+n-1 \choose n} z^n
\end{equation}
to obtain:
\ben
F^{\rm p}_g(t)&=& {(-1)^g  |B_{2g} B_{2g-2}| \over
2g (2g-2)(2g-2)!} \nonumber\\ &+&  {B_{2g} \over 2g (2g-2)} \sum_{n \in {\bf Z}} \,'
{1 \over (-it+ 2\pi n)^{2g-2}},
\label{quasi}
\een
where $'$ means that we omit $n=0$. Now we notice that, if we write
\be
F^{\rm np}=\sum_{g=0}^{\infty} F^{\rm np}_g(t) g_s^{2g-2}
\end{equation}
then for, $g\ge 2$, $F^{\rm np}_g(t)=B_{2g}/(2g(2g-2)(-it)^{2g-2}$, which
is
precisely the $n=0$ term missing in (\ref{quasi}). We then define:
\be
F_g(t)=F_g^{\rm p}(t) + F_g^{\rm np}(t).
\end{equation}
Finally, since
\be
\sum_{n \in {\bf Z}}{1 \over n+ z}={2 \pi i \over 1-e^{-2\pi i z}},
\end{equation}
by taking derivatives w.r.t. $z$ we can write
\be
F_g(t)={(-1)^g  |B_{2g} B_{2g-2}| \over
2g (2g-2)(2g-2)!} + {|B_{2g}| \over
2g (2g-2)!}{\rm Li}_{3-2g}(e^{-t}),
\label{fin}
\end{equation}
again for $g\ge 2$. If we now compare to (\ref{multibuble}), we see that 
(\ref{fin}) has precisely the structure of the free energy of a closed
topological string, with $n_1^0=1$, and the rest of the Gopakumar-Vafa
invariants being zero. Also, from the first term, which gives the
contribution of the constant maps, we find that $\chi(X)=2$. In fact,
(\ref{fin}) is precisely the $F_g$ amplitude of the resolved conifold. This
is a remarkable check of the conjecture.

\subsection{Incorporating Wilson loops}

As we have extensively discussed, most of the
wealth of Chern-Simons theory on ${\bf S}^3$ is due to the
Wilson loop operators along knots.
How do we incorporate Wilson loops in the string picture that we have just
developed? In III.C we saw that, once one has a closed string description
of the $1/N$ expansion,
Wilson loops are related to open strings in the closed string
geometry. Since the string description involves topological strings, it is
natural to assume that Wilson loops are going to be described by open
topological strings in the resolved conifold, and this means that we need a
Lagrangian submanifold specifying the boundary conditions for the strings.

These issues were addressed in an important paper by Ooguri and Vafa
(2000). In order to give boundary conditions for the open strings in the
resolved conifold, Ooguri and Vafa constructed a natural Lagrangian
submanifold ${\widetilde {\cal C}}_{\cal K}$ in $T^*{\bf S}^3$ for
any knot ${\cal K}$ in ${\bf S}^3$.
This construction is rather canonical, and it is called the conormal bundle
of ${\cal K}$. The details are as follows: suppose that a knot ${\cal K}$ is
parameterized by a curve $q(s)$, where $s\in [0, 2\pi)$. The
conormal bundle of ${\cal K}$ is the space
\be
{\widetilde {\cal C}}_{\cal K}= \Bigl\{ (q(s), p) \in T^*{\bf S}^3 |
\sum_a p_a \dot q_a=0, \,\, 0 \le s\le 2\pi \Bigr\}
\end{equation}
where $p_a$ are coordinates for the cotangent bundle, and $\dot q_a$
denote the derivatives w.r.t. $s$. This space is an
$\IR^2$-fibration of the knot itself, where the fiber on the point $q(s)$
is given by the two-dimensional subspace of $T_q^* {\bf S}^3$ of planes
orthogonal to $\dot q(s)$. $ {\widetilde {\cal C}}_{\cal K}$ has in fact
the topology of ${\bf S}^1 \times \IR^2$, and intersects ${\bf S}^3$ along
the knot ${\cal K}$.

One can now consider, together with the $N$ branes wrapping ${\bf S}^3$, a
set of $M$ probe branes wrapping ${\widetilde {\cal C}}_{\cal
K}$. As usual when we have two sets of D-branes, we have three different
types of strings: the strings with both ends on the $N$ branes
are described by $U(N)$ Chern-Simons theory on ${\bf S}^3$, as we argued before.
In the same way, the
strings with both ends on the $M$ branes are described by $U(M)$ Chern-Simons theory
on ${\widetilde {\cal C}}_{\cal
K}$. But there is a new sector due to
strings stretched between the $N$ branes and
the $M$ branes. To study these strings, we can make an analysis similar to the one
we did in section C above. First, we have to impose again that $dx^i/d\sigma=0$. Therefore, $x^i$ has to be
a constant, and since the endpoints of the string lie on different submanifolds, the only
possibility is that $x^i \in {\bf S}^3 \cap {\widetilde {\cal C}}_{\cal
K}={\cal K}$. A similar analysis holds for the Grassmann fields, and we then find that
the string functionals describing the new sector of strings are a function of a single commuting
zero mode $q$ parametrizing ${\cal K}$, and a single anticommuting zero mode $\chi$. In other
words,
\be
{\cal A}=\phi(q) + \chi \xi(q).
\end{equation}
where $\phi$ is a complex scalar field in the bifundamental
representation $(N, \overline M)$, and living in the intersection of the
two branes, ${\cal K}$. The fact that the scalar is complex is due to the fact that
our strings are oriented, and we have to consider both a real scalar in representation
$(N, {\overline M})$ together with another real scalar in representation $({\overline N}, M)$,
which we can put together as a complex scalar in one of the representations.
The $\CQ$ operator is just the exterior differential
$d$ on ${\bf S}^1$.

As we explained above, the string field is the piece of the above functional with
$U(1)_R$ charge $-1/2$. However, now the $U(1)_R$ charge assignment is different
from the one in ${\bf S}^3$, and it is given for a differential form of degree $p$
by $p-1/2$. This is because the target is now ${\bf S}^1$ with $d=1$.
Therefore, the surviving field is in this case the scalar $\phi(q)$. This is
consistent with the fact that, since the spacetime dynamics takes place now on a circle,
and since $\CQ=d$, the kinetic term for the string field action (\ref{cubicsft}) is only
non-trivial if the string field is a scalar. The full action for $\phi(q)$
is simply $\oint_{\cal K} \bar \phi d \phi$. However, there are
also two background gauge fields that interact with the Chan-Paton factors at the
endpoints of the strings. These are
the $U(N)$ gauge connection $A$ on ${\bf
S}^3$, and the $U(M)$ gauge connection $\widetilde A$ on
${\widetilde {\cal C}}_{\cal
K}$. The complex scalar 
couples to the gauge fields in the standard way,
\be
\oint_{\cal K}\;
{\rm Tr} \,\bar{\phi}\,A \phi - \phi {\widetilde A})\, \bar{\phi}.
\end{equation}
Here we regard $\widetilde A$ as a source. If we now integrate out
$\phi$ we obtain 
\be
\exp \biggl[ -\log \, {\rm det} ( {d \over ds} + \sum_a (A_a -{\widetilde A}_a) {d q_a \over ds})\biggr]
\ee
which can be easily evaluated as 
\be
\label{massiveov}
\exp\Bigl[-{\rm Tr} \; \log(1 -
 U\otimes  V)\Bigr]=
 \exp\Bigl\{\sum_{n=1}^{\infty}\; {1\over n}
\;{\rm Tr} U^{n}\; {\rm Tr}V^{n}\Bigr\},
\ee
where $U$, $V^{-1}$ are the holonomies of $A$, $\widetilde A$ around the knot
${\cal K}$, and we have dropped an overall constant . In this way we obtain the effective action
for the $A$ field
\be
S_{\rm CS}(A) + \sum_{n=1}^{\infty}{1 \over n} {\rm Tr}U^n {\rm Tr} V^{n}
\label{defact}
\end{equation}
where $S_{\rm CS}(A)$ is of course the Chern-Simons action for
$A$. Therefore, in the presence of the probe branes, the action involves an
insertion of the Ooguri-Vafa operator that was introduced in (\ref{ovop}).
Since we are regarding the $M$ branes as a probe, the holonomy $V$ is an
arbitrary source. The extra piece in (\ref{defact}) can be interpreted
as coming from an annulus of zero length interpolating between the two sets of
D-branes. Later on we will consider a simple generalization of the above
for an annulus of finite length.

Let us now follow this system through the geometric transition. The $N$
branes disappear, and the background geometry becomes the resolved
conifold. However, the $M$ probe branes are still there. It is natural
to conjecture that they are now wrapping a Lagrangian
submanifold ${\cal C}_{\cal K}$ of ${\cal O}(-1) \oplus {\cal O}(-1)
\rightarrow \IP^1$ that can be obtained from ${\widetilde {\cal C}}_{\cal
K}$ through the geometric transition. The final outcome is the existence of
a map between knots in ${\bf S}^3$ and Lagrangian
submanifolds in ${\cal O}(-1) \oplus {\cal O}(-1)
\rightarrow \IP^1$ which sends
\be
{\cal K} \rightarrow {\cal C}_{\cal K}.
\end{equation}
Moreover, one has $b_1({\cal C}_{\cal K})=1$.
This conjecture is clearly well-motivated in the physics.
Ooguri and Vafa (2000) constructed ${\cal C}_{\cal K}$ explicitly when ${\cal K}$
is the unknot, and Labastida, Mari\~no, and Vafa (2000) proposed Lagrangian submanifolds for certain
algebraic knots and links (including torus knots). Finally, Taubes (2001) has
constructed a map from knots to Lagrangian submanifolds in the
resolved conifold for a wide class of knots.

The Lagrangian submanifold ${\cal C}_{\cal
K}$ in the resolved geometry gives
precisely the open string sector that is needed in order to extend the
large $N$ duality to Wilson loops. According to Ooguri and Vafa (2000),
the free energy of open topological strings
(\ref{totalfreeopen}) with boundary
conditions specified by ${\cal C}_{\cal K}$ is identical to the free energy
of the deformed Chern-Simons theory with action (\ref{defact}), which is
nothing but (\ref{convev}):
\be
F(V)=F_{\rm CS}(V).
\label{wilsondual}
\end{equation}
Notice that, since $b_1({\cal C}_{\cal K})=1$, the topological sectors of
maps with positive winding numbers correspond to vectors $\vec k$ labelling
the connected vacuum expectation values, and one finds
\begin{equation}
\label{freelog}
i^{|\vec k|} \sum_{g=0}^{\infty} F_{g, \vec k} (t) g_s ^{2g-2 + |\vec k|}
=-{1 \over \prod_j j^{k_j}} W_{\vec k}^{(c)}.
\end{equation}
It is further assumed that there is an analytic continuation of 
$F_(V)$ from negative to positive winding numbers in such a way that
the equality (\ref{wilsondual}) holds in general. Another useful way to state
the correspondence (\ref{wilsondual}) is to use the total partition function
of topological open strings (\ref{zr}) instead of the free energy. The duality between
open string amplitudes and Wilson loop expectation values reads simply
\be
Z_R=W_R,
\label{zrwr}
\ee
where $Z_R$ was introduced in (\ref{zr}) and $W_R$ is the knot invariant in representation
$R$.

When ${\cal K}$ is the unknot in the three-sphere, the conjecture of
Ooguri and Vafa can be tested in full detail (Ooguri and Vafa, 2000; Mari\~no and
Vafa, 2002). For more
general knots and links, the open string free energy is not known, but one
can test the duality indirectly by verifying that the Chern-Simons side
satisfies the structural properties of open string amplitudes that we explained at
the end of IV.C (Labastida and Mari\~no, 2001; Ramadevi and Sharkar, 2001; Labastida,
Mari\~no, and
Vafa, 2000; Labastida and Mari\~no, 2002; Mari\~no, 2002b).

\section{String amplitudes and Chern-Simons theory}

The duality between Chern-Simons on ${\bf S}^3$ and closed topological
strings on the resolved conifold gives a very nice realization of
the gauge/string theory duality. However, from the ``gravity''
point of view we do not learn much about the closed string geometry, since
the resolved conifold is quite simple (remember that it only has one
nontrivial Gopakumar-Vafa invariant). It would be very interesting to find
a topological gauge theory dual to more complicated geometries, like the ones
we discussed in section IV, in such a
way that we could use our knowledge of gauge theory to learn about
enumerative invariants of closed strings, and about closed strings in
general.

The program of extending the geometric transition of Gopakumar and Vafa was started by
Aganagic and Vafa (2001). Their basic
idea was to construct geometries that locally contain $T^*{\bf S}^3$'s, and then follow
the geometric transitions to dual geometries where the deformed conifolds are
replaced by resolved conifolds. Remarkably, a large class of non-compact
toric manifolds can be realized in this way, as it was made clear by
Aganagic, Mari\~no, and Vafa (2004) and Diaconescu, Florea, and Grassi (2003b).
In this section we will present some examples where closed string
amplitudes can be computed by using this idea.

\subsection{Geometric transitions for toric manifolds}

The geometries that we discussed in section IV are ${\bf
T}^2 \times \IR$ fibrations of $\IR^3$ which contain two-spheres (represented by the
compact edges of the geometry). In this section we will construct geometries with the
same fibration structure that contain
three-spheres, and can be related by a geometric transition
to some of the toric geometries that we analyzed in section IV.

\begin{figure}
\scalebox{.5}{\includegraphics{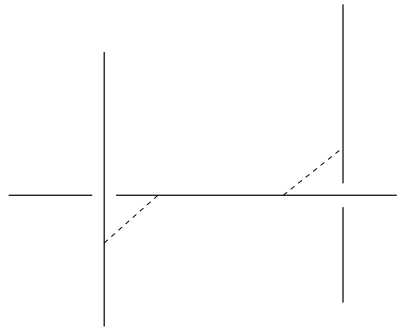}}
\caption{\label{twospheresf}
A Calabi-Yau which is a
${\bf T}^2 \times \IR$ fibration of $\IR^3$. The dashed lines
represent ${\bf S}^3$'s.}
\end{figure}

Recall from the discussion in section V that the deformed conifold has
the structure of a ${\bf
T}^2 \times \IR$ fibration of $\IR^3$ which can be
encoded in a nonplanar graph as in Fig. \ref{deformedf}.
The degeneration loci of the
cycles of the torus fiber are represented in this graph by straight lines,
while the ${\bf S}^3$ is represented by a dashed line in between
these loci. This graphical procedure can be generalized, and it is
easy to construct more
general ${\bf T}^2 \times \IR$ fibrations of $\IR^3$ by specifying degeneration loci
in a diagram that represents the ${\bf R}^3$ basis. A simple example
is shown in Fig. \ref{twospheresf}.
This geometry contains two ${\bf S}^3$'s, represented by
dashed lines. These three-spheres
are also constructed as torus fibrations over the interval, and the cycles that
degenerate at the
endpoints can be read from the graph. In fact, both are described
by a ${\bf T}^2$ fibration where the $(0,1)$ cycle collapses at one endpoint,
and the $(1,0)$ cycle collapses at the other endpoint. As we explained in
V.E, this gives a
Heegaard splitting of the three-sphere along
solid tori. These tori are glued together through the $S$ transformation that
relates one of the collapsing cycles to the other.

We can also construct geometries which contain more general
three-manifolds. If a manifold $M$ admits a Heegaard splitting
along two solid tori, it will be specified by an ${\rm SL}(2,{\bf Z})$ matrix $V_M$
mapping the $(p_L,q_L)$ cycle of one ${\bf T}^2$ to the $(p_R,q_R)$ cycle of the other
${\bf T}^2$. Equivalently, $M$ can be obtained as a torus fibration over an interval
where the $(p_L,q_L)$ and $(p_R,q_R)$ cycles degenerate at the endpoints, as
we explained in V.E in the simple case of the $(1,0)$ and $(0,1)$ cycles. The local geometry
$T^*M$ will be described by two overlapping lines with slopes $-p_L/q_L$ and
$-p_R/q_R$. The dashed line in between them will
represent the three-manifold $M$.

\begin{figure*}
\scalebox{.55}{\includegraphics{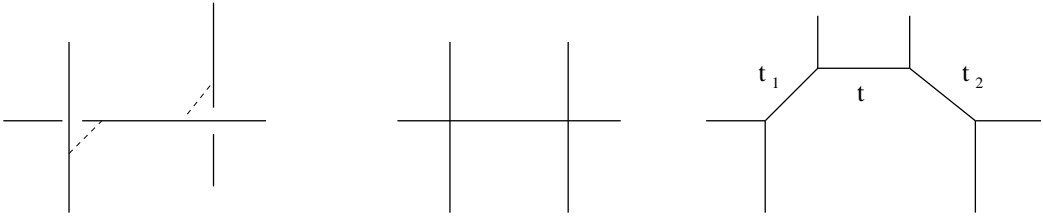}}
\caption{\label{twotransition} This figure shows the geometric transition of the
Calabi-Yau depicted in Fig. \ref{twospheresf}. In the leftmost geometry
there are two three-spheres, represented by
dashed lines.
The intermediate geometry
is singular, and the figure on the right shows the planar graph
associated to the smooth
toric Calabi-Yau after the transition. It contains three $\IP^1$s with K\"ahler
parameters $t_1$, $t_2$ and $t$.}
\end{figure*}

Given a graph like the one in Fig. \ref{twospheresf}, one can try to use the
conifold transition
``locally,'' as it was first explained by Aganagic and Vafa (2001). The
above geometry, for example, contains two deformed
conifolds with their corresponding three-spheres, therefore there is a geometric
transition where the three-spheres go to zero size and then the
corresponding singularities are blown-up to give a resolved geometry. This geometric
transition is depicted in Fig. \ref{twotransition}.
The resolved geometry is clearly toric, and it can be easily built up by
gluing four trivalent vertices, as we explained in IV.D. It has two K\"ahler
classes corresponding
to the two blown-up two-spheres, and denoted by $t_1$, $t_2$ in Fig. \ref{twotransition}.
It also contains a third two-sphere associated to the intermediate,
horizontal leg, with K\"ahler parameter $t$.

Although we have focused on the example depicted in Fig. \ref{twospheresf}, it is
clear what the general philosophy is: one considers a ``deformed'' geometry and performs
geometric transitions ``locally.'' The resulting ``resolved''
geometry will be a toric Calabi-Yau manifold of the type discussed in section IV.
The planar graph describing the resolved geometry can be easily reconstructed from the
nonplanar graph describing the deformed geometry.

\subsection{Closed string amplitudes and geometric transitions}

We will now use the generalized geometric transition that we found in the
last subsection in order to compute the topological string amplitudes.
Let us first wrap $N_i$ branes, $i=1,2$, around the two ${\bf
S}^3$'s of the deformed geometry depicted in Fig. \ref{twospheresf}.
What is the effective topological action describing the resulting
open strings? Since this geometry is not globally of the form
$T^*M$, we are in the situation described in V.D:
for open strings with both ends on the same ${\bf
S}^3$, the dynamics is described by Chern-Simons theory with gauge
group $U(N_i)$, therefore we will have two Chern-Simons theories
with groups $U(N_1)$ and $U(N_2)$. However, there is a
new sector of open strings stretched between the two three-spheres: these are
the nondegenerate instantons that we discussed in V.D following Witten (1995).

Instead of describing these open strings in geometric
terms, it is better to use the spacetime physics associated to
these strings. A similar situation was considered when we
analyzed the incorporation of Wilson loops in the geometric
transition. There we had two sets of intersecting D-branes, giving a
massless complex scalar field living in the intersection and in
the bifundamental representation of the gauge groups. In the situation depicted in
Fig. \ref{twospheresf}, the same arguments indicate that there is a
complex scalar $\phi$ in the representation
$(N_1,\overline N_2)$, corresponding to the bifundamental strings
stretched between the two sets of D-branes. The difference with the situation
that we were considering before is that this complex
scalar is now massive, since the strings have a finite length,
and its mass is proportional to the
``distance'' between the two three-spheres. This length is measured by a
complexified K\"ahler parameter that will be denoted by $r$. The kinetic
term for the complex scalar will be given by
\be
\oint_{{\bf S}^1} \bar \phi (d+A_1 - A_2 -r) \phi.
\label{scaction}
\end{equation}
We can now integrate out this complex scalar field as we did in (\ref{massiveov}) to
obtain the
correction to the Chern-Simons actions on the three-spheres due to
the presence of the new sector of open strings:
\be
\label{oop}
 {\cal O}(U_1, U_2; r) =\exp\Bigl\{\sum_{n=1}^{\infty}\; \frac{e^{-nr}}{n}
\;{\rm Tr} U_1^{n}\; {\rm Tr}U_2^{n}\Bigr\},
\ee
where $U_{1,2}$ are the holonomies of the corresponding
gauge fields around the ${\bf S}^1$ in (\ref{scaction}). The operator
${\cal O}$ can be also interpreted as the amplitude for a primitive annulus of
area $r$ together with its
multicovers, which are labelled by $n$. This annulus ``connects''
the two ${\bf S}^3$s, {\it i.e.} one of its boundaries is a circle in one
three-sphere, and the other boundary is a circle in the other sphere. The sum 
over $n$ in the exponent of (\ref{oop}) is precisely the sum over open 
string instantons in the second term of (\ref{impo}), for 
this particular geometry.

\begin{figure}
\scalebox{.5}{\includegraphics{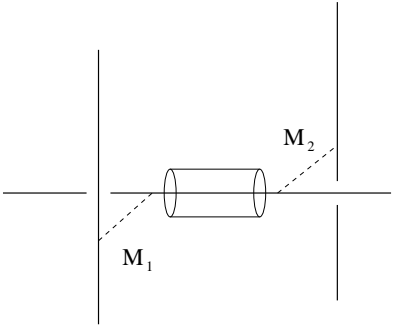}}
\caption{The only nondegenerate instantons contributing to the
geometry of Fig. \ref{twospheresf} come from an annulus stretching along the degeneracy
locus.}\label{annulus}
\end{figure}

The problem now is to determine how many configurations like this one
contribute to the full amplitude. It turns out that the only contributions
come from open strings stretching along the degeneracy locus, {\it i.e.} along
the edges of the graph that represents the geometry. This was
found by Diaconescu, Florea and Grassi (2003a, 2003b)
by using localization arguments, and
derived by Aganagic, Mari\~no, and Vafa (2004) by exploiting invariance under
deformation of complex
structures. This result simplifies the problem enormously, and gives a
precise description of all the nondegenerate instantons contributing in
this geometry: they are annuli stretching along the fixed lines of the
${\bf T}^2$ action,
together with their multicoverings, and the ${\bf S}^1$ in (\ref{scaction}) is the
circle that fibers over the edge connecting $M_1$ and $M_2$. This is illustrated in
Fig. \ref{annulus}.
The action describing the dynamics of topological D-branes in the example
we are considering above is then:
\be
\label{totalaction}
S=S_{\rm CS}(A_1) +S_{\rm CS}(A_2) + \sum_{n=1}^{\infty}\; {e^{-nr}\over n}
\;{\rm Tr} U_1^{n}\; {\rm Tr}U_2^{n},
\end{equation}
where the $A_i$ are $U(N_i)$ gauge connections on $M_i={\bf S}^3$,
$i=1,2$, and $U_i$ are the corresponding holonomies
around the ${\bf S}^1$. There is a very convenient way to write the free energy of the
theory with the above action. First notice that, by following the same
steps that led to (\ref{ovrep}), one can write the operator
(\ref{massiveov})
as
\be
{\cal O}(U_1,U_2;r)=\sum_R {\rm Tr}_R
U_1 e^{-\ell r}{\rm Tr}_R U_2,
\end{equation}
where $\ell$ denotes the number of boxes of the representation $R$. In the
situation depicted in Fig. \ref{annulus}, we see that the boundaries of
the annulus give a knot in $M_1$, and another knot in
$M_2$. Therefore, the total free energy can be written as:
\ben
\label{freetwo}
F &= &F_{\rm CS}(N_1, g_s) +F_{\rm CS}(N_2, g_s)\nonumber\\
&+&
\log  \sum_{R} e^{-\ell r }
W_{R} ({\cal K}_1)  W_{R} ({\cal K}_2),
\end{eqnarray}
where $F_{\rm CS}(N_i, g_s)$ denotes the free energy of Chern-Simons theory
with gauge group $U(N_i)$. These correspond to the degenerate
instantons that come from each of the two-spheres.
Of course, in order to compute (\ref{freetwo})
we need some extra information: we have to know
what the knots ${\cal K}_i$ are topologically, and also
if there is some framing induced by the
geometry. It turns out that these questions can be easily answered if we evaluate the
path integral by cutting the geometry into pieces. The geometry of the knots
is then encoded in the geometry of the degeneracy locus.
\begin{figure}
\scalebox{.5}{\includegraphics{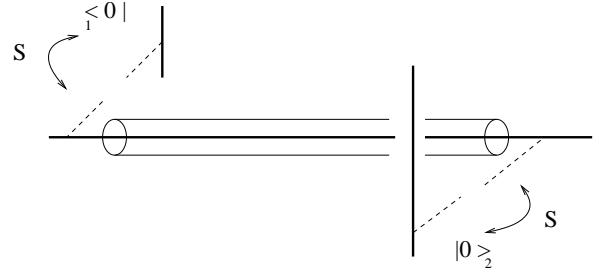}}
\caption{The geometry of Fig. \ref{twospheresf} can be cut into three pieces. The piece that
contains the annulus gives by canonical quantization the state (\ref{anulop}).}\label{mozza}
\end{figure}

The evaluation proceeds as follows: we cut the geometry into three pieces, as indicated in
Fig. \ref{mozza},
by Heegaard splitting the two three-spheres into solid tori. The first piece comes from a
solid torus embedded in the total geometry with no insertion, obtained by splitting $M_1$.
This gives the state ${}_1 \langle 0 |$ in ${\cal H}^*_1({\bf T}^2)$, where the subscript
$1$ refers to
the Hilbert space
of the $U(N_1)$ Chern-Simons theory on $M_1$. Similarly, the third piece is another
solid torus from the splitting of $M_2$, and gives the state $| 0 \rangle_2$.
The path integral with the insertion of
${\cal O}(U_1, U_2;r)$ produces the following operator in the
canonical formalism:
\begin{equation}
{\cal O} = \sum_R |R\rangle_1 \, e^{-\ell r}{}_2\langle
R| \in {\cal H}_1({\bf T}^2) \otimes {\cal H}_2^* ({\bf T}^2)
\label{anulop}
\end{equation}
where $| R \rangle$ is the Chern-Simons state that we constructed in section 2, 
and we have introduced subscripts for the labels of the different Hilbert spaces.
The gluing is made, as before, through the $S$ transformation in both sides,
and the total partition function is then given by
${}_1 \langle 0 | S {\cal O} S | 0 \rangle_2$, so we find:
\be
Z(g_s, N_{1,2}, r)=
\sum_R {}_1 \langle 0 | S |R\rangle_1 e^{-\ell r}{}_2\langle
R|S | 0 \rangle_2.
\label{twospheres}
\ee
Comparing to (\ref{freetwo}), we see that
\be
\label{qdimdef}
W_{R}({\cal K}_i) ={S_{0R} \over S_{00}} (g_s, t_i), \quad i=1,2.
\ee
where $g_s$ is the open string coupling constant $2\pi/(k_i + N_i)$ (which is the same for
the two Chern-Simons theories, see (\ref{equalg})) and $t_i=g_s N_i$ are
the 't Hooft parameters of the $U(N_i)$ Chern-Simons theories.
This means that ${\cal K}_{1,2}$ are unknots in the three-spheres $M_{1,2}$, respectively.
Geometrically, each of the boundaries of the annulus in Fig. \ref{annulus} creates a
Wilson line along the noncontractible cycle of the solid torus along which we split
the three-sphere.

What happens now if we go through the geometric transition of
Fig. \ref{twotransition}? As in the case originally studied by Gopakumar and
Vafa, the `t Hooft parameters become the K\"ahler parameters $t_1$, $t_2$
in the toric diagram of Fig.
\ref{twotransition}. There is a third K\"ahler parameter $t$ in the
toric geometry after the transition. It turns out that this parameter is related
to the parameter $r$ appearing in (\ref{twospheres}) as follows:
\be
t=r- {t_1 + t_2 \over 2}.
\label{renorm}
\ee
This relation was first suggested by Diaconescu, Florea and Grassi (2003a). It is clearly
needed in order
to obtain a free energy of the expected form, with a well-defined limit as
$t_{1,2} \rightarrow \infty$. The total free energy of the
resulting toric manifold can be obtained from (\ref{freetwo}) and (\ref{qdimdef}),
and it can be written in closed form as
\begin{widetext}
\be
\label{firstf}
F= \sum_{d=1}^{\infty} {1\over  d \Bigl(2 \sin {dg_2\over 2}\Bigr)^{2}}
\biggl\{ e^{-d t_1} + e^{-d t_2} + e^{-d t} (1- e^{-d t_1})(1- e^{-dt_2})
\biggr\}.
\ee
\end{widetext}
From this expression we can read the Gopakumar-Vafa invariants of the toric manifold. Notice
that (\ref{firstf}) gives the free energy of closed topological strings at all genera.
In other words, the nonperturbative solution of Chern-Simons theory (which
allows to compute (\ref{qdimdef}) exactly) gives us the nonperturbative answer for the
topological string amplitude. This is one of the most important aspects of this
approach to topological string theory.

One can consider other noncompact Calabi-Yau manifolds and obtain different
closed and open string amplitudes
by using these generalized geometric transitions (Aganagic, Mari\~no, and
Vafa, 2004; Diaconescu, Florea, and Grassi, 2003b). However, this procedure becomes
cumbersome, since in some cases one has to take appropriate limits of the amplitudes
in order to reproduce the sought-for answers. The underlying problem of this approach is that
we are taking as our basic
building block for the resolved geometries the tetravalent vertex that corresponds to the
resolved conifold. It is clear however that the true building block is the trivalent
vertex shown in Fig. \ref{vgraph}, which
corresponds to ${\bf C}^3$. In the next section, we will see how one can define
an amplitude associated to this trivalent vertex that allows one to recover any open or
closed topological string amplitude for noncompact, toric
geometries.

\section{The topological vertex}

\subsection{Framing of topological open string amplitudes}

\begin{figure}
\scalebox{.5}{\includegraphics{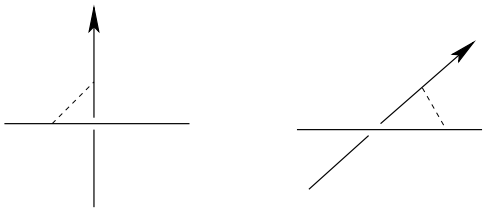}}
\caption{In this figure we show two different ways of compactifying
a Lagrangian submanifold located at the horizontal edge. They are specified by
additional lines (drawn as arrows) where the
torus fibration degenerates. The dashed lines represent the compactified
submanifolds, which have now the topology of ${\bf S}^3$.}\label{framings}
\end{figure}

\begin{figure}
\scalebox{.5}{\includegraphics{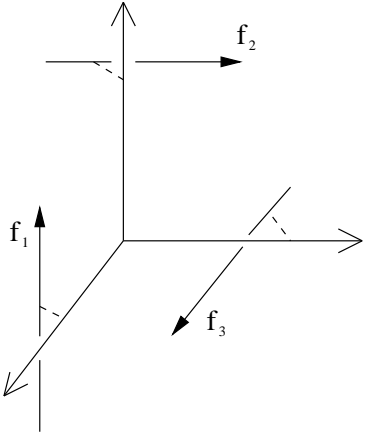}}
\caption{\label{canframing}The canonical choice of framing for the topological
vertex.}
\end{figure}

Since the topological vertex is an open string amplitude, we have
to discuss one aspect that we have not addressed yet:
the framing ambiguity of topological open string amplitudes.
The framing ambiguity was discovered by Aganagic, Klemm, and Vafa (2000).
They realized that when the boundary conditions are
specified by noncompact Lagrangian submanifolds
like the ones described in (\ref{lagcthree}), the corresponding topological
open string amplitudes
are not univocally defined: they
depend on a choice of an integer number (more precisely, one integer number
for each boundary). For the Lagrangian submanifolds studied in IV.D, the framing
ambiguity can be specified by modifying the geometry in an appropriate
way. These Lagrangian submanifolds simply correspond to points in the
edges of the trivalent graphs. Their geometry can be modified by
introducing additional locations in
the base ${\bf R}^3$ where the ${\bf T}^2$ fiber degenerates, as we did before
when we considered general deformed geometries. In this way
the Lagrangian submanifolds become compact ${\bf S}^3$ cycles in the geometry, exactly
as in Fig. \ref{deformedf}. The additional lines are labelled by a vector $f=(p,q)$ where
the $(-q,p)$ cycle degenerates. This procedure is illustrated in Fig \ref{framings}.
It is useful to introduce the symplectic product of
two vectors $v=(v_1, v_2)$ and $w=(w_1, w_2)$ as
\be
v\wedge w = v_1w_2 -v_2 w_1.
\ee
This product is invariant under ${\rm Sl}(2, {\bf Z})$ transformations.
If the original Lagrangian submanifold is located at an edge $v$, the
condition for the compactified cycle to be a nondegenerate ${\bf S}^3$ is
\be
f\wedge v =1.
\label{fr}
\ee
Clearly, if $f$ satisfies (\ref{fr}), so does $f-n v$ for any
integer $n$. The choice of the integer $n$ is precisely the
framing ambiguity found by Aganagic, Klemm, and Vafa (2002). In
the case of the Lagrangian submanifolds
of ${\bf C}^3$ that we constructed in IV.D, a particular choice of compactification
(therefore of framing) that will be very important in the following is shown in Fig.
\ref{canframing}.

What is the effect of a change of framing on open topological string amplitudes? A
proposal for this was made by Aganagic, Klemm, and Vafa (2002) and further
studied by Mari\~no and Vafa (2002),
based on the duality with Chern-Simons theory. As we explained in V.G, vacuum expectation
values of Wilson loops in Chern-Simons theory on ${\bf S}^3$ compute open string
amplitudes, as stated in (\ref{zrwr}). On the
other hand, we explained
in II.D that Wilson loop correlation functions depend on a choice of framing. This
indicates that the framing ambiguity of Chern-Simons theory
corresponds to the ambiguity of topological open string amplitudes that we have
just described. This correspondence also suggests a very precise prescription to compute the
effect of a change of framing for open string amplitudes. Let us consider
for simplicity an open string amplitude
involving a single Lagrangian submanifold, computed for a framing $f$. If we now consider
the framing $f-nv$, the coefficients $Z_R$ of the total partition function (\ref{zr})
change as follows
\be
Z_R \rightarrow (-1)^{n \ell(R)} q^{n \kappa_R\over 2} Z_R,
\label{changefr}
\ee
where $\kappa_R$ was defined in (\ref{kapar}), and $q=e^{ig_s}$. This is essentially
the behavior of Chern-Simons invariants under change of framing spelled out in (\ref{unframing}).
The extra sign in (\ref{changefr}) is crucial to guarantee integrality of the
resulting amplitudes, as it was verified in Aganagic, Klemm, and Vafa (2002) and Mari\~no and
Vafa (2002). If the open string amplitudes involves $L$ boundaries, one has to
specify $L$ different framings, and (\ref{changefr}) is generalized to
\be
Z_{R_1 \cdots R_L} \rightarrow (-1)^{\sum_{\alpha=1}^L n_{\alpha} \ell(R_{\alpha})}
q^{\sum_{\alpha=1}^L n_{\alpha} \kappa_{R_{\alpha}}/2} Z_{R_1 \cdots R_L}.
\label{zrs}
\ee

\subsection{Definition of the topological vertex}

In section IV we considered ${\bf C}^3$ with one Lagrangian
submanifold in each of the vertices of the toric diagram. Since each of these
submanifolds has the topology of ${\bf C}\times {\bf S}^1$, we can consider
the topological {\it open} string amplitude associated to this geometry. The total
open string partition function will be given by
\be
Z(V_i) = \sum_{R_1, R_2, R_3} C_{R_1 R_2 R_3} \prod_{i=1}^3
{\rm Tr}_{R_i} V_i
\label{zvertex}
\ee
where $V_i$ is a matrix source associated to the $i$-th Lagrangian
submanifold. The amplitude $C_{R_1 R_2 R_3}$ is naturally a
function of the string coupling constant $g_s$ and, in the genus
expansion, it contains information about maps from Riemann
surfaces of arbitrary genera into ${\bf C}^3$ with boundaries on
$L_i$. This open string amplitude is called the {\it topological vertex},
and it is the basic object from which, by gluing, one
can obtain closed and open string amplitudes on arbitrary toric
geometries. Since the vertex is an open string amplitude, it will depend on a choice
of three different framings. As we explained in the previous section, this choice
will be given by three different vectors $f_1$, $f_2$ and $f_3$ that specify
extra degeneration loci and lead to a compactification of the $L_i$.

As we mentioned in V.D, the ${\bf C}^3$ geometry can be represented by graphs involving
three vectors $v_i$ obtained from the set in Fig. \ref{vgraph}
by an ${\rm Sl}(2, {\bf Z})$ transformation, and satisfying (\ref{sumvzero}).
We will then introduce a topological vertex amplitude
$C^{(v_i, f_i)}_{R_1 R_2 R_3}$ which depends on both, a choice of three
vectors $v_i$ for the edges, and a choice of three vectors $f_i$ for the framings.
Due to (\ref{fr}) we require
$$f_i\wedge v_i =1.$$
We will
orient the edges $v_i$ in a clockwise way. Since wedge products are
preserved by ${\rm Sl}(2, {\bf Z})$,
we also have
\be
v_2 \wedge v_1 =v_3\wedge v_2 = v_1 \wedge v_3=1.
\ee
However, not all of these choices give independent
amplitudes. First of all, there is an underlying ${\rm Sl}(2,{\bf Z})$
symmetry relating the choices: if $g \in {\rm Sl}(2,{\bf Z})$, then the
amplitudes are invariant under
$$
(f_i,v_i)\rightarrow (g\cdot f_i,g\cdot v_i)
$$
Moreover, if the topological vertex amplitude $C^{(v_i, f_i)}_{R_1 R_2 R_3}$
is known for a set of framings $f_i$, then it can be
obtained for any set of the form $f_i-n_iv_i$, and it is given by the general
rule (\ref{zrs})
\ben
\label{framedc}
& & C^{(v_i, f_i-nv_i)}_{R_1R_2R_3} \nonumber \\ &=&
(-1)^{\sum_i n_i \ell(R_i)}q^{\sum_i n_i \kappa_{R_i}/2}C^{(v_i, f_i)}_{R_1R_2R_3},
\een
for all admissible choices of the vectors $v_i$.
Since any two choices of framing can be related through (\ref{framedc}),
it is useful to pick a convenient set of $f_i$ for any
given choice of $v_i$ which we will define as the {\it canonical framing}
of the topological vertex. This canonical framing turns out to be
$$(f_1,f_2,f_3)=(v_2,v_3,v_1).$$
Due to the ${\rm Sl}(2, {\bf Z})$ symmetry and the transformation
rule (\ref{framedc}), any topological vertex amplitude can be
obtained from the amplitude computed for a {\it fixed} choice of $v_i$ in the
canonical framing. A useful choice of the $v_i$ is
$v_1= (-1,-1),v_2 = (0,1), v_3 =(1,0)$, as in Fig. \ref{vgraph}.
The vertex
amplitude for the canonical choice of $v_i$ and
in the canonical framing will be simply denoted by
$C_{R_1 R_2 R_3}$. Any other choice of framing will be
characterized by framing vectors of the
form $f_i- n_i v_i$, and the corresponding vertex amplitude will be
denoted by
$$C^{n_1,n_2,n_3}_{R_1 R_2 R_3}.$$
Notice that $n_i = f_{i}\wedge v_{i+1}$ (where $i$ runs mod $3$).

One of the most important properties of $C_{R_1 R_2 R_3}$ is its
{\it cyclic symmetry}.
To see this, notice that the ${\rm Sl}(2, {\bf Z})$ transformation
$g= TS^{-1}$ takes
$$ (v_i, f_i) \rightarrow (v_{i+1}, f_{i+1})$$
where again $i$ runs mod $3$. It then follows that
\be
\label{syma}
C_{R_1 R_2 R_3}= C_{R_3 R_1 R_2}= C_{R_2 R_3 R_1}.
\ee
Finally, it will be sometimes useful to consider the vertex in the
basis of conjugacy classes $C_{\vec k^{(1)} \vec k^{(2)}\vec k^{(3)}}$, which
is obtained from $C_{R_1 R_2 R_3}$ by
\be
C_{\vec k^{(1)} \vec k^{(2)}\vec k^{(3)}}=\sum_{R_i} \prod_{i=1}^3
\chi_{R_i}(C(\vec k^{(i)})) C_{R_1 R_2 R_3}.
\ee

\subsection{Gluing rules}

We saw in section IV that any
noncompact toric geometry can be encoded in a planar graph
which can be obtained by gluing trivalent vertices. It is then natural to
expect that the string amplitudes associated to such a diagram can be computed by
gluing the open topological string amplitudes associated to
the trivalent vertices, in the same way that one computes
amplitudes in perturbative quantum field theory by gluing vertices
through propagators. This idea was suggested by Aganagic, Mari\~no, and Vafa (2004)
and Iqbal (2002), and was developed into a complete set of rules by Aganagic, Klemm,
Mari\~no, and Vafa (2003).
The gluing rules for the topological vertex turn out to be
quite simple. Here we will state three rules (for a change of orientation in one
edge, for the propagator, and for the matching of framings in the gluing)
which make possible to compute any closed string amplitude on toric, noncompact
Calabi-Yau threefolds. They make also possible to compute open string amplitudes
for Lagrangian submanifolds on edges that go to infinity. The case
of Lagrangian submanifolds in inner edges is also very easy
to analyze, but we refer the reader to the paper by Aganagic, Klemm,
Mari\~no, and Vafa (2003)
for the details. A mathematical point of view on the gluing rules can be found
in Diaconescu and Florea (2003).

1) {\it Orientation}. Trivalent vertices are glued along their edges, and this corresponds
to gluing curves with holes along their boundaries. In order to do
that, the boundaries must have opposite orientations. This change of orientation
will be represented as an inversion of the edge vector, therefore in gluing
the vertices we will have an outgoing edge on one side, say $v_1$, and an ingoing edge on
the other side, $-v_1$. What is the corresponding effect on the amplitude
$C_{\vec k^{(1)} \vec k^{(2)} \vec k^{(3)}}$? Changing the
orientation of $h$ boundaries along the first edge
gives rise to a relative factor $(-1)^h$, where
$h = | \vec k^{(1)}| $. In the language
of topological D-branes, this means that we are gluing
branes to antibranes (Vafa, 2001b). If we denote by $Q^t$ the representation
whose Young tableau is transposed to the Young tableau of $Q$ ({\it i.e.}
is obtained by exchanging rows and columns), then one has the following
relation between characters
\be
\label{tran} \chi_{Q^t}(C(\vec{k}))=(-1)^{|{\vec{k}}|+\ell(Q)}\chi_Q(C(\vec{k})),
\ee
and from here one can easily deduce
$$C_{R_1 R_2 R_3}\rightarrow
(-1)^{\ell(R_1)} C_{R_1^t R_2 R_3}$$
as we invert the orientation of $v_1$. Of course, a similar equation follows
for the other $v_i$.

2) {\it Propagator}. Since gluing the edges corresponds to gluing
curves with holes along their boundaries,
we must have matching number of holes and winding numbers
along the edge. Therefore, the propagator must be diagonal in the
$\vec k$ basis. After taking into account the change of
orientation discussed above, and after dividing by the
order of the automorphism group associated to $\vec k$ (which
is nothing but $z_{\vec k}$), we find that the propagator
for gluing edges with representations $R_1$, $R_2$ is given by
\be
(-1)^{\ell (R_1)} e^{-\ell (R_1)t} \delta_{R_1 R_2^t},
\ee
where $t$ is the K\"ahler parameter that corresponds to the
$\IP^1$ represented by the gluing edge.

3) {\it Framing}. When gluing two vertices, the framings of the
two edges involved in the gluing have to match. This means
that in general we will have to change
the framing of one of the vertices. Let us consider the case in which we glue
together two vertices with outgoing vectors $(v_i, v_j, v_k)$ and $(v_i', v_j', v_k')$,
respectively, and let us assume that we glue them through the vectors $v_i$, $v_i'=
-v_i$. We also assume that both vertices are canonically framed, so that
$f_i=v_j$, $f_i'=v'_j$. In order to match the framings we have to change the framing
of, say, $v_i'$, so that the new framing is $-f_i$ (the opposite sign is again
due to the change of orientation). There is an integer $n_i$ such that
$f_i'-n_i v_i' = -f_i$
(since $f_i\wedge v_i=f_i'\wedge v_i'=1$, $f_i + f_i'$ is parallel to $v_i$), and it is
immediate to check that
$$
n_i=v'_j \wedge v_j.$$
The gluing of the two vertex amplitudes is then given by
\be
\label{gluecs}
 \sum_{R_i} C_{R_jR_k R_i} e^{- \ell(R_i) t_i}(-1)^{(n_i+1) \ell(R_i)}
q^{-n_i \kappa_{R_i}/2}
  C_{R_i^t R_j'R_k'}
\ee
where we have taken into account the change of
orientation in the $(v_i', v_j', v_k')$ to perform the
gluing, and $t_i$ is K\"ahler parameter associated to the edge.

Given then a planar trivalent graph representing a
noncompact Calabi-Yau manifold without D-branes, we can compute 
the closed string amplitude as follows: we give a presentation 
of the graph in terms of vertices glued together, as we did 
in IV.D. We associate the appropriate
amplitude to each trivalent vertex (labelled by representations),
and we use the above gluing rules. The edges
that go to infinity carry the trivial representation, and
we finally sum over all possible representations along the inner
edges. The resulting quantity is the total partition
function $Z_{\rm closed}=e^F$ for closed string amplitudes. We can slightly 
modify this rule to compute open string amplitudes associated to 
D-branes, in the simple case in which the Lagrangian 
submanifolds are located at the outer edges of the graph ({\it i.e.} the 
edges that go to infinity). 
In this case, we compute the amplitude by associating the representations 
$R_1, \cdots, R_L$ to the outer edges with D-branes. The result is 
$Z_{\rm closed}Z_{R_1 \cdots R_L}$, where $Z_{R_1 \cdots R_L}$ is the open string 
amplitude that appears in (\ref{zmanyv}).    

We will present some concrete examples of this procedure
in a moment. Before doing that, we will derive an explicit
expression for the topological vertex amplitude by using a
geometric transition.

\subsection{Derivation of the topological vertex}

\begin{figure*}
\scalebox{.5}{\includegraphics{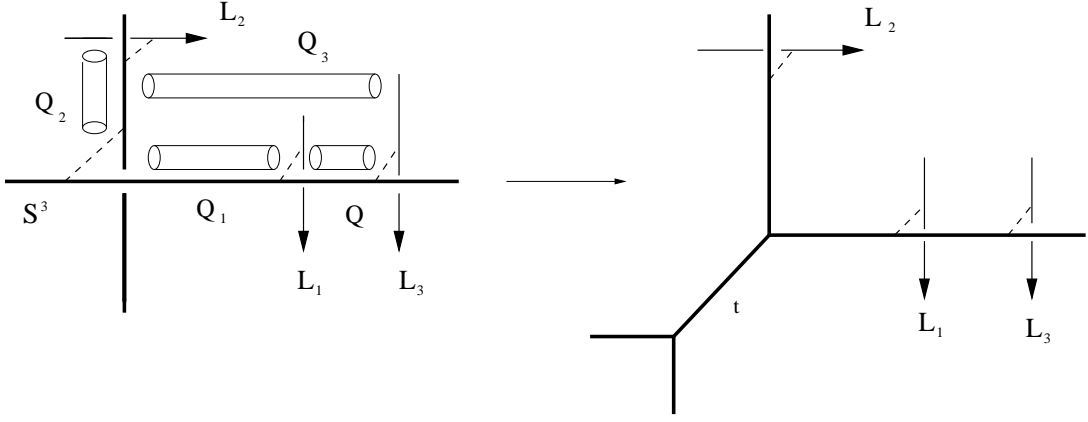}}
\caption{\label{vertexsetting}The configuration used to derive the
topological vertex amplitude. The figure on the left shows the ``deformed" geometry.
The figure
on the right shows the ``resolved" geometry obtained by
geometric transition. It contains a resolved conifold and a $\IP^1$ of size $t$.
We have also depicted in the figure on the left the open strings stretched among the
different branes which contribute to the amplitude. }
\end{figure*}

In order to derive the expression for the vertex, we will
consider the configuration drawn in the first picture in Fig. \ref{vertexsetting},
which represents a geometry with an ${\bf S}^3$ together with
three Lagrangian submanifolds $L_1$, $L_2$ and $L_3$. We also make
a choice of framing for these Lagrangian submanifolds, indicated by
arrows. The world-volumes of the ${\bf S}^3$ and of $L_1$, $L_3$ are parallel,
and we consider topological D-branes wrapped on ${\bf S}^3$ and the $L_i$.
The branes wrapping the $L_i$ are probe (spectator) branes,
and the large $N$ transition of the three-sphere leads to a geometry
with a resolved conifold and three framed Lagrangian submanifolds. As we have
seen in the examples above, the K\"ahler parameter of the $\IP^1$ of the conifold
$t$ is the 't Hooft parameter of the Chern-Simons theory on ${\bf S}^3$. The
resulting configuration is
shown in the second picture of Fig. \ref{vertexsetting}, and can be related to the
topological vertex of Fig. \ref{canframing} by (a) taking the K\"ahler parameter $t$
of the $\IP^1$ in the resolved conifold to infinity (so that the extra trivalent vertex
disappears), and by (b) moving the Lagrangian submanifold $L_1$ to the outgoing
edge along the direction $(-1,-1)$. We will
first compute the total open string amplitude by using the
geometric transition, and then we will implement (a) and (b).

The open string theory on the ${\bf S}^3$ is $U(N)$ Chern-Simons
theory with some matter fields coming from the three non-compact
Lagrangian submanifolds $L_i$. As we discussed in VI,
there are strings stretching between the ${\bf S}^3$ and $L_{1,2,3}$, and also
strings between $L_1$ and $L_3$. These stretched strings are annuli along
the degeneracy locus, and they are depicted in Fig. \ref{vertexsetting}.
The only spacetime excitation
associated to these strings is a
matter field in the bifundamental representation, and integrating it out corresponds to
inserting an annulus operator like (\ref{oop}). When the two branes
intersect on a circle (like the branes considered in VI,
or like the ${\bf S}^3$ and $L_2$ in this situation) the matter field
is a boson (a scalar field). When the branes are parallel, however, it is a fermion. This
is because we can turn the two parallel branes into a brane and an antibrane
intersecting along a circle. This leads to a Grassmann field, as explained
in Vafa (2001b), and the resulting operator turns out to be
\ben
& & \exp \bigg\{  - \sum_n {e^{-nt}\over n}{\rm Tr}\, U^n {\rm Tr}\, V^n \bigg\}
\nonumber\\ &=&\sum_R {\rm Tr}_R \, U e^{-\ell(R) t} (-1)^{\ell(R)}
{\rm Tr}_{R^t}V.
\een
In Fig. \ref{vertexsetting} the probe branes associated to $L_2$ and the
dynamical branes on ${\bf S}^3$ intersect on a circle, while the
probe branes associated to $L_1$, $L_3$ are parallel to each other and to the dynamical
branes. We then have the following operators:
\ben
& & \sum_{Q_1} {\rm Tr}_{Q^t_1} \, U_1 e^{-\ell(Q_1) t} (-1)^{\ell(Q_1)}
{\rm Tr}_{Q_1} {\widehat V}_1, \nonumber\\
& & \sum_R {\rm Tr}_{Q_2} \, U_2 e^{-\ell(Q_2) t} 
{\rm Tr}_{Q_2}V_2, \nonumber\\
& & \sum_{Q_3} {\rm Tr}_{Q^t_3} \, U_1 e^{-\ell(Q_3) t} (-1)^{\ell(Q_3)}
{\rm Tr}_{Q_3} V_3, \nonumber\\
& & \sum_{Q} {\rm Tr}_{Q^t} \, V_1 e^{-\ell(Q) t} (-1)^{\ell(Q)}
{\rm Tr}_{Q} V_3, 
\een
which correspond to the annuli labelled with representations $Q_1$, $Q_2$, $Q_3$ and 
$Q$ in Fig. \ref{vertexsetting}. The matrices $V_2$ and $V_3$ are sources corresponding
to D-branes wrapping $L_2$, $L_3$, while $\widehat V_1$, $V_1$ are sources
for branes wrapping $L_1$ with opposite orientations, and represent
Chan-Paton factors for open strings ending on opposite sides of $L_1$. $U_1$, $U_2$ are holonomies of the
gauge connection on ${\bf S}^3$ around the boundaries of the annuli with representations 
$Q_1$ and $Q_2$ (the boundary of the annulus carrying the representation $Q_3$ is geometrically 
identical to the boundary of the annulus associated to $Q_1$, and it gives the holonomy $U_1$ 
as well). Putting all these ingredients together, we find 
that the open string amplitude on the deformed geometry is given by
\begin{widetext}
\be
Z(V_1, V_2, V_3) ={1\over S_{00}}
\sum_{Q_1,Q_2,Q_3,Q} (-1)^{\ell(Q_1)+\ell(Q_3)+\ell(Q)}
\langle {\rm Tr}_{Q_2}U_2 \; {\rm Tr}_{Q_1^t}U_1 {\rm Tr}_{Q_3^t} U_1\rangle
{\rm Tr}_{Q_1} {\widehat V}_{1}\;{\rm Tr}_{Q^t} V_1\;{\rm Tr}_{Q_2} V_2\;{\rm
Tr}_{Q\otimes Q_3} V_3
\label{vertexdef}
\ee
\end{widetext}
where we have factored out $1/S_{00}$,
the partition function of ${\cal O}(-1)\oplus
{\cal O}(-1)\rightarrow \IP^1$. The above amplitude is an
open string amplitude with three boundaries, and
$V_i$ are the corresponding sources. Notice that the annuli that that carry the 
representations $Q_1$, $Q_3$ are supported
on the horizontal edge, while the annulus connecting $L_2$ to ${\bf S}^3$ lies on the
vertical edge. The horizontal and the vertical edge are related by an $S$ transformation,
therefore
\ben
\langle {\rm Tr}_{Q_2}U_2 \; {\rm Tr}_{Q_1^t}U_1 {\rm Tr}_{Q_3^t} U_1\rangle &=&
\sum_{Q'} N_{Q_1^t Q_3^t}^{Q'} \langle Q_2| S|Q' \rangle \nonumber \\
&=&\sum_{Q'} N_{Q_1^t Q_3^t}^{Q'} S^{-1}_{Q_2 Q'},
\een
where we have fused together the $U_1$ holonomies.
From a geometric point of view, this means that the
boundaries of the annuli give a link in ${\bf S}^3$ with the
topology depicted in Fig. \ref{fusion} (where the representations 
$R, R_1, R_2$ in Fig. \ref{fusion} are now $Q_2$, $Q_3$ and $Q_1$, 
respectively), and the above expression is nothing but 
(\ref{threefus}). We can also use the direct sum formula
(\ref{dsum}) for this invariant, and we finally arrive at the
following expression for (\ref{vertexdef}):
\begin{widetext}
\be
\label{wf}
Z(V_1, V_2, V_3) = \sum_{Q_1,Q_2,Q_3,Q}
 (-1)^{\ell(Q_1)+\ell(Q_3)+\ell(Q)}{{\cal W}_{Q_1^t Q_2}
 {\cal W}_{Q_3^t  Q_2}\over
 {\cal W}_{ Q_2}}
\;{\rm Tr}_{Q_1} {\widehat V}_1 \;{\rm Tr}_{Q^t} V_1
\; {\rm Tr}_{Q_2}V_2
\;{\rm Tr}_{Q \otimes Q_3} V_3,
\ee
\end{widetext}
where ${\cal W}_{R_1 R_2}$ is the Hopf link invariant defined in (\ref{hopfplus}) and
evaluated in (\ref{mlfor}).  (\ref{wf}) gives the answer
for the open topological string amplitude on the geometry
depicted on the left in Fig. \ref{vertexsetting}. We now
incorporate the two modifications which are needed in order to
obtain the topological vertex. First of all, we have to take $t\rightarrow \infty$. As
we pointed out in the last section, in order to have a well-defined limit it is crucial
to renormalize the Chern-Simons expectation values. The relation (\ref{renorm}) suggests
the definition
\be
W_{R_1 R_2} =\lim_{t\rightarrow \infty} e^{-{\ell(R_1) + \ell (R_2) \over 2}t}
{\cal W}_{R_1 R_2}.
\label{largens}
\ee
This limit exists, since ${\cal W}_{R_1 R_2}$ is of the form
$\lambda^{{\ell (R_1)+\ell (R_2)\over 2}} W_{R_1 R_2} + {\cal O} (e^{-t})$ (remember that
$\lambda=e^t$). The
quantity $W_{R_1 R_2}$, which is the ``leading" coefficient of the
Hopf link invariant (\ref{hopfplus}), is the building block of the topological
vertex amplitude. It is
a rational function of $q^{\pm {1\over 2}}$, therefore it only depends on the
string coupling constant. We will also denote $W_R=W_{R 0}$.
The limit (\ref{largens}) was first considered by Aganagic, Mari\~no, and Vafa (2004).

\begin{figure*}
\scalebox{.5}{\includegraphics{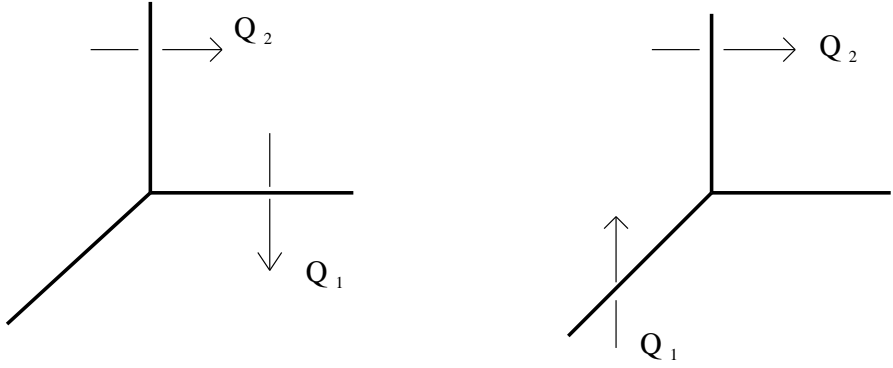}}
\caption{\label{move}Moving the Lagrangian submanifold with representation
$Q_1$ to the outgoing edge.}
\end{figure*}

In order to implement the second modification, we have to understand what is the
effect on the amplitude of moving $L_1$ to the outgoing edge along $(-1, -1)$. In
order to do that, we consider the simplified situation depicted in Fig. \ref{move}
where we only have two stacks of
D-branes wrapping $L_{1,2}$. The amplitude can be easily computed following the arguments
that led to (\ref{wf}), and one immediately obtains
$$Z(V_1, V_2) = \sum_{Q_1, Q_2} W_{Q_2 Q_1^t}~ (-1)^{\ell(Q_1)}~ {\rm Tr}_{Q_1}
V_1 ~ {\rm Tr}_{Q_2}V_2.$$
On the other hand, this is a particular case of the topological vertex
amplitude with $R_1$ trivial, $f_2=v_3$ and $f_3=(0,-1)$, so there is a noncanonical
framing on $v_3$ which corresponds to $n=-1$. We deduce
\be
\label{twopoint}
C_{0 R_2 R_1}=W_{R_2 R_1^t} q^{\kappa_{R_1}/2}.
\ee
On the other hand, the amplitude on the right-hand side of Fig. \ref{move} is
the canonically framed
vertex $C_{R_1 R_2 0}$, but by cyclic symmetry this is equal to (\ref{twopoint}) with
$R_2 \leftrightarrow R_1$.
We conclude that in going from the left to the right hand side of
Fig. \ref{move} we must replace
\ben
& & (-1)^{\ell(Q_1)} W_{Q_2 Q_1^t}\; {\rm Tr}_{Q_1} {\widehat V}_1\; {\rm Tr}_{Q_2} V_2
\nonumber\\
& & \rightarrow W_{Q_2^t Q_1} q^{\kappa_{Q_2}/2}\; {\rm Tr}_{Q_1} V_1
\;{\rm Tr}_{Q_2}V_2.\een
After moving $L_1$ to the outgoing edge, all strings end on the same side of the
corresponding branes, and this explains why we have
replaced ${\widehat V}_1$ by $V_1$ in the above formula. Collecting
the coefficient of ${\rm
Tr}_{R_1} V_1\; {\rm Tr}_{R_2} V_2\; {\rm Tr}_{R_3} V_3$ in the
partition function we compute $C^{0,0,-1}_{R_1R_2R_3}$.
We then get the following expression for the
topological vertex amplitude in the canonical framing:
\be
\label{vertex}
C_{R_1 R_2 R_3}=
q^{\kappa_{R_2} + \kappa_{R_3}\over 2} \sum_{Q_1,Q_3, Q} N_{Q Q_1}^{~~R_1} N_{Q
Q_3}^{~~R_3^t}
{W_{R_2^t Q_1}W_{R_2 Q_3}\over W_{R_2}}.
\ee
This is the final expression for the topological vertex amplitude.

Using (\ref{mlfor}) it is possible to give an explicit expression for $W_{R_1 R_2}$
which is useful in
computations. It is easy to see that the leading coefficient
in $\lambda$ in (\ref{mlfor}) is obtained by taking the $\lambda$-independent
piece in (\ref{cnml}). The
generating function of elementary symmetric polynomials (\ref{eser}) becomes then
\be
S(t)
\prod_{j=1}^{c_R} { {1 + q^{l^R_j -j} t}
\over {1 + q^{- j} t}},
\ee
where
\be
S(t)=\prod_{j=1}^{\infty} (1 + q^{-j}t)=1+\sum_{r=1}^{\infty} {q^{-{r (r+1)\over 2}}
t^r  \over \prod_{m=1}^r [m]}.
\ee
In terms of Schur polynomials, we find:
\be
W_{R_1 R_2}(q)= s_{R_2}(x_i=q^{-i+{1\over 2}})s_{R_1}(x_i=q^{l_i^{R_2}-i+{1\over 2}}),
\ee
where there are now an infinite number of variables $x_i$ with $i=1,2, \cdots$. 
One can also write (\ref{vertex}) in terms of skew Schur
polynomials (Okounkov, Reshetikhin, and
Vafa, 2003), and by using the properties of these polynomials one finds identities
for the topological vertex that are very useful
in computations (Hollowood, Iqbal, and Vafa, 2003;
Eguchi and Kanno, 2004)

\subsection{Some applications}

We will now present some examples of computation of topological
string amplitudes by using the topological vertex.

1) {\it Resolved conifold}. The toric diagram for the
resolved conifold geometry is depicted in Fig. \ref{res}. Our rules give immediately:
\be
Z_{\IP^1}=\sum_{R} C_{00  R^t} (-1)^{\ell(R)}
e^{-\ell(R)  t} C_{R 00}.
\ee
Since $C_{R00}=W_R=s_R(x_i=q^{-i+{1\over 2}})$, we can use the
well-known formula (see for example Macdonald, 1995; Fulton and Harris, 1991)
\be
\sum_R s_R(x) s_R(y)= {1\over \prod_{i,j}(1-x_i y_j)}
\ee
to obtain
\be
Z_{\IP^1}=\exp \biggl\{ -\sum_{d=1}^{\infty}
{e^{-dt}\over d (q^{d\over 2} - q^{-{d\over 2}})^2} \biggr\},
\ee
in agreement with the known result (\ref{resf}).

2) {\it Framed unknot}. Let us now consider an open string amplitude, corresponding
to the resolved conifold with a Lagrangian brane in one of the external legs, and in
arbitrary framing $p$. The open string amplitude is given by
\be
Z(V,p)=\sum_Q Z_Q (p) {\rm Tr}_Q\, V,
\ee
where
\be
Z_Q (p)= {1 \over Z_{\IP^1}} (-1)^{\ell(Q)p}q^{\kappa_Q p\over 2}
\sum_{R} C_{0 Q  R^t}
(-e^{-t})^{\ell(R)} C_{R 0 0}.
\ee
One can use various identities involving symmetric polynomials to
show that
\be
Z_Q (p) =(-1)^{\ell(Q)p}q^{\kappa_Q p/2} e^{-\ell(Q)t/2}({\rm dim}_q Q).
\ee
The r.h.s. is essentially the Chern-Simons invariant of the unknot.
The open string free energy is given by $F(V,p)=\log\, Z(V,p)$,
which can be written as in (\ref{freevk}). It turns out that the leading term
of $F_{w,g}(p,t)$ as $t\rightarrow \infty$ (which we will simply denote by
$F_{w,g}(p)$) can be computed in open Gromov-Witten theory
by using Hodge integrals (Katz and Liu, 2002; see also Li and Song, 2002). The result is:
\begin{widetext}
\be
\label{klform}
F_{w, g}(p) =(-1)^{p\ell+1}
(p(p+1))^{h-1}\biggl( \prod_{i=1}^h { \prod_{j=1}^{w_i-1}
(j+w_i p) \over (w_i-1)! } \biggr)
 {\rm Res}_{u=0}
\int_{{\overline M}_{g,h}}
{c_g (\DE ^{\vee}(u))c_g(\DE ^{\vee} ((-p-1)u)) c_g (\DE ^{\vee} (p u))
u^{2h-4} \over \prod_{i=1}^h (u- w_i \psi_i)}.
\ee
\end{widetext}
In this formula, $\overline M_{g,h}$ is the Deligne-Mumford moduli space,
$\DE$ is the Hodge bundle over $\overline M_{g,h}$, and its dual
is denoted by $\DE ^{\vee} $. We have also written
\begin{equation}\label{serieshod}
c_g(\DE ^{\vee} (u))= \sum_{i=0}^g c_{g-i} (\DE ^{\vee}) u^i,\end{equation}
where $c_j(\DE ^{\vee})$ are Chern classes,
and similarly for the other two factors. On the other hand, the $t\rightarrow \infty$ limit
of $Z_Q$ is
\be
 (-1)^{\ell(Q)p}q^{\kappa_Q p/2}W_Q.
\ee
By equating the open Gromov-Witten result with the Chern-Simons result one finds
a highly non-trivial identity that expresses the
Hodge integrals appearing in (\ref{klform}) in terms
of the $W_Q$'s, as first noticed by Mari\~no and Vafa (2002). The explicit
expression that one finds is
\begin{widetext}
\begin{eqnarray}
\label{moreor}
\sum_{g=0}^\infty F_{\vec k, g} g_s^{2g-2 + |\vec k|}
&= & (-1)^{p \ell} i^{-|\vec k| -\ell} \prod_j k_j! \sum_{n\ge 1}
{(-1)^{n} \over n}
\sum_{\vec k_1, \cdots, \vec k_n} \delta_{\sum_{\sigma=1}^n \vec k_\sigma, \vec k}
\sum_{R_{\sigma}} \prod_{\sigma =1}^n { \chi_{R_{\sigma}} (C(\vec k_{\sigma}))
\over z_{\vec k_{\sigma}} }\nonumber\\ & &\cdot
{\rm e}^{i (p +{1\over 2})\kappa_{R_{\sigma}}g_s/2}
\prod_{ 1\le i < j \le c_{R_{\sigma}}}
{\sin \Bigl[ (l^{\sigma}_i -l^{\sigma}_j +j-i) g_s/2\Bigr]
\over \sin \Bigl[ (j-i) g_s /2\Bigr]}
\prod_{i=1}^{c_{R_{\sigma}}}
\prod_{v=1}^{l^{\sigma}_i} {1\over 2 \sin
\Bigl[ (v-i+c_{R_{\sigma}}) g_s /2\Bigr]},\nonumber\\
\end{eqnarray}
\end{widetext}
where we have relabelled $w\rightarrow \vec k$ for positive winding numbers, as explained
in IV.C. This equality is a very explicit mathematical prediction of the
duality between Chern-Simons theory and topological string theory.
It has been rigorously proved by Liu, Liu, and Zhou (2003a) and by
Okounkov and Pandharipande (2004), and
shown to have many applications in Gromov-Witten theory
(Liu, Liu and Zhou, 2003b; Zhou, 2003).

2) {\it Local} $\IP^2$. The toric diagram is depicted in Fig. \ref{figptwo}.
Using again the rules explained above, we find
the total partition function
\ben
Z_{\IP^2}& =& \sum_{R_1,R_2,R_3}
(-1)^{\sum_i \ell(R_i)}e^{-\sum_i \ell(R_i) t}
q^{-\sum_i \kappa_{R_i}} \nonumber \\ & & \times C_{0 R_2^t R_3} C_{0 R_1^t R_2}
C_{0 R_3^t R_1},
\label{ptwo}
\een
where $t$ is the K\"ahler parameter corresponding to the hyperplane
class in $\IP^2$. Using that $C_{0 R_2 R_3^t}= W_{R_2 R_3}q^{-\kappa_{R_3}/2}$
one recovers the expression for $Z_{\IP^2}$ first obtained by
Aganagic, Mari\~no, and Vafa (2004) by using the method
of geometric transition explained in VI. Notice that the free energy
has the structure
\be
\label{fp}
F_{\IP^2}=\log \biggr\{ 1 + \sum_{\ell=1}^{\infty} a_{\ell} (q) e^{-\ell
t}\biggr\}= \sum_{\ell=1}^\infty a_{\ell}^{(c)}(q) e^{-\ell t}.
\ee
The coefficients $a_{\ell} (q)$,
$a_{\ell}^{(c)}(q)$  can be easily
obtained in terms of $W_{R_1 R_2}$. One finds,
for example,
\ben
\label{asf}
a_1^{(c)}(q) &=&a_1(q)= - { 3 \over (q^{{1\over2}} -q^{-{1\over2}})^2}, \nonumber\\
a_2^{(c)}(q)&=&
{ 6 \over (q^{{1\over2}} -q^{-{1\over2}})^2} + {1 \over 2}a_1 (q^2).
\een
If we compare to (\ref{gvseries}) and
take into account the effects of multicovering, we find the
following values for the Gopakumar-Vafa invariants of ${\cal O}(-3)
\rightarrow \IP^2$:
\ben
\label{gvfirst}
n_1^0=3, \,\,\,\,\,\,\,\,\,\,\,\,\,\,\,\,\,\,\,  n_1^g =0 \,\,\, {\rm
for}\, g>0,\nonumber\\
 n_2^0 = -6, \,\,\,\,\,\,\,\,\,\,\,\,\,\,\,\,\,\,\,  n_2^g =0 \,\,\, {\rm
for}\, g>0,
\een
in agreement with the results listed in (\ref{pgv}). In fact, one can
go much further with this method and compute the Gopakumar-Vafa invariants
to high degree. We see again that the use
of nonperturbative results in Chern-Simons theory leads to
the topological string amplitudes to {\it all genera}.
A complete listing of the Gopakumar-Vafa invariants up to
degree 12 can be found in Aganagic, Mari\~no, and Vafa (2004). The partition
function (\ref{ptwo}) can be also computed
in Gromov-Witten theory by using localization techniques, and one
finds indeed the same result (Zhou, 2003).
\begin{figure}
\scalebox{.5}{\includegraphics{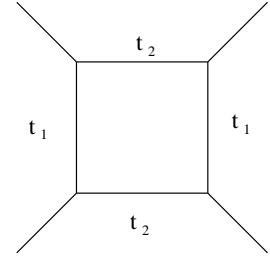}}
\caption{The toric diagram of local $\IP^1\times \IP^1$.}\label{Pone}
\end{figure}

3) {\it Local} $\IP^1 \times \IP^1$. The local $\IP^1 \times \IP^1$ geometry 
is the noncompact Calabi-Yau manifold given by the 
four-manifold $\IP^1 \times \IP^1$ together with its anticanonical bundle. 
It also admits a symplectic quotient description of the form (\ref{xquot}), this 
time with $N=2$ and two K\"ahler parameters $t_1$, $t_2$. The charges 
$Q_{1,2}^j$, $j=1, \cdots, 5$ can be grouped in two vectors 
\ben
Q_1& =& (-2,1,1,0,0), \nonumber\\
Q_2& =& (-2,0,0,1,1). 
\een
The toric diagram for this geometry can be easily worked out from 
this description, and it is represented in Fig. \ref{Pone}.  
Using the gluing rules we find the closed string partition function
\ben
Z_{\IP^1 \times \IP^1}& =&
\sum_{R_i}e ^{-(\ell(R_1)+ \ell(R_3)) t_1 - (\ell(R_2)+ \ell(R_4)) t_2}
q^{\sum_i \kappa_{R_i}/2} \nonumber\\
& & \times C_{0 R_4 R_1^t} C_{0 R_1 R_2^t} C_{0 R_2 R_3^t}
C_{0 R_3 R_4^t}.
\label{ponebis}
\een
This amplitude can be written as
\ben
Z_{\IP^1 \times \IP^1}& =&
\sum_{R_i}e^{-(\ell(R_1)+ \ell(R_3)) t_1 - (\ell(R_2)+ \ell(R_4)) t_2}
\nonumber\\ & & \times W_{R_4 R_1} W_{R_1 R_2} W_{R_2 R_3}
W_{R_3 R_4}.
\label{zpone}
\een
This is the expression first obtained by Aganagic, Mari\~no, and Vafa (2004), and it has
been shown to agree with Gromov-Witten theory by Zhou (2003). The local $\IP^1 \times \IP^1$
geometry is interesting since it gives a string realization of the Seiberg-Witten
solution (1994) of $SU(2)$, ${\cal N}=2$ Yang-Mills theory, as first shown by
Katz, Klemm and Vafa (1997). In particular, the prepotential of local $\IP^1 \times \IP^1$
gives, in a certain limit, the Seiberg-Witten prepotential. This was explicitly
verified by Iqbal and Kashani-Poor (2003a) by using the expression (\ref{zpone}) and
the results of Nekrasov (2002) for the Seiberg-Witten prepotential. Further applications of
the topological vertex to the computation of supersymmetric gauge theory amplitudes
can be found in Iqbal and Kashani-Poor (2003b), Eguchi and Kanno (2003) and Hollowood, Iqbal,
and Vafa (2003).

\subsection{Further properties of the topological vertex}

Since it was first introduced by Aganagic, Klemm, Mari\~no, and Vafa (2003), the topological
vertex has been
shown to satisfy three remarkable properties:
it has an underlying integrable structure (Aganagic, Dijkgraaf, Klemm, Mari\~no, and Vafa, 2003), it has a
natural combinatorial
interpretation in terms of counting tridimensional Young tableaux (Okounkov,
Reshetikhin, and Vafa, 2003), and it can
be also reinterpreted in terms of an appropriate counting of sheaves on ${\bf C}^3$
(Iqbal {\it et al.}, 2003; Maulik {\it et al.},
2003). We briefly review each of these properties.

1) {\it Integrable structure}. If we put ${\rm Tr}\, V^n_i =t_n^i/n$ in (\ref{zvertex}),
the resulting function of three infinite sets of ``times'' $Z(t_n^i)$ turns out to be
a tau function of the 3-KP hierarchy as constructed for example by Kac and van de
Leur (2003). This integrability
property is better understood in the context of mirror symmetry, where the
computation of the vertex
can be seen to reduce to a theory of free fermions in a Riemann surface
with three punctures (Aganagic, Dijkgraaf, Klemm, Mari\~no, and Vafa, 2003).

2) {\it Combinatorial interpretation}. Consider the problem of enumerating three-dimensional
Young tableaux $\pi$ (also called plane partitions) with the following boundary condition: along the
edges $x$, $y$, $z$ they end up in two-dimensional Young tableaux with the shapes
$R_1$, $R_2$ and $R_3$, respectively. Let us introduce the partition function
\be
{\cal C}_{R_1 R_2 R_3}=\sum_{\pi} q^{|\pi|},
\ee
where $|\pi|$ is the number of boxes in $\pi$, and the sum is over plane
partitions satisfying the above boundary conditions. It can be shown that, up to an overall
factor independent of the $R_i$, the above partition function equals the topological vertex
$C_{R_1 R_2 R_3}$ (Okounkov, Reshetikhin, and Vafa, 2003). This combinatorial
interpretation of the topological vertex
makes possible to establish a precise correspondence between quantum
topological strings on local, toric
Calabi-Yau manifolds and the classical statistical mechanics of
melting crystals (Okounkov, Reshetikhin, and Vafa, 2003; Iqbal {\it et al.}, 2003;
Saulina and Vafa, 2004).

3) {\it Relation to the counting of ideal sheaves}. Let $X$ be a Calabi-Yau threefold. An ideal 
sheaf ${\cal I}$ defines a closed subscheme $Y$ through ${\cal O}_Y ={\cal O}_X/{\cal I}$. This means, 
roughly speaking, that there is a subvariety $Y$ of $X$ defined by the zero locus 
of the equations that generate the ideal ${\cal I}$. 
Given a two-homology class $\beta$, one can consider the moduli space of ideal sheaves
$I_n(X, \beta)$ such that the holomorphic Euler characteristic of $Y$ is $n$ and with 
${\cal O}_Y$ supported on curves in the homology class $\beta$. This is a space of virtual dimension zero, and
by counting the number of points with appropriate signs one can define the so-called
Donaldson-Thomas invariant ${\widehat N}_{n, \beta}$. The Donaldson-Thomas partition
function is given by
\be
Z_{DT}(X)=\sum_{\beta}\sum_{n \in {\bf Z}} {\widehat N}_{n, \beta}\, Q^{\beta}(-q)^n,
\label{dtz}
\ee
where $q$ is interpreted here as a formal expansion parameter, and the notation for
$Q^{\beta}$ is identical to the one in (\ref{threepoint}).
Maulik {\it et al.} (2003) have shown that, when $X={\bf C}^3$ (so that $Z_{DT}$ only
depends on $q$), the Donaldson-Thomas partition function naturally depends on three
sets of representations, and agrees indeed with the topological vertex $C_{R_1 R_2 R_3}(q)$,
where $q$ in (\ref{dtz}) is identified with $e^{ig_s}$.
They have also shown that the Donaldson-Thomas partition
function satisfies the same gluing rules as the topological vertex, leading to the
identification of the Donaldson-Thomas partition function $Z_{DT}(X)$ with the
topological string all-genus partition function $Z(X)=e^F$ for all noncompact, toric
Calabi-Yau manifolds $X$. They also conjecture that the equality holds for all
Calabi-Yau threefolds. Iqbal {\it et. al.} rephrase the Donaldson-Thomas
partition function for ${\bf C}^3$ (which computes the topological vertex)
in terms of the counting of noncommutative $U(1)$ instantons in six dimensions. These
developments seem to indicate that the topological vertex, apart from
providing a powerful computational tool, plays a central logical role
in the theory of Gromov-Witten invariants and opens the way to connections to other
moduli problems in algebraic geometry.

\section{Conclusions and future directions}

The correspondence between Chern-Simons theory and topological strings provides one of the most
fascinating examples of the string theory/gauge theory correspondence. It has deep
mathematical
implications that hold a lot of promise for the theory of knot and link invariants, as well
as for the theory
of Gromov-Witten invariants. The physical point of view, which culminated in the
idea of the topological vertex,
has allowed to obtain a complete solution to topological string theory on a wide class of Calabi-Yau threefolds.
There are however many questions and problems that remain open and will no doubt give us
further insights on these
connections. We conclude the review with some of these problems.

1) From a mathematical point of view, some of the ingredients and results in the topological string
side need further study. For example, a rigorous construction of open Gromov-Witten invariants has not been
given yet, although the formal use of localization tecnhiques leads to sensible results in agreement
with the predictions of physics. One important problem is to derive the explicit expression for the
topological vertex (\ref{vertex}) in the context of Gromov-Witten theory. This will put the physical predictions on
a firmer ground. Some steps in this direction have been already taken by Diaconescu and Florea (2003), and Li {\it et al.} (2004) 
have given a mathematical 
treatment of the topological vertex by using relative Gromov-Witten invariants. 

2) The correspondence between knot invariants and open Gromov-Witten invariants on the resolved conifold
that was explained in V.G is still very much uncharted. Although there are proposals for Lagrangian
submanifolds in the resolved conifold associated to nontrivial knots (Labastida, Mari\~no, and Vafa, 2000; Taubes, 2001), no
results have been obtained for open Gromov-Witten invariants with those Lagrangian boundary
conditions. This correspondence is potentially very interesting
from a mathematical point of view, since it gives a dictionary between two important and very different
sets of invariants. It is likely that the unveiling of this correspondence will lead to deep results
in the theory of knot invariants.

3) It would be very interesting to see if Chern-Simons theory on other three-manifolds has a string
theory description as well, since this would lead in particular to a fascinating reformulation of the
theory of finite-type invariants which has been so vigorously developed in the last years. So far 
only small steps have been taken in this direction. The geometric
transition of Gopakumar and Vafa was extended to lens spaces by Aganagic, Klemm, Mari\~no, and Vafa (2004), and
further studied by Okuda and Ooguri (2004). One of the problems faced by
the extension of the correspondence to lens spaces is a typical one also encountered in the AdS/CFT correspondence:
the Chern-Simons side is easily computed for small 't Hooft coupling, while the topological string theory
side is better computed for large 't Hooft coupling. This makes the
comparison of observables a difficult task, and
in that respect more techniques are needed in order to evaluate the field theory and string theory results in other
't Hooft coupling regimes. The formulation of Chern-Simons theory in terms of a matrix model
given by Mari\~no (2002a), which
was very useful for the tests performed by Aganagic, Mari\~no, Klemm, and Vafa (2004), may be
also useful in doing Chern-Simons field theory computations at large 't Hooft coupling.

4) The connection between Gromov-Witten and Donaldson-Thomas invariants found
by Maulik {\it et al.} (2003) and
Iqbal {\it et al.} (2003) may shed a new light on many aspects of
Gromov-Witten theory, and seems to be a very promising avenue in the mathematical
understanding of the Gopakumar-Vafa invariants.

\section*{Acknowledgments}
I would like to thank Mina Aganagic, Vincent Bouchard, Robbert Dijkgraaf, Bogdan Florea,
Albrecht Klemm, Jose Labastida and
Cumrun Vafa for enjoyable collaborations on the topics
discussed in this review. I would also like to thank Vincent Bouchard, Brenno Carlini Vallilo and 
Arthur Greenspoon for a careful 
reading of the manuscript.

\appendix

\section{Symmetric polynomials}

In this brief Appendix we summarize some useful ingredients of the elementary
theory of symmetric functions. A standard
reference is Macdonald (1995).

Let $x_1, \cdots, x_N$ denote a set of $N$ variables. The {\it elementary
symmetric polynomials} in these variables, $e_m (x)$, are defined as:
\begin{equation}
\label{el}
e_m (x) =\sum_{i_1< \cdots <i_m} x_{i_1} \cdots x_{i_m}.
\end{equation}
The generating function of these polynomials is given by
\begin{equation}
\label{esgen}
E(t) = \sum_{m\ge 0} e_m(x) t^m =\prod_{i=1}^N (1 + x_i t).
\end{equation}
The {\it complete symmetric function} $h_m$ can be defined in terms of
its generating function
\begin{equation}
H(t)= \sum_{m\ge 0} h_m t^m = \prod_{i=1}^N (1 - x_i t)^{-1},
\end{equation}
and one has
\begin{equation}
E(t) H(-t)=1.
\end{equation}
The products of elementary symmetric polynomials and of complete symmetric functions
provide two different basis for the
symmetric functions of $N$
variables.

Another basis is given by the {\it
Schur polynomials}, $s_{R}(x)$, which are labelled by representations $R$. We will
always express these representations in terms of Young tableaux, so $R$ is given by
a partition $(l_1, l_2, \cdots, l_{c_R})$, where $l_i$ is the number
of boxes of the $i$-th row of the tableau, and we have
$l_1 \ge l_2
\ge \cdots \ge l_{c_R}$. The total number of boxes of a tableau will be
denoted by $\ell(R)=\sum_i l_i$.
The Schur polynomials are defined as quotients of determinants,
\begin{equation}
\label{schur}
s_{R}(x)={ {\rm det}\, x_j^{l_i + N-i} \over
{\rm det}\,  x_j^{ N-i} }.
\end{equation}
They can be written in terms of the
symmetric polynomials $e_i(x_1,\cdots,x_N)$, $i\ge 1$, as follows:
\be
\label{jt}
s_{R} = {\rm det} M_R
\ee
where
$$M_R^{ij}=(e_{l^t_i +j-i})$$
$M_R$ is an $r \times r$ matrix, with
$r= c_{R^t}$, and $R^t$ denotes the transposed Young tableau
with row lengths $l^t_i$.
To evaluate $s_{R}$ we put $e_0=1$, $e_k=0$
for $k<0$. The expression (\ref{jt}) is known
as the Jacobi-Trudy identity.

A third set of symmetric functions is given by the {\it Newton
polynomials} $P_{\vec k}(x)$. These are labelled by vectors
$\vec k=(k_1, k_2, \cdots)$, where the $k_j$ are nonnegative
integers, and they are defined as
\begin{equation}
\label{newt}
P_{\vec k}(x) = \prod_{j}^p P_j^{k_j}(x) ,
\end{equation}
where
\begin{equation}
\label{pos}
P_j(x) =\sum_{i=1}^N x_i^j,
\end{equation}
are power sums. The Newton polynomials are homogeneous of degree
$\ell=\sum_j jk_j$ and give a basis for the symmetric functions
in $x_1, \cdots, x_N$ with rational coefficients. They
are related to the Schur polynomials
through the Frobenius formula
\begin{equation}
\label{frobpol}
P_{\vec k}(x)= \sum_{R} \chi_{R}(C(\vec k)) s_{R}(x),
\end{equation}
where the sum is over all tableaux such that $\ell(R) = \ell$.

\bibliographystyle{myrmp}
\bibliography{all}

\end{document}